\documentclass{aa}  
\usepackage{natbib}
\usepackage{graphicx}
\usepackage[dvipsnames]{xcolor}
\usepackage{tikz}
\usepackage{subcaption}
\usepackage{gensymb}
\usepackage{titlesec} 
\usepackage{placeins} 
\usepackage{txfonts}
\usepackage{dblfloatfix} 
\newcommand{\as}{$^{\prime\prime}$}

\newcommand{\angstrom}{\mbox{\normalfont\AA}}
\usepackage{float}
\usepackage[hidelinks,colorlinks=true,linkcolor=blue,citecolor=blue]{hyperref}
\usepackage{multicol, blindtext}
\begin{document} 
\title{J-PLUS: Spectroscopic validation of H$\alpha$ emission line maps in spatially resolved galaxies}

   \author{Rahna, P. T.\thanks{rahna.payyasseri@gmail.com}\inst{\ref{CEFCA}}
          \and
          Akhlaghi, M. \inst{\ref{CEFCA}}
          \and 
          L\'opez-Sanjuan, C. \inst{\ref{CEFCA}}
          \and
          Logroño-García, R. \inst{\ref{CEFCA}}
          \and
          Muniesa, D. J.\inst{\ref{CEFCA}}
          \and          
          Domínguez-Sánchez, H.\inst{\ref{CEFCA}}
          \and
          Fernández-Ontiveros, J. A.\inst{\ref{CEFCA}}
          \and
          Sobral, David \inst{\ref{FFCUL},\ref{BNP}}
          \and
          Lumbreras-Calle, A. \inst{\ref{CEFCA}}
          \and           
          Chies-Santos, A. L.\inst{\ref{UFRGS}} 
          \and          
          Rodríguez-Martín, J. E. \inst{\ref{IAA}}
          \and
          Eskandarlou, S.\inst{\ref{CEFCA}}
          \and
          Ederoclite, A.\inst{\ref{CEFCA}}
          \and
          Alvarez-Candal, A.\inst{\ref{IAA}}
        \and 
        H.~V\'azquez Rami\'o\inst{\ref{CEFCA}}
\and A.~J.~Cenarro\inst{\ref{CEFCA}}
\and A.~Mar\'{\i}n-Franch\inst{\ref{CEFCA}}
\and J.~Alcaniz\inst{\ref{ON}}
\and R.~E.~Angulo\inst{\ref{DIPC},\ref{ikerbasque}}
\and D.~Crist\'obal-Hornillos\inst{\ref{CEFCA}}
\and R.~A.~Dupke\inst{\ref{ON},\ref{MU},\ref{Alabama}}
\and C.~Hern\'andez-Monteagudo\inst{\ref{IAC},\ref{ULL}}
\and M.~Moles\inst{\ref{CEFCA}}
\and L.~Sodr\'e Jr.\inst{\ref{USP}}
\and J.~Varela\inst{\ref{CEFCA}}
          }

\institute{Centro de Estudios de F\'{\i}sica del Cosmos de Arag\'on (CEFCA), Plaza San Juan 1, 44001 Teruel, Spain\label{CEFCA}
\\
\email{rahna.payyasseri@gmail.com}
\and
Departamento de F\'isica, Faculdade de Ci\^encias, Universidade de Lisboa, Edif\'icio C8, Campo Grande, PT1749-016 Lisbon, Portugal\label{FFCUL}
\and
BNP Paribas Corporate \& Institutional Banking, Torre Ocidente Rua Galileu Galilei, 1500-392 Lisbon, Portugal\label{BNP}
\and
 Instituto de F\'isica, Universidade Federal do Rio Grande do Sul (UFRGS), Av. Bento Goncalves, 9500 Porto Alegre, RS, Brazil\label{UFRGS}
\and
Instituto de Astrofísica de Andalucía, CSIC, Apt 3004, E18080 Granada, Spain\label{IAA}
\and
Observat\'orio Nacional - MCTI (ON), Rua Gal. Jos\'e Cristino 77, S\~ao Crist\'ov\~ao, 20921-400 Rio de Janeiro, Brazil\label{ON}
\and
Donostia International Physics Centre (DIPC), Paseo Manuel de Lardizabal 4, 20018 Donostia-San Sebastián, Spain\label{DIPC}
\and
IKERBASQUE, Basque Foundation for Science, 48013, Bilbao, Spain\label{ikerbasque}
\and
University of Michigan, Department of Astronomy, 1085 South University Ave., Ann Arbor, MI 48109, USA\label{MU}
\and
University of Alabama, Department of Physics and Astronomy, Gallalee Hall, Tuscaloosa, AL 35401, USA\label{Alabama}
\and
Instituto de Astrof\'{\i}sica de Canarias, La Laguna, 38205, Tenerife, Spain\label{IAC}
\and
Departamento de Astrof\'{\i}sica, Universidad de La Laguna, 38206, Tenerife, Spain\label{ULL}
\and
Instituto de Astronomia, Geof\'{\i}sica e Ci\^encias Atmosf\'ericas, Universidade de S\~ao Paulo, 05508-090 S\~ao Paulo, Brazil\label{USP}  
     }


  \abstract
   {}
   {We present a dedicated automated pipeline to construct spatially resolved emission H$\alpha$+[NII] maps and to derive the spectral energy distributions (SEDs) in 12 optical filters (five broad and seven narrow and medium) of H$\alpha$ emission line regions in nearby galaxies (z $<$  0.0165) observed by the Javalambre Photometric Local Universe Survey (J-PLUS).}
   {We used the $J0660$ filter of $140$ \AA~width centered at $6600$ \AA~to trace H$\alpha$ + [NII] emission, and $r$ and $i$ broad bands were used to estimate the stellar continuum. We created pure emission line images after the continnum subtraction, where the H$\alpha$ emission line regions were detected. This method was also applied to integral field unit (IFU) spectroscopic data from PHANGS-MUSE, CALIFA, and MaNGA surveys by building synthetic narrow bands based on J-PLUS filters. The studied sample includes the cross-matched catalog of these IFU surveys with the J-PLUS third data release (DR3), amounting to two PHANGS-MUSE, $78$ CALIFA, and $78$ MaNGA galaxies at $z < 0.0165$, respectively.}
   {We compared the H$\alpha$+[NII] radial profiles from J-PLUS and the IFU surveys, finding good agreement within the expected uncertainties. We also compared the SEDs from the emission line regions detected in J-PLUS images, reproducing the main spectral features present in the spectroscopic data. Finally, we compared the emission fluxes from the J-PLUS and IFU surveys accounting for scale differences, finding a difference of only 2\% with a dispersion of 7\% in the measurements.}
   {The J-PLUS data provide reliable spatially resolved H$\alpha$+[NII] emission maps for nearby galaxies. We provide the J-PLUS DR3 catalog for the $158$ galaxies with IFU data, including emission maps, SEDs of star-forming clumps, and radial profiles.}


   \maketitle
%

\section{Introduction}
Spatially resolved studies of galaxies have significantly advanced our understanding of galaxy formation and evolution. Unlike previous techniques that gather an integrated flux from entire galaxies, spatially resolved studies reveal local variations in different regions of galaxies, allowing the detailed understanding of different characteristics such as stellar population properties, star formation, gas kinematics, and the chemical composition in galaxies. In particular, spatially resolving galaxies has provided new opportunities for detailed emission line studies in galaxies. For example, H$\alpha$ (6563 \AA) is a prominent emission line in the optical, used as an important star formation tracer  \citep[$<$ 10 Myr,][]{Kennicutt1998}, which is produced from hydrogen recombination in nebular regions ionized by the radiation field of young, short-lived massive stars, and also due to the contribution from active galactic nuclei (AGN). Spatially resolved H$\alpha$ emission line maps (hereafter "emap") are crucial for understanding the star formation activity, gas dynamics, and feedback processes in galaxies. 

Narrow band (NB) imaging and integral field spectroscopy (IFS) are two key techniques that have allowed spatially resolved emission line studies of galaxies. The NB technique dedicated to H$\alpha$ imaging surveys \citep[e.g.,][]{2004James, 2013Sobral, 2015Stroe, 2015Sobral} has greatly improved our understanding of spatially resolved information on star formation in galaxies. However, these post-NB surveys were usually limited to a few square degrees or a selected number of galaxies (e.g., 100 to 1000).

The introduction of IFS in astronomy has been a game changer for spatially resolved studies of galaxies, enabling astronomers to simultaneously gather spectra from different regions of a galaxy in the form of 3D data cubes. The TIGER (Traitement Intégral des Galaxies par l’Étude de leurs Raies) instrument \citep{1995Bacon} developed at the Observatoire de Lyon, France mounted on the Canada-France-Hawaii Telescope (CFHT) was the first instrument that demonstrated the power of IFS in astronomy. Then, more advanced optical integral field unit (IFUs) such as the Spectroscopic Areal Unit for Research on Optical Nebulae \citep[SAURON;][]{2000Miller}, VIsible MultiObject Spectrograph \citep[VIMOS;][]{2003Oliver}, Sydney-AAO Multi-object Integral-field spectrograph \citep[SAMI;][]{2012Croom}, The Multi Unit Spectroscopic Explorer \citep[MUSE;][]{2014Bacon}, Mapping Nearby Galaxies at Apache Point Observatory \citep[MaNGA;][]{2015Bundy}, and The PPaK Calar Alto Legacy Integral Field Area Survey \citep[CALIFA;][]{2012Sanches} were introduced into the field and these have significantly improved our understanding of galaxies, nebula, and star-forming regions by offering a resolved perspective that allows for detailed analysis of their internal structures and physical processes. 
The earlier IFS systems such as SAURON or TIGER were lenslet-based, MaNGA and PPaK/CALIFA are fiber-based IFS, while MUSE is image slicer-based. Recently, IFUs have expanded to space-based missions such as the Near-Infrared Spectrograph \citep[NIRSpec;][]{2022Jakobsen} on the James Webb Space Telescope (JWST), enabling unprecedented high-resolution spectroscopy of distant galaxies. These IFSs have studied H$\alpha$ emission line maps of thousands of galaxies that allow for statistical studies of galaxy evolution, environmental effects, and the relationship between star formation and galaxy morphology \citep{2019Novais,2023Barrera-Ballesteros}. However, these IFUs are time-expensive and limited due to the small field of view (FoV) and limited number of observed targets. 

Multiband photometric surveys such as the Advanced Large, Homogeneous Area Medium-Band Redshift Astronomical survey \citep[ALHAMBRA;][]{2008Moles} covering 4 deg$^{2}$ and Physics of the Accelerating Universe \citep[PAU;][]{2009Ben} covering 100 deg$^{2}$ have significantly contributed to wide-field surveys and bridge the gap between broad-band photometry and IFS with their multiple filters.
The Javalambre Photometric Local Universe Survey \citep[J-PLUS;][]{2014Cenarro, 2019Cenarro} and Javalambre-Physics of the Accelerating universe astrophysical Survey \citep[J-PAS;][] {2014Benitez, 2016Dupke} are the two ongoing optical surveys in the northern sky at the Observatorio Astrofísico de Javalambre (OAJ).
The large FoV, high spatial resolution, and contiguous NB filters covering a wide wavelength range (3300-11000 \AA) of J-PAS act as a low-resolution IFU (R $\sim$ 60), which is suitable for IFU-like science such as spatially resolved studies of galaxies planned over 8500 deg$^{2}$ (by the end of 2030). J-PLUS includes some of the narrow bands from J-PAS covering prominent emission lines such as [OII]$\lambda 3725$ and H$\alpha$+[NII]$\lambda \lambda 6548, 6583$ at z<0.0165 and also [OIII]$\lambda 5007$ at 0.006<z<0.0473 \citep{2022Lumbreras-Calle} allowing us to study the spatially resolved properties of galaxies in these spectral lines. In comparison to the FoV of other time-expensive IFUs such as MaNGA (12\as - 32\as), SAMI (15\as), PPaK/CALIFA (60\as), and MUSE (60\as), the uniqueness of J-PLUS is its large FoV ($1.4^\circ$) offering a large contiguous observing area to understand the resolved structure of galaxies, galaxy groups, galaxy clusters, and thus the ability to probe their environment. 

The method of measuring H$\alpha$ fluxes and star formation rates (SFRs) from J-PLUS photometry was first introduced by \cite{2015Vilella-Rojo} and \cite{2019Garcia}. \cite{2021Vilella-Rojo} applied their method to estimate the star formation main sequence in the local Universe using J-PLUS DR1. Here, we aim to extend that work to validate the H$\alpha$ flux extraction for spatially resolved studies using J-PLUS DR3. We selected galaxies with redshifts up to z $\leq$ 0.0165, ensuring that the J0660 NB filter covers the H$\alpha$ emission line within this range. This leads to thousands of galaxies at D $<$70 Mpc within the 3192 deg$^{2}$ observed area of J-PLUS DR3. At this redshift range, J-PLUS is able to resolve star-forming regions on the physical scale of several hundred parsecs ($\sim$ 168 pc).
We validated our method of building the emaps by comparing them with other well-known IFU observations of the same galaxies. This is the first of a series of works where we aim to use the H$\alpha$ map to study the spatially resolved star formation main sequence of J-PLUS galaxies and ultraviolet (UV) and H$\alpha$ studies of J-PLUS galaxies.

This paper is organized as follows. In Sect.~\ref{sec:data} we detail the J-PLUS, PHANGS-MUSE, CALIFA, and MaNGA data. Sect.~\ref{sec:sample} describes the sample selection. Our methods and the pipeline developed are described in Sect.~\ref{sec:methods}, and the results are discussed in Sect.~\ref{sec:results}. Our final conclusions are given in Sect.~\ref{sec:conclusions}.  

\section{Data} \label{sec:data}
In this section, we discuss various data sets used for this study. We chose three different IFU surveys (CALIFA, MaNGA and PHANGS-MUSE) for the comparison of our method with J-PLUS. CALIFA and MaNGA were selected for their large number of local galaxies and PHANGS-MUSE was selected for its higher resolution and depth. More details about these datasets are given below. 
\subsection{J-PLUS DR3}
\begin{table}
\caption{Filter characteristics of J-PLUS}
    \centering
    \begin{tabular}{llllll}
    \hline \hline

 \textbf{Name} & \textbf{$\lambda_{cent}$} & \textbf{$\lambda_{pivot}$} & \textbf{FWHM} & \textbf{Spec.line} & m$_{lim} ^{(a)}$\\
 & (\AA) & (\AA) & (\AA) & & (mag) \\ \hline
        uJAVA & 3485 & 3523 & 508 & – & 20.8\\ 
        J0378 & 3785 & 3786 & 168 & [OII] & 20.8\\ 
        J0395 & 3950 & 3951 & 100 & CaH+K & 20.8 \\ 
        J0410 & 4100 & 4101 & 200 & H$\delta$  & 21\\
        J0430 & 4300 & 4300 & 200 & – & 21\\ 
        gSDSS & 4803 & 4745 & 1409 & – & 21.8\\ 
        J0515 & 5150 & 5150 & 200 & [OIII] & 21\\ 
        rSDSS & 6254 & 6230 & 1388 & – & 21.8 \\ 
        J0660 & 6600 & 6600 & 138 & H$\alpha$+[NII] & 21 \\ 
        iSDSS & 7668 & 7677 & 1535 & – & 21.3\\ 
        J0861 & 8610 & 8603 & 400 & Ca Triplet & 20.4 \\ 
        zSDSS & 9114 & 8922 & 1409 & - & 20.5 \\ \hline
    \end{tabular}
\footnotesize{Note: $^{(a)}$ J-PLUS DR3 typical 5$\sigma$ depths on 3\as aperture \citep{2024lopez}}
\label{tab:jplus}
\end{table}
\begin{table*}[h]
\caption{Comparison of the instrument properties of different IFU surveys with J-PLUS.}
\resizebox{\textwidth}{!}{%
\begin{tabular}{lllll}
\hline\hline
\textbf{Parameters} & \textbf{J-PLUS} & \textbf{eCALIFA} & \textbf{MaNGA} & \textbf{PHANGS-MUSE}  \\
\hline
Telescope (m) & 0.8 & 3.5 & 2.5 & 8   \\

Wavelength (\AA) & 3000 -10000 & 3700 - 7500 & 3600 - 10000 & 4750 - 9350 \\
Spatial Sampling ("/pixel) & 0.55 & 0.5 & 0.5 & 0.2 \\
Spectral Sampling (\AA/slice) & -- & 2 & 1 & 1.25  \\
PSF (") & 1.1 & 1 & 2.54 & 0.7 -1 \\
FoV Area (arcsec²) & 25,401,600 ({$1.4^\circ$}$\times$ {$1.4^\circ$}) & 3552 (74" $\times$ 64") & 93.5 (12") to 665.8 (32") &   3600 (59.9" $\times$ 60.0")  \\

Spectral Resolution (\angstrom) & 60 (for H$\alpha$) & 850 & 2000  & 1770 -- 3590 \\
\hline

\end{tabular}%
}

\label{tab:IFUs_inst}
\end{table*}
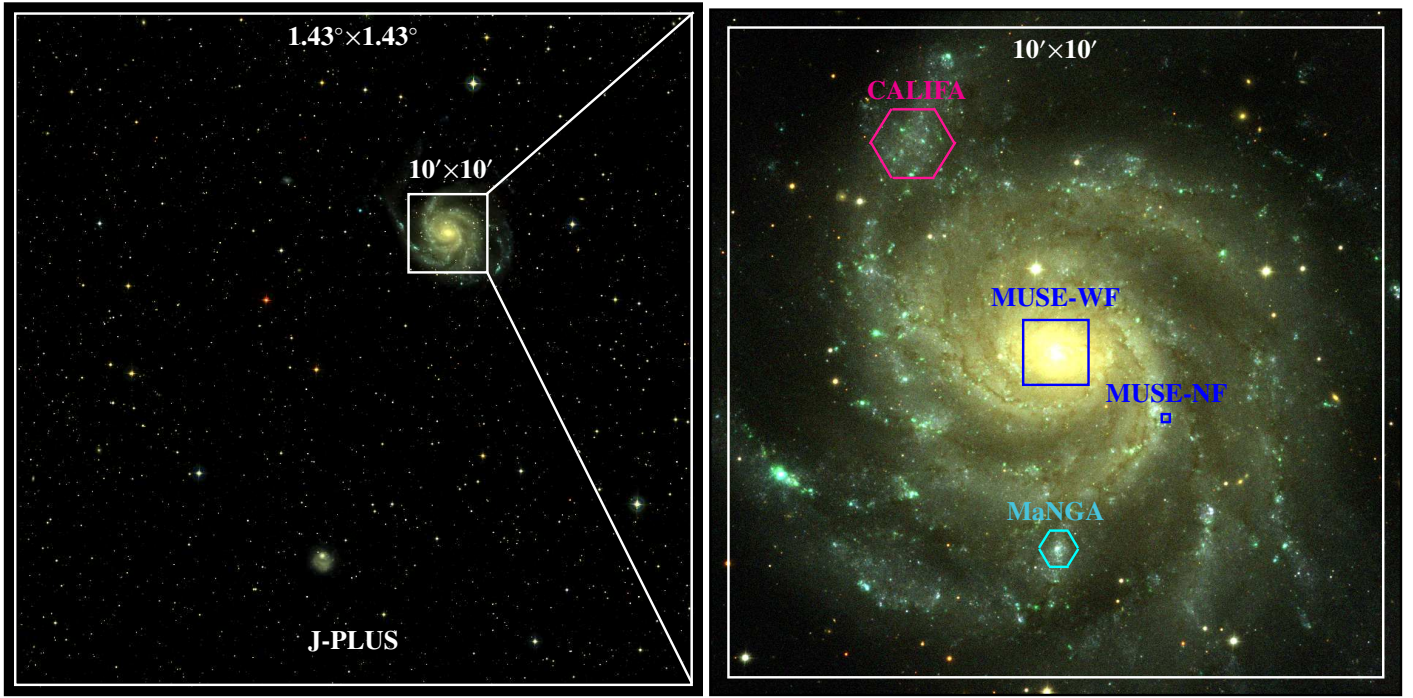
\begin{figure*}[htbp]
    \centering
    \begin{subfigure}[b]{0.5\textwidth}
        \centering
        \begin{tikzpicture}
            \node[anchor=south west,inner sep=0] (image) at (0,0) {\includegraphics[width=\textwidth]{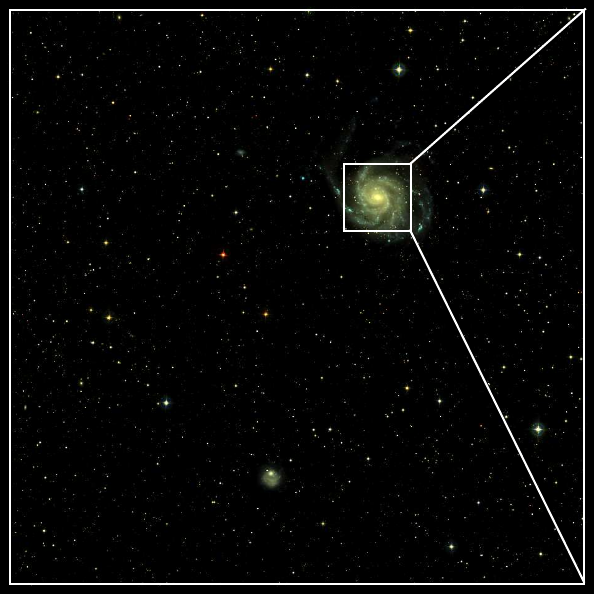}};
            \begin{scope}[x={(image.south east)},y={(image.north west)}]
\node[font=\bfseries, text=white] at (0.5,0.95) {\textbf{\textcolor{white}{1.43\degree$\times$1.43\degree}}};
\node[font=\bfseries, text=white] at (0.5,0.08) {\textbf{\textcolor{white}{J-PLUS}}};
\node[font=\bfseries, text=white] at (0.64,0.76) {\textbf{\textcolor{white}{10$^{\prime}\times$10$^{\prime}$}}};
            \end{scope}
        \end{tikzpicture}
        \label{fig:fov1}
    \end{subfigure}
    \hfill
    \begin{subfigure}[b]{0.4956\textwidth}
        \centering
        \begin{tikzpicture}
            \node[anchor=south west,inner sep=0] (image) at (0,0) {\includegraphics[width=\textwidth]{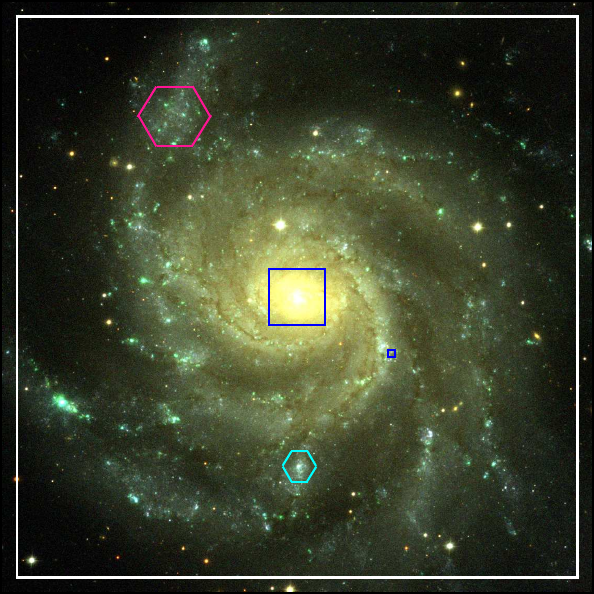}};
            \begin{scope}[x={(image.south east)},y={(image.north west)}]
        \node[font=\bfseries, text=white] at (0.5,0.94) {\textbf{\textcolor{white}{10$^{\prime}\times$10$^{\prime}$}}};        
        \node[font=\bfseries, text=blue] at (0.5,0.58) {\textbf{\textcolor{blue}{MUSE-WF}}};
        \node[font=\bfseries, text=white] at (0.66,0.44) {\textbf{\textcolor{blue}{MUSE-NF}}};
        \node[font=\bfseries, text=white] at (0.3,0.88) {\textbf{\textcolor{magenta}{CALIFA}}};
        \node[font=\bfseries, text=white] at (0.5,0.27) {\textbf{\textcolor{SkyBlue}{MaNGA}}}; 
            \end{scope}
        \end{tikzpicture}
        \label{fig:fov2}
    \end{subfigure}
    \caption{Comparison of J-PLUS FoV for a single pointing with other IFUs. Left panel: The full FoV of J-PLUS (pointing 02433 including M101). Right panel: 10$^{\prime}$ $\times$ 10$^{\prime}$ field of J-PLUS observation of M101 and the FoVs of MUSE in its wide-field mode (59.9$^{\prime\prime}$ $\times$ 60$^{\prime\prime}$), and narrow-field mode(7.42$^{\prime\prime}$ $\times$ 7.43$^{\prime\prime}$), CALIFA (78$^{\prime\prime}$) and MaNGA (32$^{\prime\prime}$) are marked.}
    \label{fig:fov}
\end{figure*}

\begin{figure}
    \centering
    \includegraphics[width=9cm]{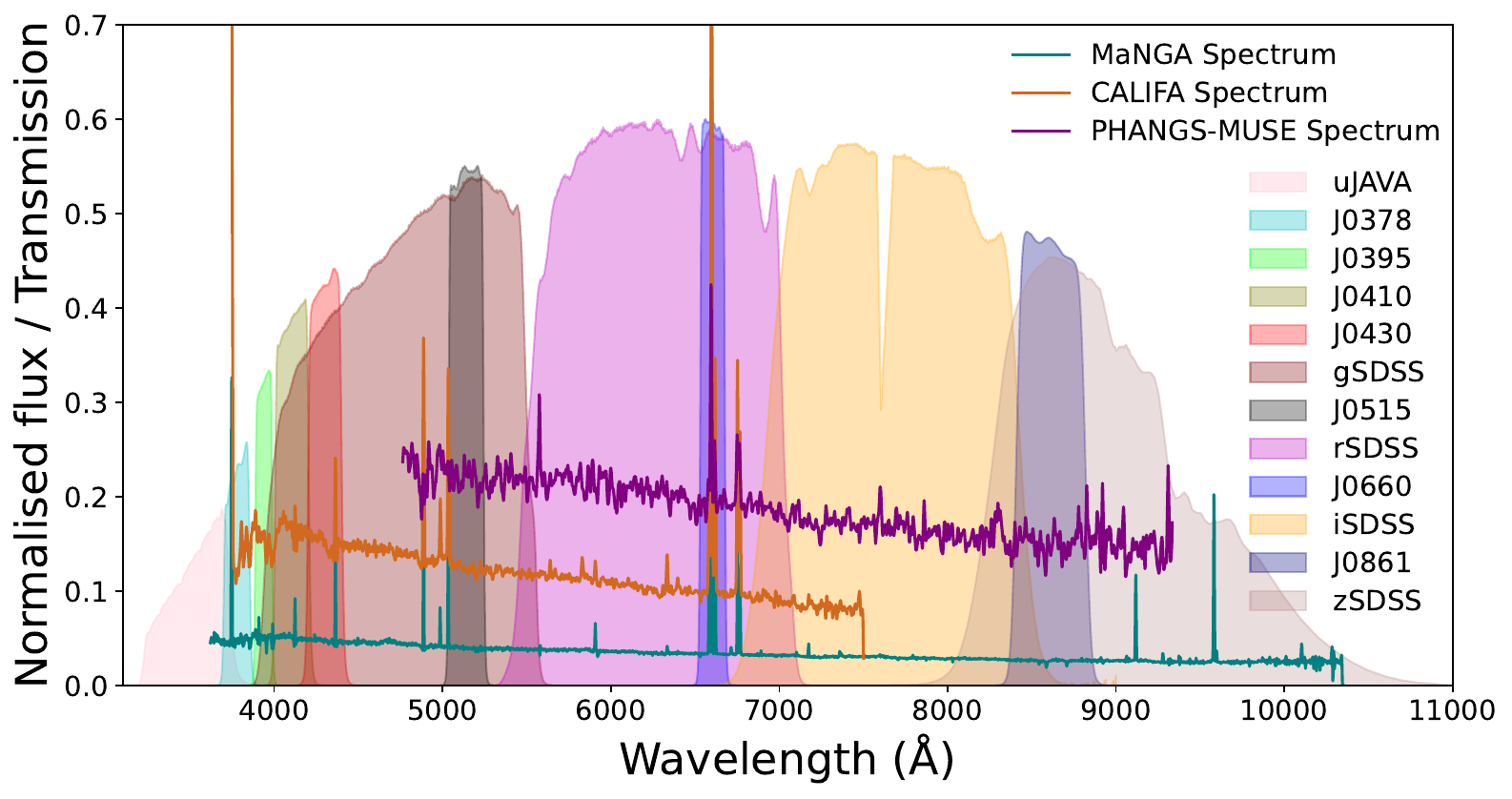}
    \caption{The Filter curves of J-PLUS compared to the spectra of PHANGS-MUSE (NGC1087 at 307, 457 xy position), CALIFA (NGC3395 spectra at 79, 72 xy position), and MaNGA (MaNGA--8150-6103 spectra at 26, 27 xy position).}
    \label{fig:jfilt-ifuspec}
\end{figure} 

\begin{figure*}
    \centering
    \includegraphics[width=15cm]{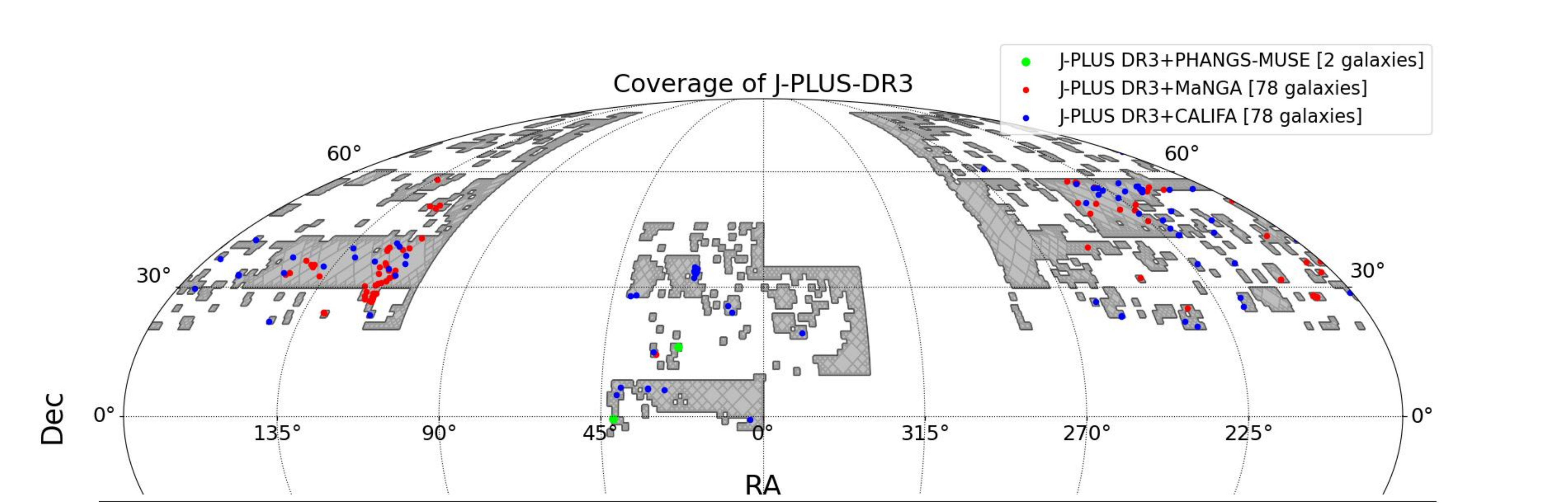}
    \caption{Coverage of J-PLUS DR3 overplotted with the distribution of cross matched catalogs of J-PLUS DR3 with PHANGS-MUSE, CALIFA, and MaNGA.}
    \label{fig: xmatch}
\end{figure*}

\begin{figure*}
    \centering
    \includegraphics[width=12cm]{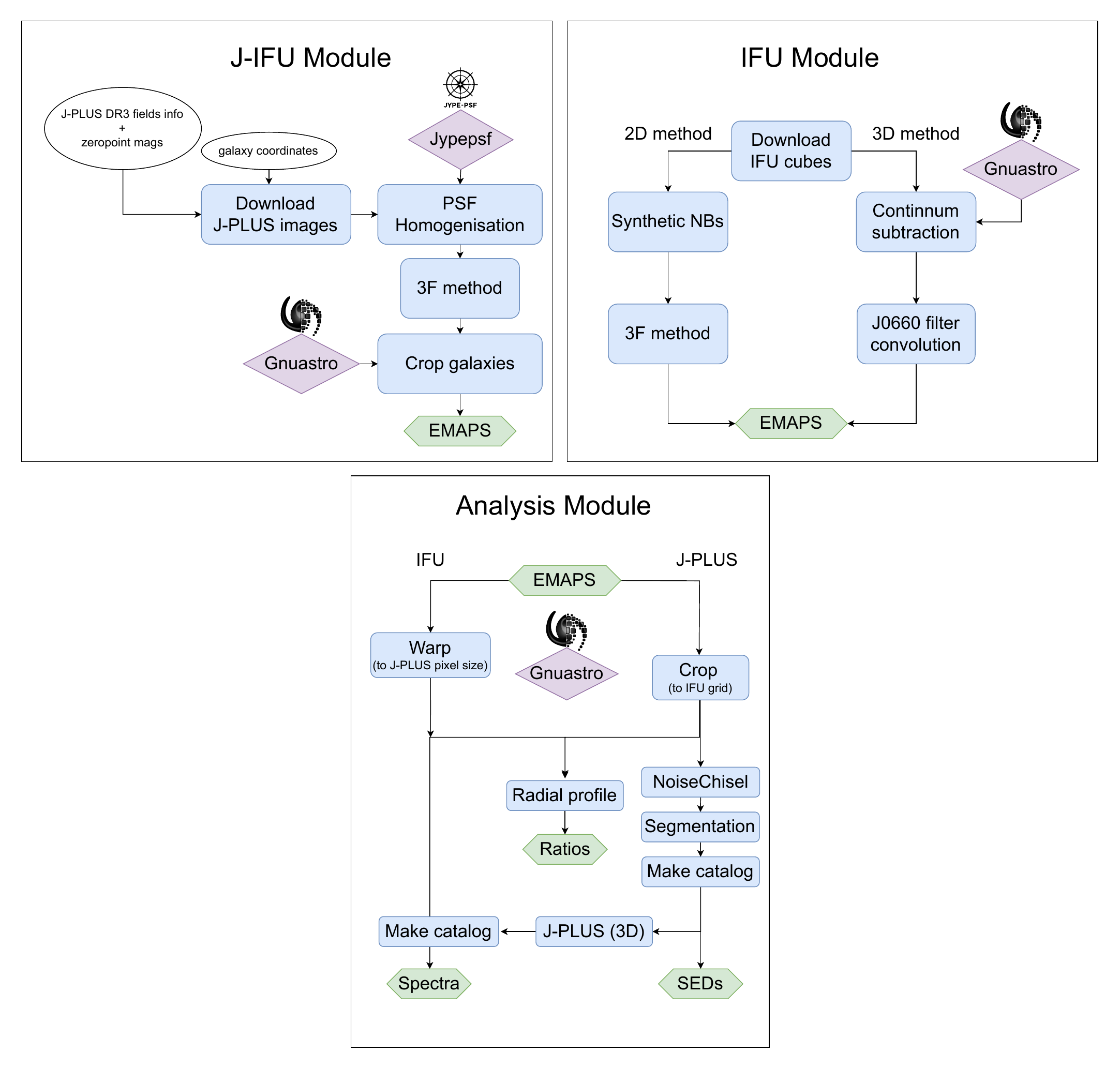}
    \caption{Flowchart of the J-SHE modules. Top left: J-IFU module. Top right: IFU module. Bottom: Analysis module. White ellipses are inputs, purple blocks are external packages, blue boxes indicate different processes, and green hexagons represent outputs.}
    \label{fig: pip}
\end{figure*}

J-PLUS \citep{2014Cenarro, 2019Cenarro} is a multi band photometric survey designed to observe the northern sky in several thousand square degrees using the 80cm Javalambre Auxiliary Survey Telescope (JAST80) at OAJ, Teruel. It has 12 broad, intermediate, and narrow band optical filters with 2 deg$^{2}$ FoV. The properties of J-PLUS filters are given in Table~\ref{tab:jplus}. After two data releases (DR1 on July 2018, and DR2 on July 2020), J-PLUS had its third data release in July 2022, containing 1642 J-PLUS fields covering 3192 deg$^{2}$. The pixel scale of J-PLUS is 0.555\as /pixel. The point spread function (PSF) of J-PLUS observation is $\sim$1.1 arcsecs in the r band. The DR3 images that we have used here are available from the OAJ database \footnote{https://archive.cefca.es/catalogues/jplus-dr3}. The calibration and zeropoints are taken from \cite{2019Sanjuan, 2024lopez}.

\subsection{PHANGS-MUSE}
The MUSE \citep{2010Bacon} is an IFS at the Very Large Telescope (VLT) of the European Southern Observatory (ESO) Paranal Observatory, Chile. It is composed of 24 identical integral-field units (IFUs). In its wide field mode (WFM; FoV of $1^{\prime}\times1^{\prime}$) 90000 spectra are taken with a spatial resolution of 0.2\as /pixel covering a wavelength range of 4750 - 9350 \angstrom~. The spectral resolution of MUSE is 2.5 \angstrom~ at 7000 \angstrom ~and the spectral sampling of 1.25\angstrom~ /slice. 

The PHANGS (Physics at High Angular resolution in Nearby GalaxieS) is the survey designed to observe nearby galaxies with high spatial resolution utilizing various telescopes, such as ALMA \citep{2021Leroy}, Hubble \citep{2022Lee}, VLT/MUSE, and JWST \citep{2023Lee} for the multiscale, multiphase study of star-formation and the interstellar medium.
The PHANGS-MUSE survey \citep{2021Kreckel} is a program with MUSE to map 19 nearby (D$\le$20 Mpc), massive star-forming galaxies. It provides their central star-forming disk by combining 5 to 15 MUSE pointings. 
This was the first IFS view of nearby star formation in different local environments. The PHANGS-MUSE observations were carried out in WFM mode. The PSF of the PHANGS-MUSE observations is between 0.7\as and 1\as. The PHANGS-MUSE pipeline-processed data cubes are downloaded from the ESO archive\footnote{https://archive.eso.org/scienceportal/home?data\_collection=PHANGS}. The MUSE data are not corrected for foreground Galactic extinction.
\subsection{CALIFA}
CALIFA \citep{2012Sanches} is a large project to obtain spatially resolved spectra of $\sim$600 nearby galaxies with a size of 64$^{\prime\prime}$ (diameter) in the redshift range of 0.005 $<$ z $<$ 0.03 using a PMAS/PPAK integral field spectrophotometer, mounted on the Calar Alto 3.5 m telescope. It utilises an hexagonal field-of-view (FoV) of $74 ^{\prime\prime} \times 64^{\prime\prime}$. It has two observing modes: Intermediate spectral resolution (V1200, R$\sim$1550) and low resolution (V500, R$\sim$850). The CALIFA observations were carried out between 2010 and 2016.

We used version 2.3 from the extended data release (eDR) of CALIFA \citep{2023Sanchez} that includes 895 galaxies. This version has significantly improved spatial resolution and image quality. The PSF is improved from $\sim$2.4 arcsec to $\sim$1 arcsec of full width at half-maximum (FWHM). The configuration includes the low-resolution setup (V500, R$\sim$850), integration time of 900 s per pointing, with a wavelength range of 3745 and 7500 \angstrom~. The spatial sampling of the eDR CALIFA data is 0.5$^{\prime\prime}$/pixels with a wavelength resolution of 2\angstrom~/slice.
The eCALIFA cubes are downloaded from the CALIFA website \footnote{https://ifs.astroscu.unam.mx}. The downloaded CALIFA cubes are already corrected for Galactic extinction using the method by \citet{1998Schlegel} and \citet{1989Cardelli}.

\subsection{MaNGA}
MaNGA \citep{2015Bundy} is a program under the fourth phase of the Sloan Digital Sky Surveys \citep[SDSS-IV;][]{2017Blanton}. MaNGA has mapped the kinematics and composition of 10,100 nearby galaxies in the redshift range of 0.01 $<$ z $<$ 0.15 with an average of z$\sim$0.03 using 17 hexagonal shaped IFUs mounted on Sloan Foundation 2.5m telescope at Apache Point Observatory from 2014 to 2020. The IFU size ranges from 12 arcsecs (19 fibers) diameter to 32 arcsecs (127 fibers) with a wavelength coverage of 3600 - 10000 \angstrom~across $\sim$2700 deg$^{2}$ and a spectral resolution of R$\sim$2000. The medium spatial resolution of MaNGA is $\sim$ 2.54$^{\prime\prime}$. The pixel scale of MaNGA is 0.5\as /pixel with a wavelength resolution of 0.834 \angstrom~/slice.
 The MaNGA cubes are downloaded from DR17 archive\footnote{https://data.sdss.org/sas/dr17/manga/spectro/redux/v3\_1\_1}. The MaNGA fluxes are not corrected for Galactic extinction.

The comparison of the instrumental properties of MUSE, MaNGA, and CALIFA with J-PLUS is given in Table~\ref{tab:IFUs_inst} and Figure~\ref{fig:fov} clearly demonstrate the capability of J-PLUS's wide field compared to these IFUs to cover the complete size of nearby galaxies and their surrounding environment. Additionally, J-PLUS offers an advantage in wavelength coverage (compared to CALIFA and PHANGS), as illustrated in Fig.~\ref{fig:jfilt-ifuspec}, which presents J-PLUS filter curves alongside spectra from PHANGS-MUSE (NGC1087), CALIFA (NGC3395), and MaNGA (MaNGA-8150-6103) galaxies for spectral coverage comparison.

\section{Sample selection} \label{sec:sample}
The sample of nearby galaxies was selected from the J-PLUS DR3 survey at redshift of z $<$ 0.0165 (or about 70 Mpc) based on the width ($\sim$140 Å) of the $J0660$ filter centered at 6600\AA~ (covering H$\alpha$ 6562.8 \AA). For the validation of our method of building the H$\alpha$ emaps with spectroscopy we selected a sample based on the cross-matched catalogs of MUSE-PHANGS, CALIFA, and MaNGA with J-PLUS DR3. Based on their IFU galaxy spectra, we have verified that the redshift range of all the selected galaxies ensures that the H$\alpha$ emission line is completely covered by this filter. The [N II] 6583 \AA~line is outside the $J0660$ filter at z > 0.014. 

Within PHANGS-MUSE, two galaxies overlap with J-PLUS DR3: NGC0628 at z=0.0056 and NGC1087 at z=0.00219. The full disk of the galaxy NGC 1087 has been observed in MUSE with 6 pointings, and the center of NGC 628 is observed in MUSE with 12 pointings. The total exposure time of MUSE observation of NGC628 is 36750 seconds (10.2 hours) and 15957.46 seconds (4.4 hours) for NGC1087.

There are 176 galaxies cross-matched with J-PLUS DR3 and eCALIFA, with 78 galaxies in the redshift range of z$<$0.0165. Among them, 20 galaxies are early-type (E/SO), nine galaxies show dust-lane features, and 48 galaxies show clear evidence of star formation activity (based on GALEX-UV and H$\alpha$ + [NII] images) in the visual morphology and exhibit H$\alpha$ emission line at 6562.8 \AA~rest frame wavelength in CALIFA spectra. Among the 48 galaxies, the [NII]$\lambda 6583$ line lies outside the filter window for redshifts greater than 0.014 for 10 galaxies. Above redshifts greater than 0.0165, the H$\alpha$ line begins to fall outside the filter curve, and galaxies with such redshifts were excluded from the sample.

From the 10,000 MaNGA galaxies, 3,509 galaxies are in the J-PLUS DR3. Within a redshift range of z $<$0.0165, the final cross-matched catalog amount to 78 J-PLUS-MaNGA galaxies. Among these, 65 galaxies have H$\alpha$ emission line (i.e., star- forming) in their MaNGA spectra and 13 galaxies do not show H$\alpha$ line. Seven galaxies are with the morphology of early type (E/S0). From the MaNGA spectra, we noticed that for 38 galaxies, the [NII] line is partially or completely outside the filter curve at z>0.014 and 30 galaxies have only the H$\alpha$ line within the filter curve, while the [NII] line is completely outside the filter curve.

The total sample includes 158 galaxies with 3D data cubes from either of the aforementioned IFUs. For the final analysis, we used 115 galaxies that exhibit H$\alpha$ line in their spectra and have coverage of the full H$\alpha$ line within the J0660 filter curve. Fig.~\ref{fig: xmatch} shows the coverage of J-PLUS DR3 with the distribution of the final sample selected from the cross-matched catalog. The information of the final cross-matched sample is given in Table A1. 
\section{The J-SHE Pipeline for building emaps of J-PLUS and IFUs} \label{sec:methods}
In this section, we describe the methodology followed to build the emaps and radial profile of galaxies, spectral energy distributions (SEDs), and spectra of emission line regions from J-PLUS and IFUs. The J-PLUS Spatially resolved H$\alpha$ Emission line (\texttt{J-SHE}) pipeline is written in \textsc{GNU Make} mainly utilizing GNU Astronomy Utilities \citep[Gnuastro;][]{2015Akhlaghi} and Python packages. Gnuastro is an official GNU package for the manipulation and analysis of astronomical data. Our pipeline consists of three modules (see Fig.~\ref{fig: pip}). The first and second modules are dedicated to building emaps for J-PLUS (hereafter, J-IFU module) and IFUs (hereafter, IFU module) respectively. The third module generates the galaxy radial profiles, SEDs, and spectra of emission line regions from the emaps (hereafter, analysis module). 

\subsection{J-IFU module} \label{sec:jifumodule}
The J-IFU module builds the H$\alpha$ +[NII] emaps of J-PLUS galaxies. The input of the pipeline are the J-PLUS observation field tile information (tile id and zero points), and galaxy coordinates (RA and Dec). The initial step in the pipeline involves downloading the J-PLUS images of the tile in $12$ filters that contain the requested galaxies. This is followed by PSF homogenization on all images of the field. This process is done using \texttt{JYPE-PSF}, which is described in the next section. After the PSF homogenization, we construct the emaps by following the recipe of three filter (3F) method (see Sect. \ref{sec:3Fmethod}) from \cite{2015Vilella-Rojo}. After building the emap of each J-PLUS observation field, we cropped the image to a small size by including the extent of the galaxy for further analysis. The flow chart of the module is given in Fig.~\ref{fig: pip} (top-left flowchart) and more details are described below.

\subsubsection{PSF homogenization}
\textsc{JYPE-PSF} is a Python module for PSF modeling, analysis, and normalization within the OAJ data reduction pipeline developed by the Department of Data Archival and Processing (DPAD) at the CEFCA. This module can perform an analysis of the 2D variations of the PSF of J-PLUS coadded images using SEXtractor \citep{1996Bertin} and PSFEx \citep{2011Bertin}. PSFEx computes a space-varying kernel that transforms the PSF of an image into an arbitrary PSF. Each J-PLUS tile consists of 12 coadded images generated at the same position for each filter. This program is responsible for normalizing the 2D variations of the PSF of each tile to a target PSF.

\textsc{JYPE-PSF} first analyze the FWHM of the entire tile and select an optimal target PSF as the highest FWHM measure in that tile. Using PSFEx, it derives the PSF homogenization kernel for the input coadded images. Then the convolution kernel K(x) is applied to the PSF model, and minimizes (based on $\chi^{2}$) the difference with the target PSF for the entire tile. The homogenization kernel components are stored as a FITS data cube. The target shape is a perfectly round analytical function of the Moffat profile.
Finally, each coadded image is convolved using Fast Fourier Transform (FFT) with this kernel to homogenize the variable PSF to a constant arbitrary shape to the target PSF.

\subsubsection{Zero point map}
There is a (X,Y) variation of the zero point with the position of the source on the CCD FoV for every J-PLUS observation. This observed variation can be influenced by several factors, including changes in airmass across the observation, scattered light in the focal plane, and positional dependencies in the effective filter transmission curves as discussed in \cite{2019Sanjuan, 2024lopez}, and references therein. It is not corrected in the final DR3 images and the correction is estimated in \cite{2019Sanjuan} with the following equation,
\begin{equation*}
 ZP = ZPT + A\_CALIB  \times X + B\_CALIB  \times Y + C\_CALIB,
\end{equation*}
where the parameters A\_CALIB, B\_CALIB, and C\_CALIB, define the position dependence of the zero point, which is downloaded from the J-PLUS database using ADQL. More details are given in \cite{2019Sanjuan}. We applied this correction in the final flux calibration of emaps.

\subsubsection{3F method} \label{sec:3Fmethod}
In the 3F method \citep{2015Vilella-Rojo}, two broad-band filters (rSDSS and iSDSS in the case of H$\alpha$ are used to trace the continuum and one NB filter (H$\alpha$ covering J0660) is used to trace the emission line. Then the H$\alpha$ + [NII] flux is derived using Eq.(3) from \cite{2015Vilella-Rojo} as shown below:

\begin{equation*}
\text{F}_{H\alpha + [\text{NII}]} = 
\frac{
(\overline{F}_{r'} - \overline{F}_{i'}) - \left( \frac{\alpha_{r'} - \alpha_{i'}}{\alpha_{F660} - \alpha_{i'}} \right)(\overline{F}_{F660} - \overline{F}_{i'})
}{
\beta_{F660} \left( \frac{\alpha_{i'} - \alpha_{r'}}{\alpha_{F660} - \alpha_{i'}} \right) + \beta_{r'}
}.
\end{equation*}

The \( \overline{F} \) values represent the observed average fluxes inside the filters, while \( \alpha \) and \( \beta \) are defined as

\begin{equation*}
\alpha_x \equiv \frac{\int \lambda^2 P_x(\lambda) \, d\lambda}{\int P_x(\lambda) \lambda \, d\lambda}, \quad 
\beta_x \equiv \frac{\lambda_s P_x(\lambda = \lambda_s)}{\int P_x(\lambda) \lambda \, d\lambda},
\end{equation*}

where, $ \lambda_s = \lambda_{\text{H}\alpha} = 6562.8 \, \text{\AA}
$.

\cite{2015Vilella-Rojo} have demonstrated that using two filters for continuum estimation provides a more accurate representation of the continuum's true shape compared to assuming a flat continuum with a single filter.

\subsection{IFU module}  \label{sec:ifumodule}
The IFU module constructs the H$\alpha$ + [NII] emaps from other IFUs (here it is PHANGS-MUSE, CALIFA, and MaNGA) used for validation. The three data cubes are in different formats. The flowchart of this module is given in the upper right of Fig.~\ref{fig: pip}. The initial step of the module is downloading the IFU cubes. We followed two different methods to build the IFU emaps for comparison.

In the first method (called "2D" in the flow chart), we initially build synthetic NBs from IFU cubes by convolving with J-PLUS filter curves. This creates rSDSS, iSDSS and J0660 synthetic images. Then, we followed the 3F method (as in the J-IFU module) for the continuum subtraction. For filter convolution, we use the MPDAF package of Python \citep{2019Piqueras}.  It is a Python package used for the analysis of VLT/MUSE IFU cubes.

In the second method (called "3D" in the flow chart), we first estimate and subtract the continuum from the 3D cube and then convolve the continuum subtracted cube with the filter covering the emission line (e.g., $J0660$). For 3D continuum subtraction, we used Gnuastro: \texttt{astarithmetic} command. We estimate the continuum flux at every slice of the IFU cube by taking the sigma-clipped median value of N/2 (N=100) slices or wavelength steps before and after it. The steps are given in the section of continuum subtraction in the Gnuastro reference manual \citep{gnuastrobook}. After the 3D continuum subtraction, the 3D cube is collapsed to a 2D image by convolving the cube with the J0660 filter transmission curve.

After constructing emaps, the 2D images from the IFUs are warped to the pixel grid of the J-PLUS image of the galaxy using \texttt{astwarp} from Gnuastro. The warping pixel scale is set by the highest pixel scale size among the IFUs and J-PLUS. The one with the highest pixel scale size will be used to rescale the other. PHANGS-MUSE (0.2"/pixel), CALIFA (0.5"/pixel) and MaNGA (0.5"/pixel) have higher spatial sampling compared to J-PLUS (0.55"/pixel), thus the pixel scale of the IFUs are changed to J-PLUS size. Afterwards, the image size of J-PLUS is resampled to the same image size of IFUs. This step is necessary for the pixel-to-pixel comparison between the emaps as we do in later sections.

\begin{figure}
    \centering
    \includegraphics[width=9.1cm]{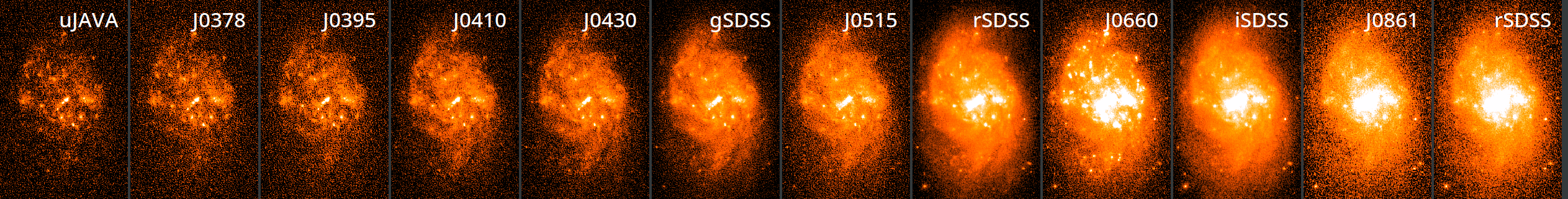}
    \includegraphics[width=9.1cm]{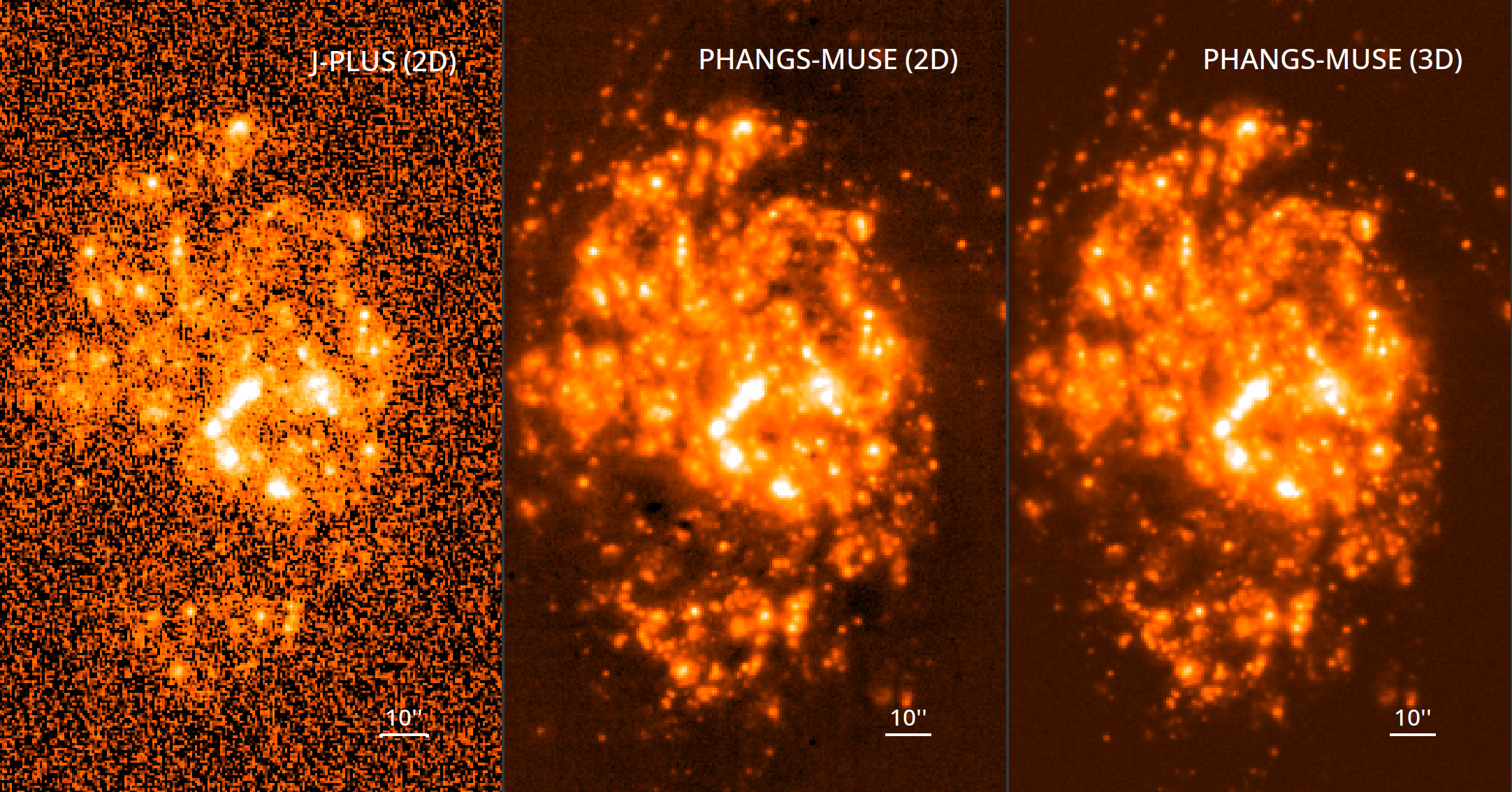}
    \caption{H$\alpha$+[NII] emap of NGC 1087 in J-PLUS and in PHANGS-MUSE (2x3 MUSE pointings) constructed in Sect.~\ref{sec:jifumodule} and Sect.~\ref{sec:ifumodule}. The J-PLUS images in 12 filters are shown in the top panel, plotted with the same surface brightness level.}
    \label{fig:emap-ngc1087}
\end{figure}

\begin{figure*}
    \centering    
    \includegraphics[width=9.1cm]{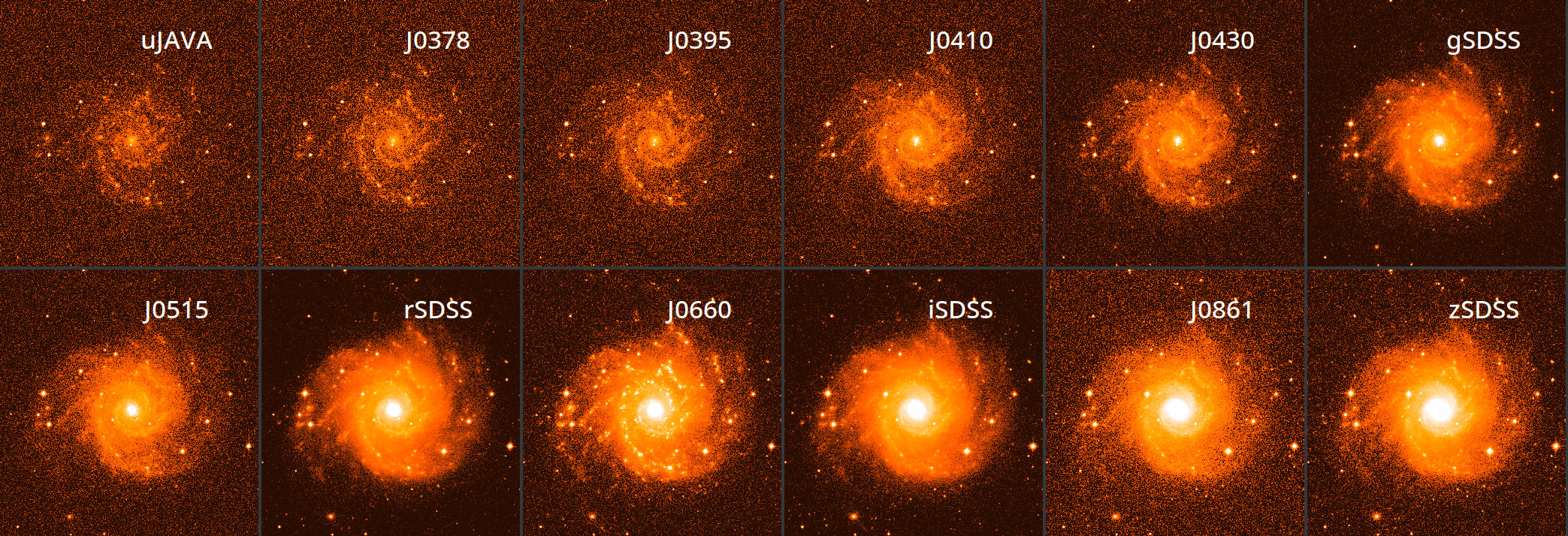}
    \includegraphics[width=9.1cm]{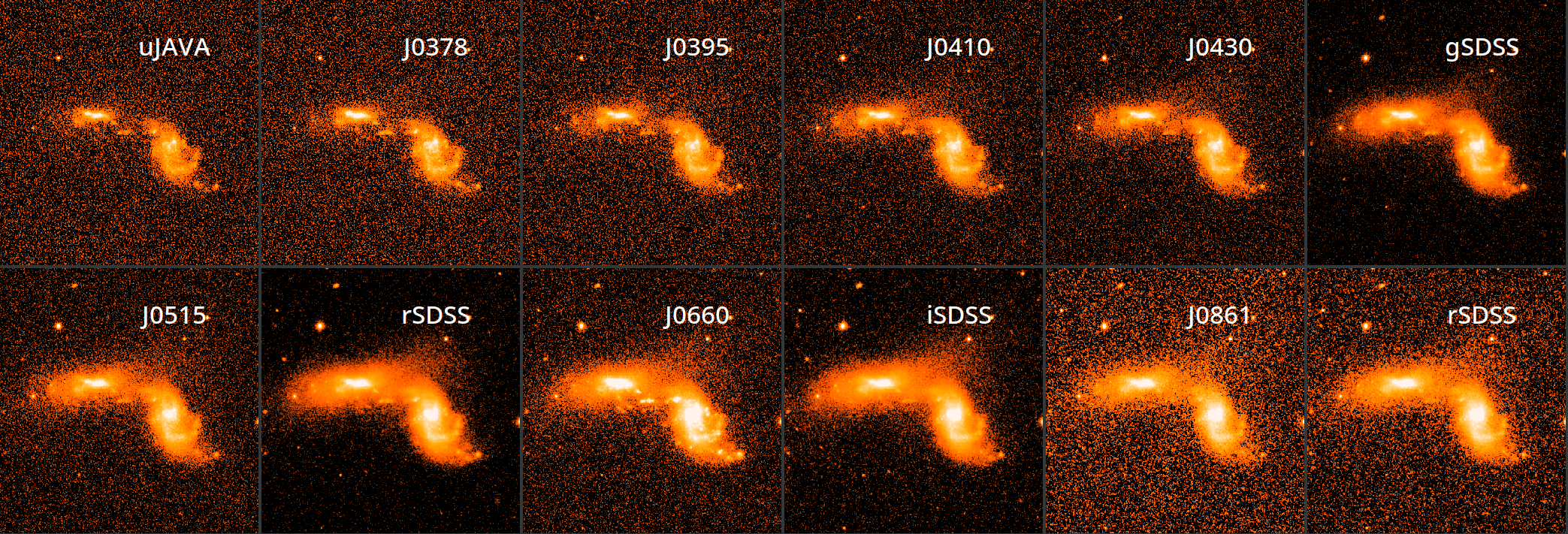}    
    \includegraphics[width=6.7cm]{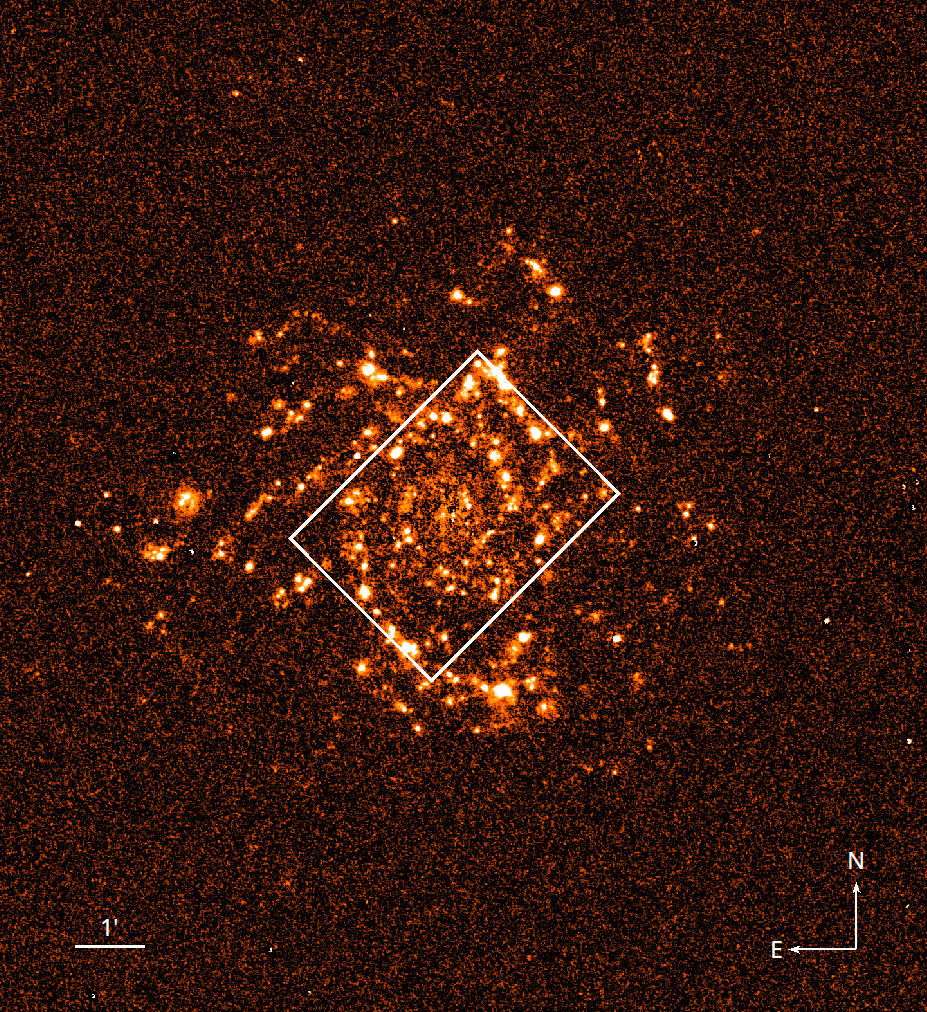}
    \includegraphics[width=2.4cm]{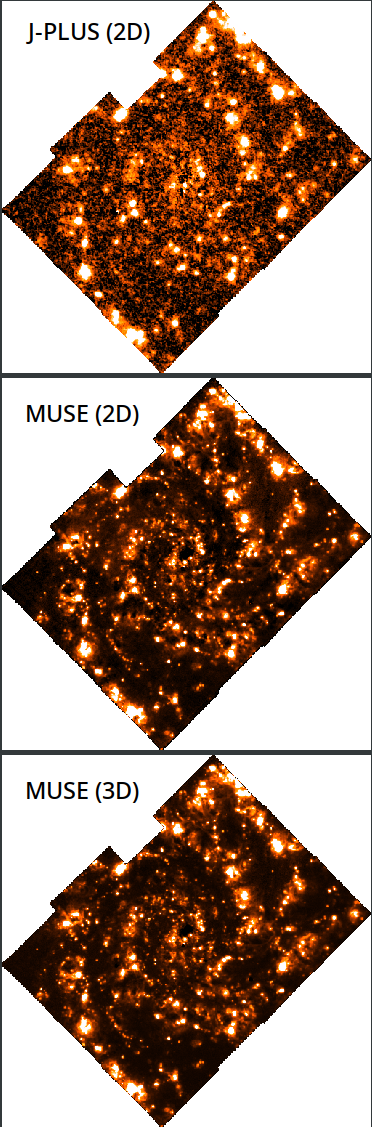}
     \includegraphics[width=6.2cm]{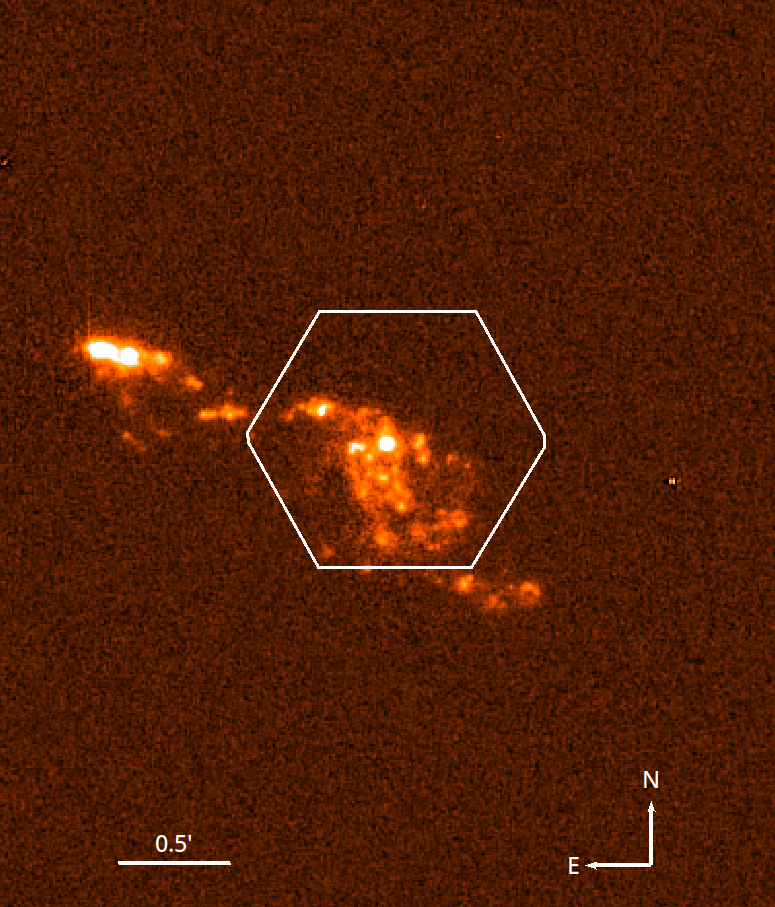}
    \includegraphics[width=2.8cm]{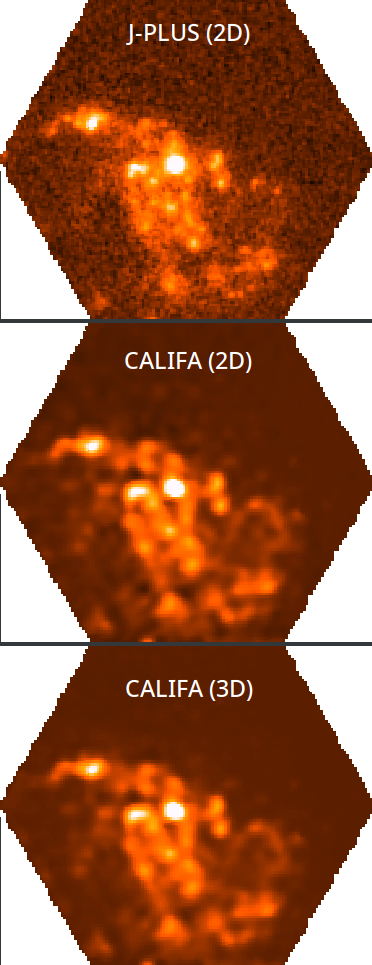}
  \caption{Top: J-PLUS images in 12 filters. Left: Full extent of H$\alpha$+[NII] emap of NGC 628 in visible in a crop of the J-PLUS tile with a label of PHANGS-MUSE coadd size (consisting of 3x4 MUSE pointings; white square) of same galaxy, and emaps constructed in Sect.~\ref{sec:jifumodule} and Sect.~\ref{sec:ifumodule}.
  Right: Full extent of H$\alpha$+[NII] emap of NGC 3395 in J-PLUS with a label of CALIFA FoV (white), and emaps constructed in Sect.~\ref{sec:jifumodule} and Sect.~\ref{sec:ifumodule}}
    \label{fig: emap-ngc628-ngc3395}
\end{figure*}

\begin{figure}
\includegraphics[width=9cm]{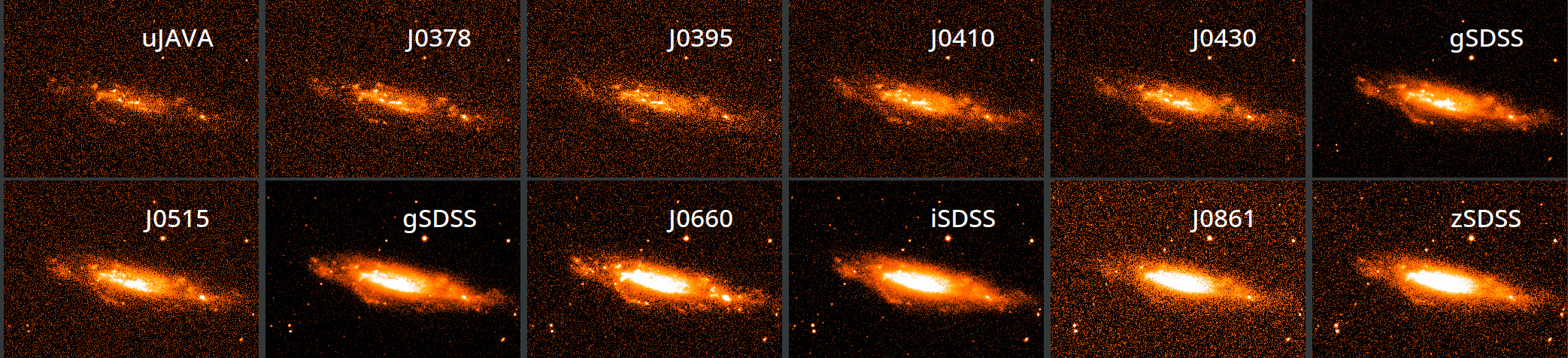}
\includegraphics[width=9cm]{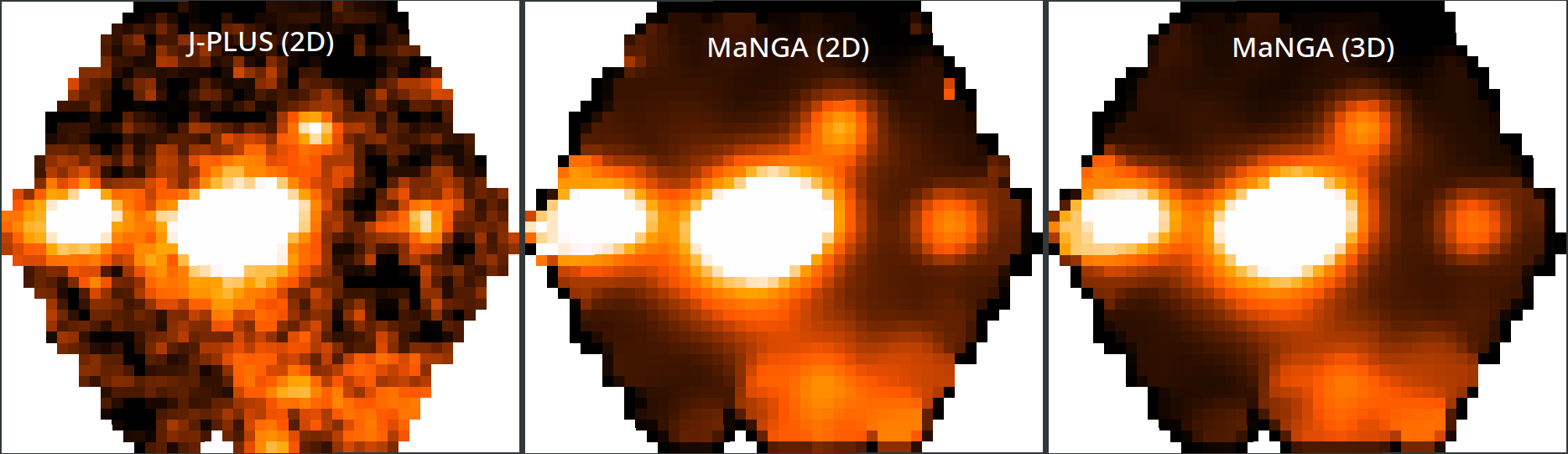}
\includegraphics[width=9cm]{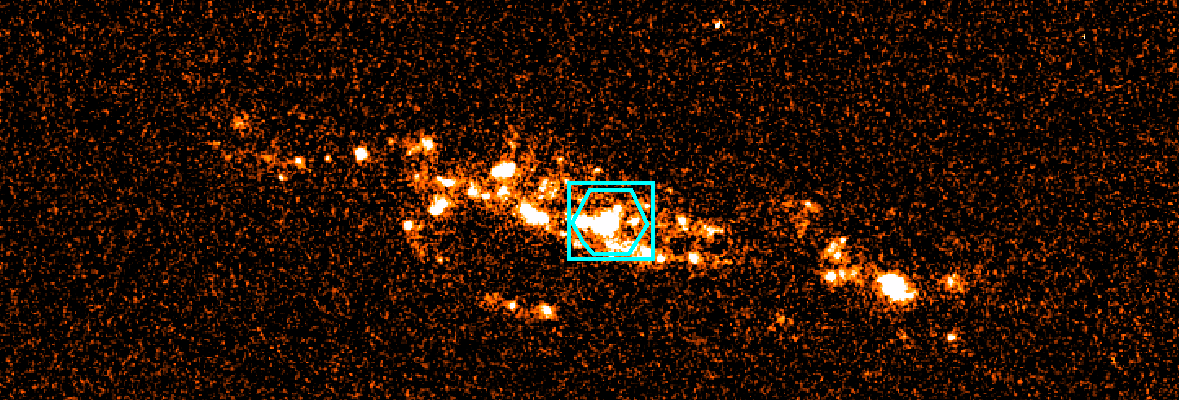}
\caption{Top: J-PLUS images in 12 filters. Middle: Comparison of J-PLUS emap with MaNGA emaps constructed in different methods. Lower: Full extent of H$\alpha$ + [NII] emap of MaNGA-8150-6103 in J-PLUS with a label of MaNGA FoV (blue)}
 \label{fig: emap-manga-8309-3703}
\end{figure}

The examples of output 2D emaps from these two modules and their comparison are given from Figures~\ref{fig:emap-ngc1087} to~\ref{fig: emap-manga-8309-3703}. The figures also show the full extent of the galaxy emap and the cropped images based on the IFU size. The size of the IFU FoV is labeled in the figure. 

\subsection{Analysis module}
The analysis module produces the radial profile from galaxy emaps (see Sect.\ref{sec:amodule-rad}) identifies and constructs the SEDs and spectra of emission line regions (see Sect.\ref{sec:amodule-sed}).  We extensively used Gnuastro packages inside this module for analysis. 
\begin{figure*}
    \centering
    \includegraphics[width=9cm]{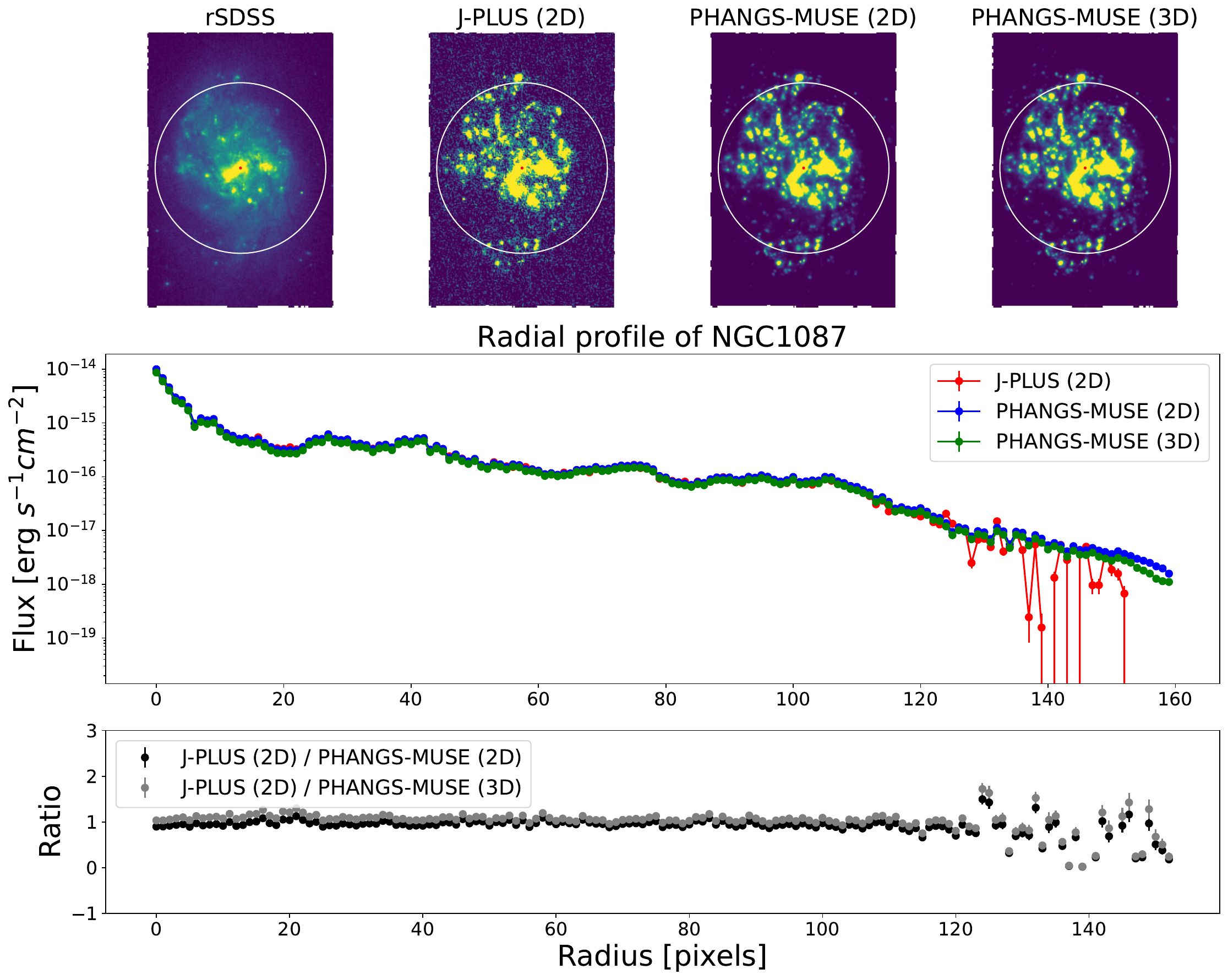}
    \includegraphics[width=9cm]{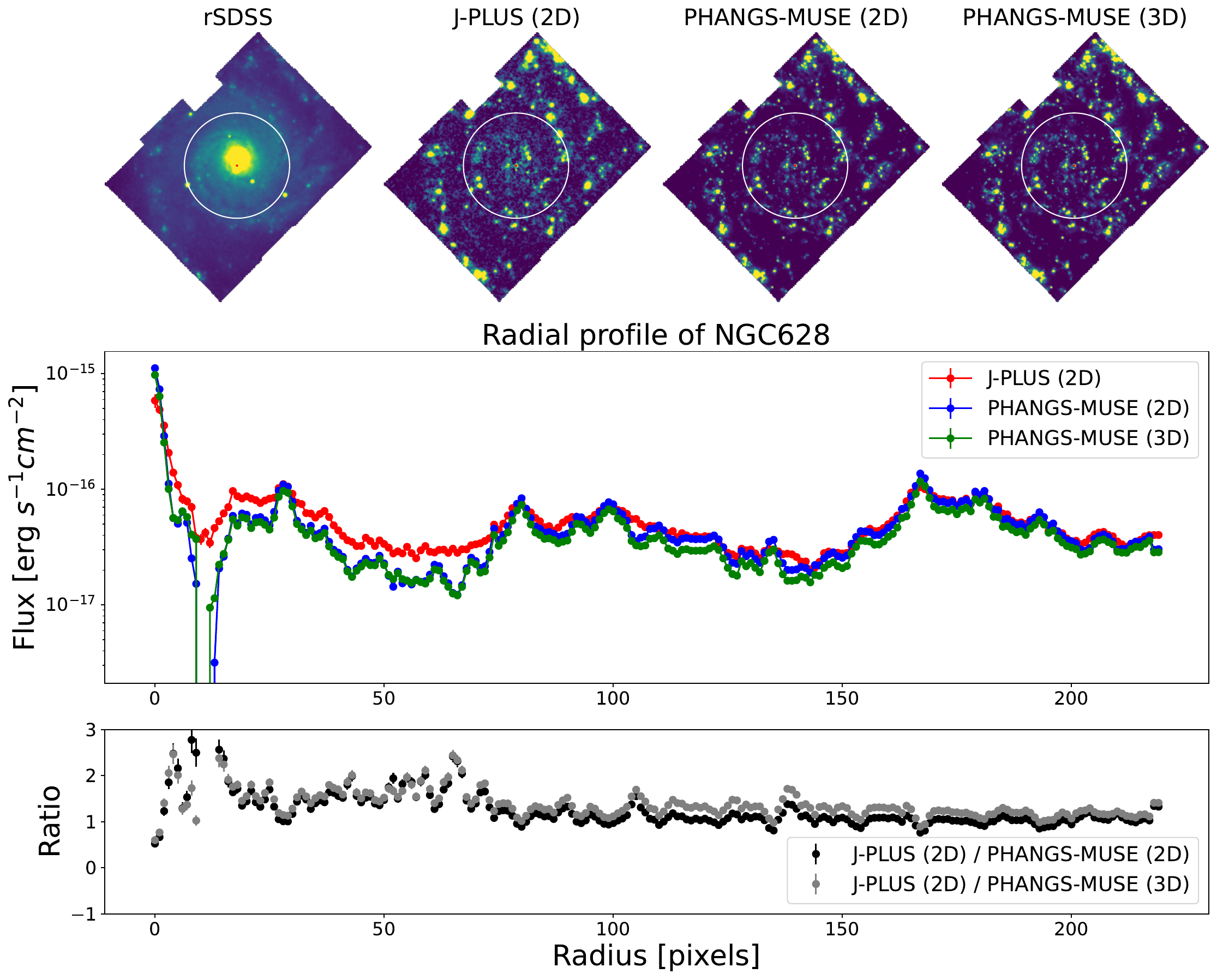}
    \caption{Upper: rSDSS (J-PLUS), J-PLUS (2D), PHANGS-MUSE (2D), and PHANGS-MUSE (3D) H$\alpha$+[NII] map of NGC 1087 (left) and NGC 628 (Right). Middle: Radial profiles of H$\alpha$+[NII] maps from J-PLUS (2D), PHANGS-MUSE (2D) and PHANGS-MUSE (3D). Lower: The ratio of the radial profiles between J-PLUS and PHANGS-MUSE H$\alpha$+[NII] maps. The center and radius of 100 pixels are labeled with red and white circles respectively. Note that emaps are in the same flux scale except rSDSS.}
   \label{fig:rprof-ngc1087}
\end{figure*}
\begin{figure}
    \centering    
    \includegraphics[width=0.45\textwidth]{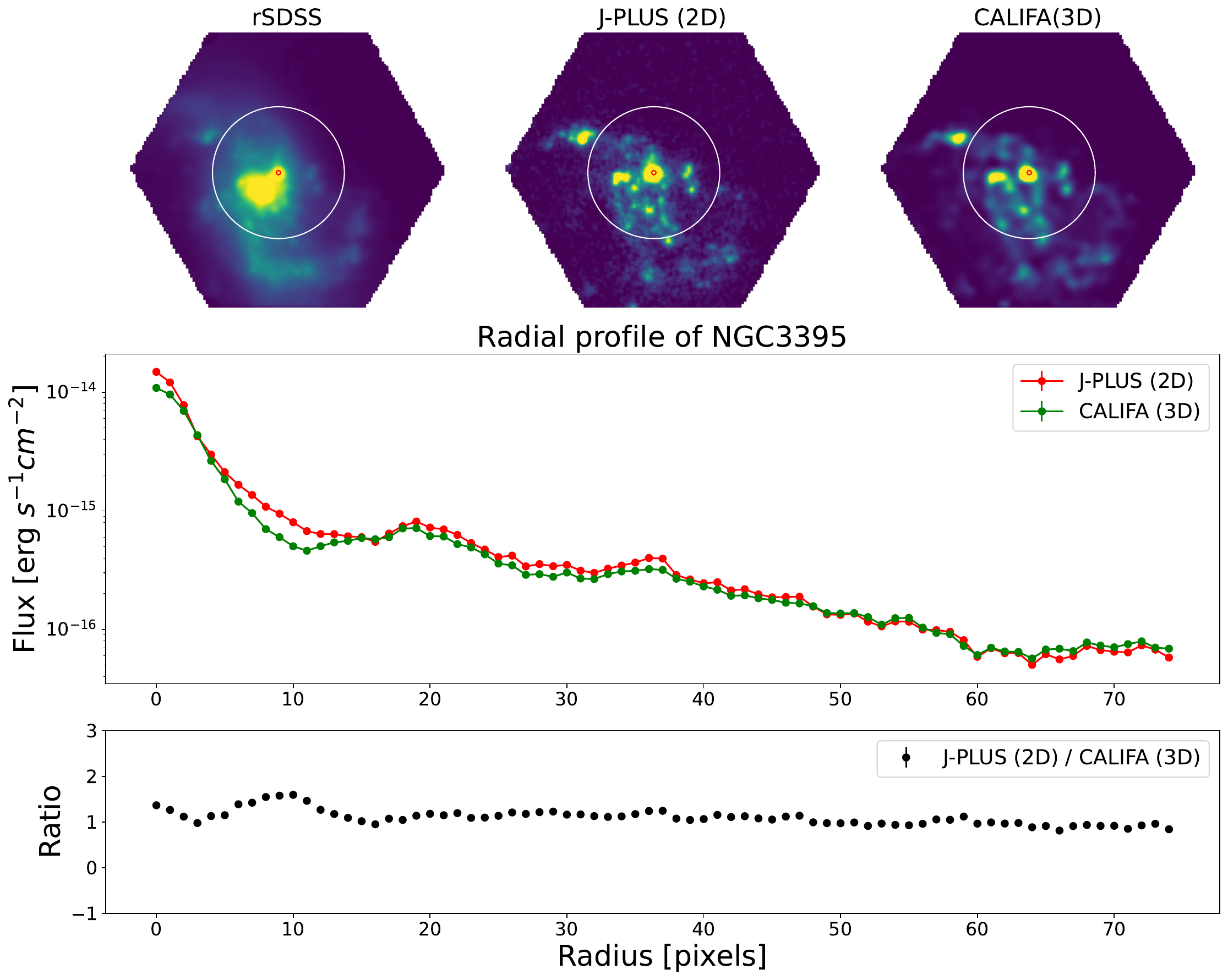} 
    \caption{Upper: rSDSS (J-PLUS), J-PLUS (2D), and CALIFA (3D) H$\alpha$+[NII] map of NGC 3395 (the white circle is in a 30 pixel radius). Middle: Radial profiles of H$\alpha$+[NII] maps from J-PLUS (2D), and CALIFA (3D). Lower: The ratio of the radial profiles between J-PLUS and CALIFA H$\alpha$+[NII] maps. Note that emaps are in the same scale except rSDSS.}
   \label{fig:rprof-ngc3395}
    \end{figure}
    \hspace{0.05\textwidth}  
\begin{figure}
    \centering    
    \includegraphics[width=9cm]{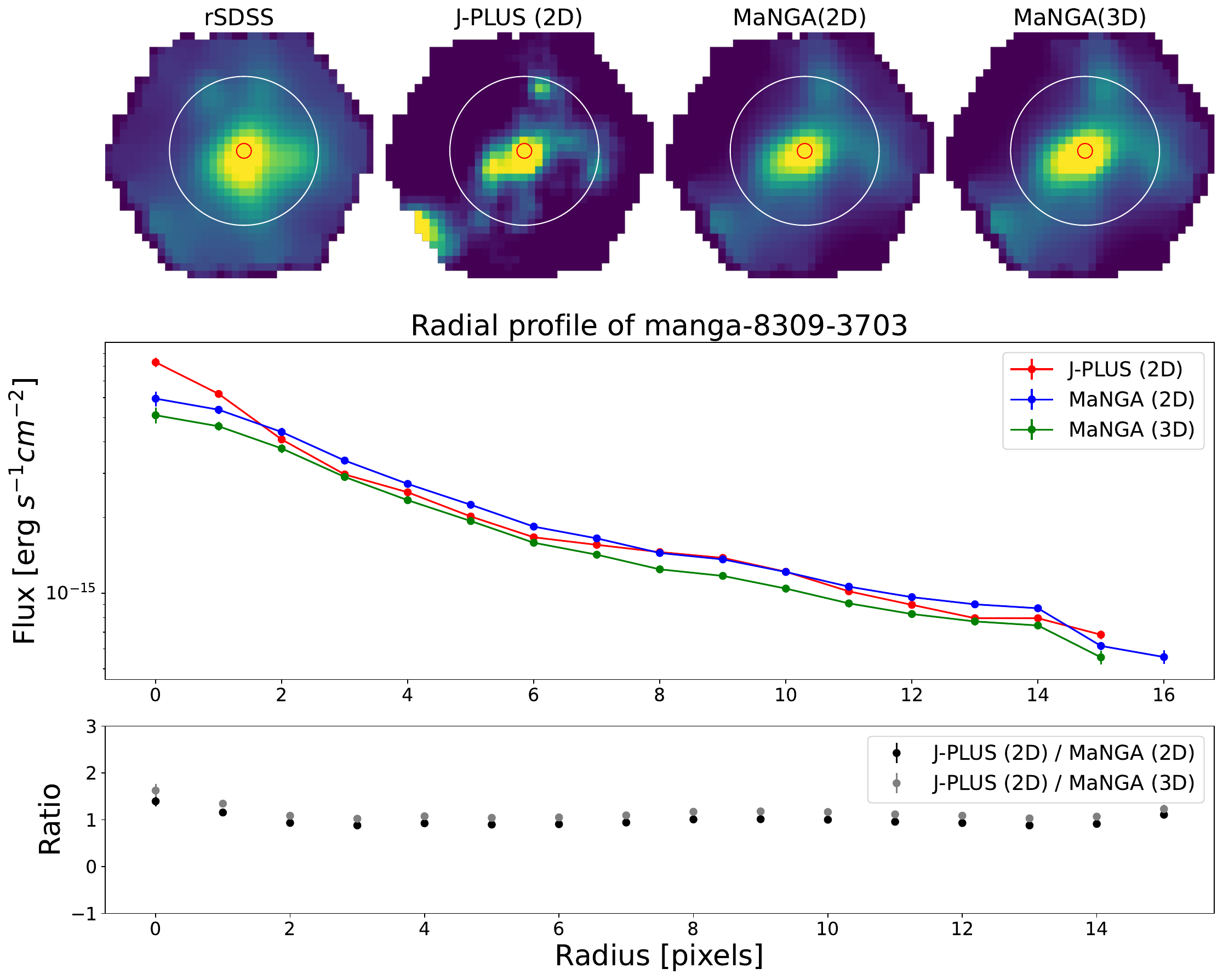}        
    \caption{Upper: rSDSS, J-PLUS (2D), MaNGA (2D), and MaNGA (3D) H$\alpha$+[NII] map of MaNGA-8309-3703 (the white circle is in a 10 pixel radius). Middle: Radial profiles of H$\alpha$+[NII] maps from J-PLUS (2D), MaNGA (2D) and MaNGA (3D). Lower: The ratio of the radial profiles between J-PLUS and MaNGA H$\alpha$+[NII] maps. Note that emaps are in the same scale except rSDSS.}
   \label{fig:rprof-manga-8309-3703}
\end{figure}
\subsubsection{Estimation of the radial profiles} \label{sec:amodule-rad}
The warped emaps are used to plot radial profiles of galaxies. We use Gnuastro's \texttt{astscript-radial-profile} \citep{2024Infante-Sainz} for this step. We estimated the axis ratio and position angle of the galaxies from rSDSS images using Gnuastro's MakeCatalog, \texttt{astmkcatalog} \citep{2019Akhlaghi}. Then the final estimation of these parameters was based on the shape of H$\alpha$+[NII] images. For MaNGA images, we used the circular aperture because of their small image size.

We show some examples of radial profile of J-PLUS galaxies in comparison from MUSE (Fig.~\ref{fig:rprof-ngc1087}), CALIFA (Fig.~\ref{fig:rprof-ngc3395}) and MaNGA (Fig.~\ref{fig:rprof-manga-8309-3703}). The upper panels are the emaps and their radial profiles are given in the middle panels. We discuss the main results from this comparison in Sect.~\ref{sec:results}.

\begin{figure*}
    \centering
    \includegraphics[width=4cm]{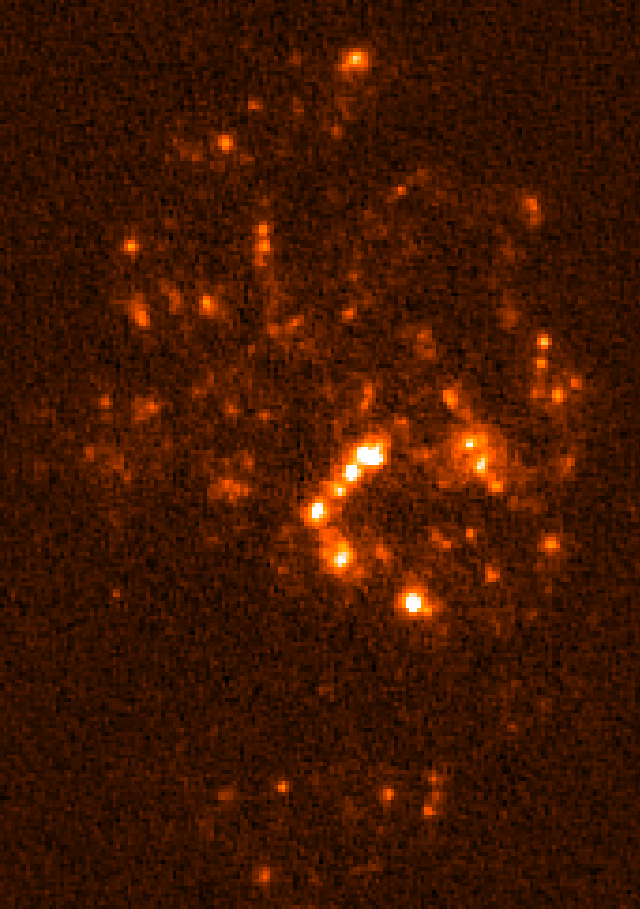}
    \includegraphics[width=2cm]{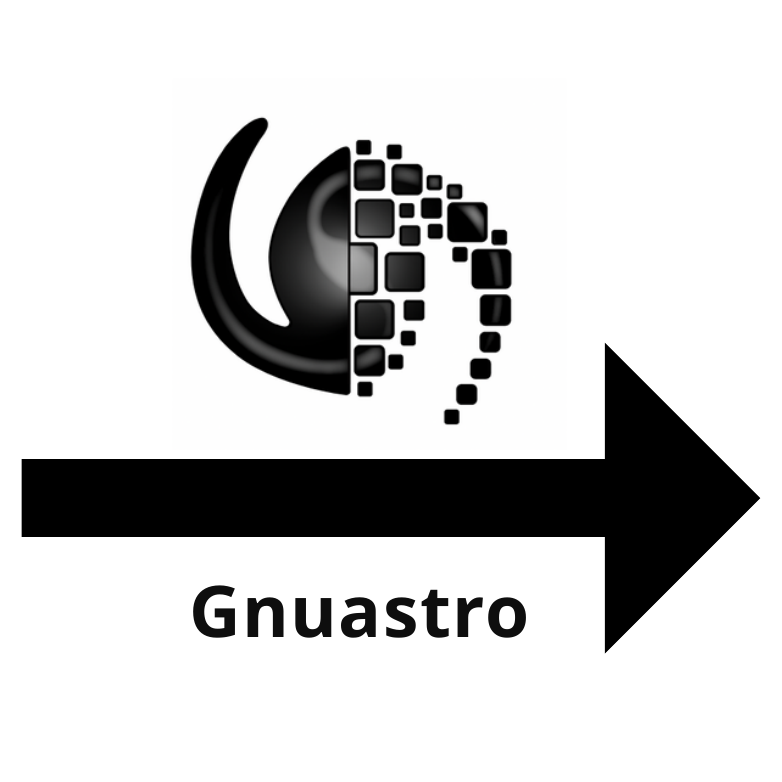}
    \includegraphics[width=4cm]{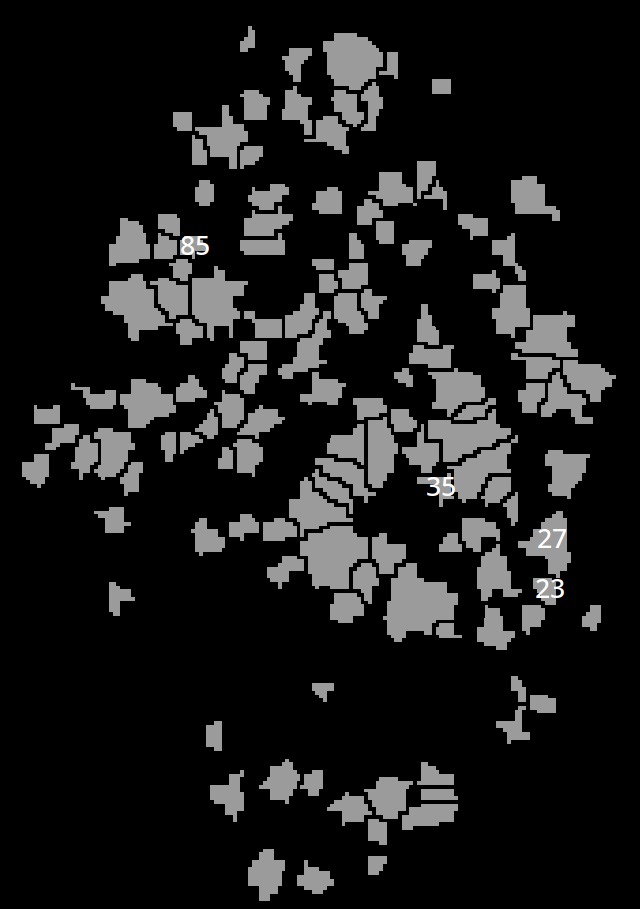}
    \includegraphics[width=9cm]
    {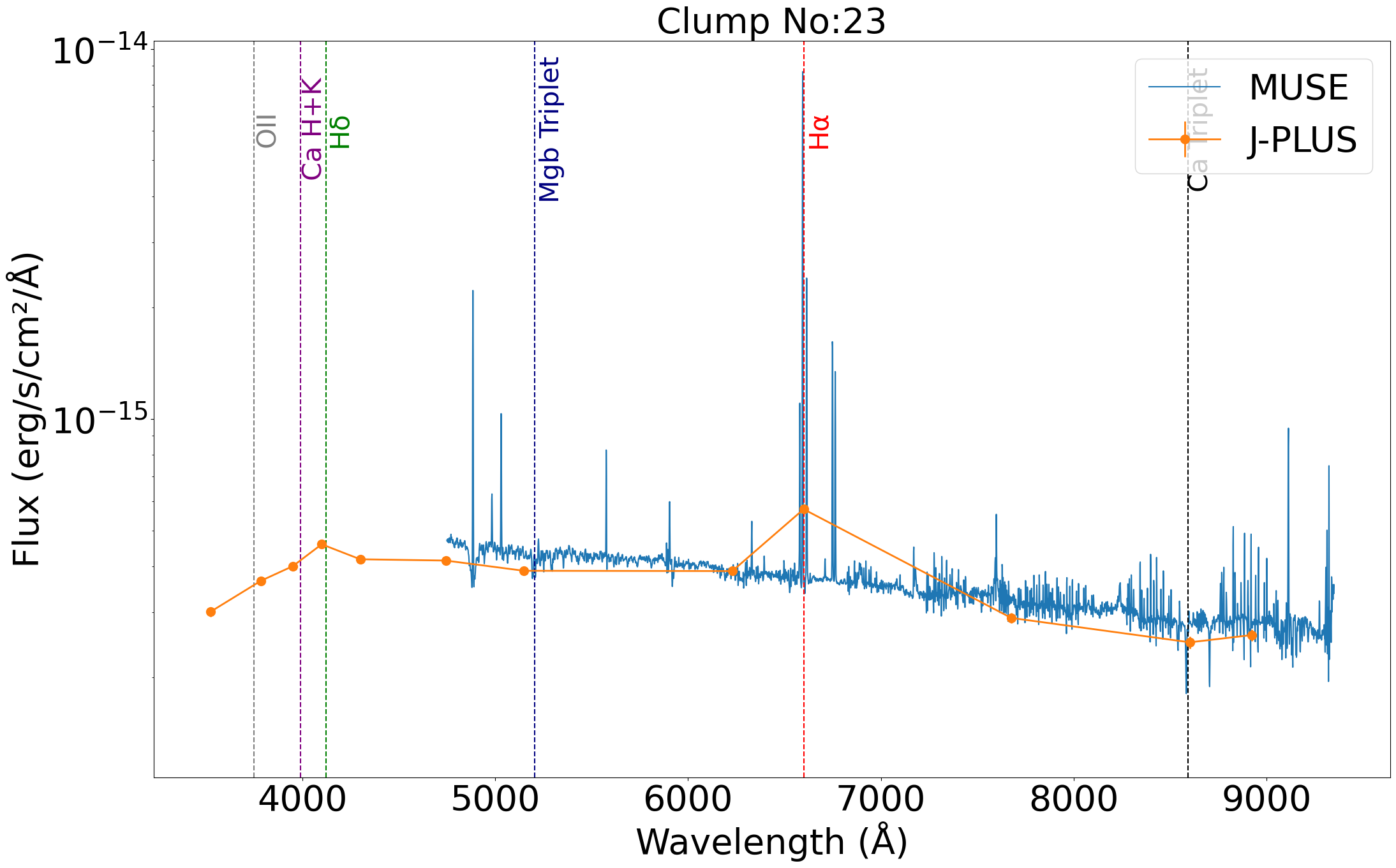}
    \includegraphics[width=9cm]
    {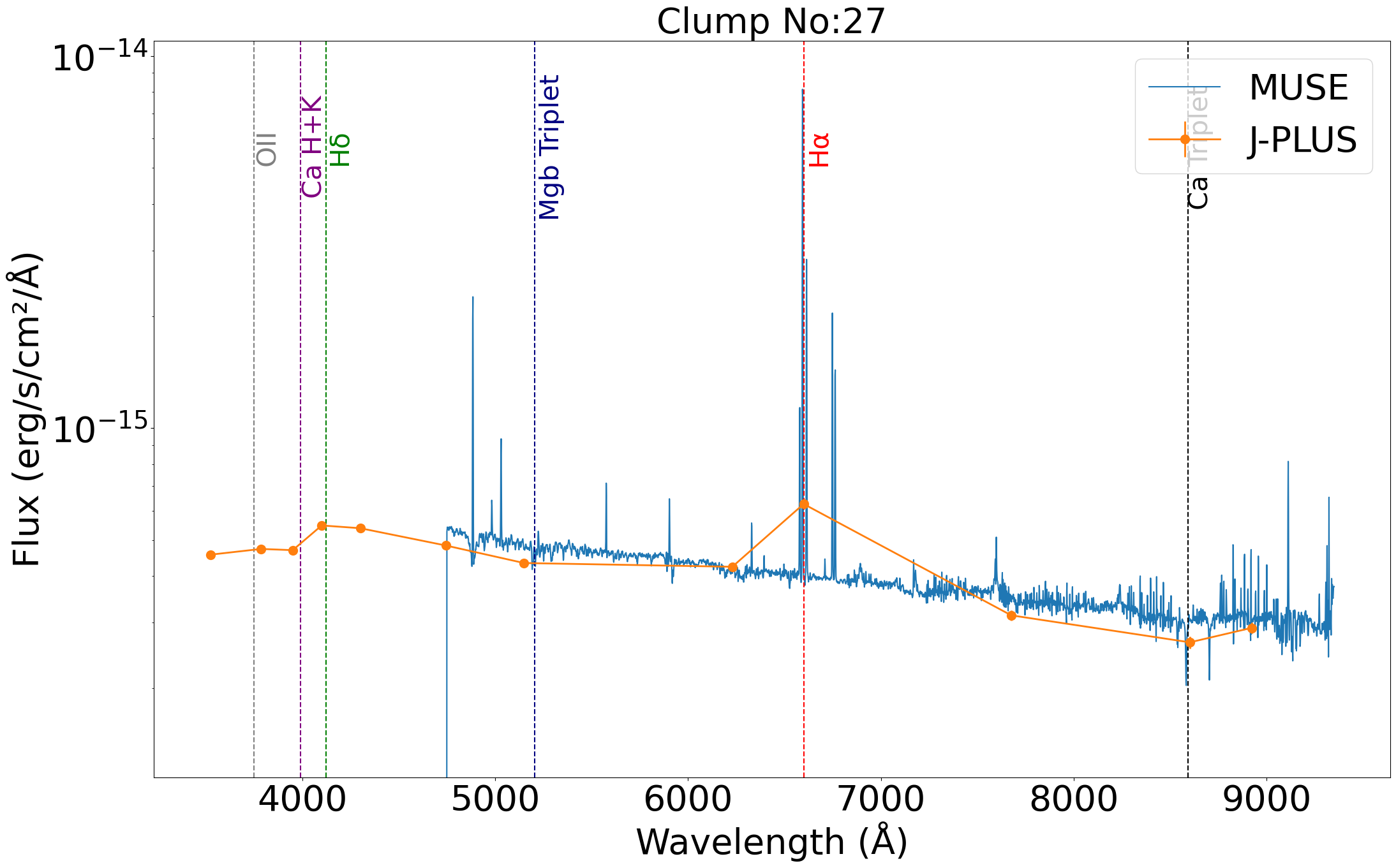}
    \includegraphics[width=9cm]
    {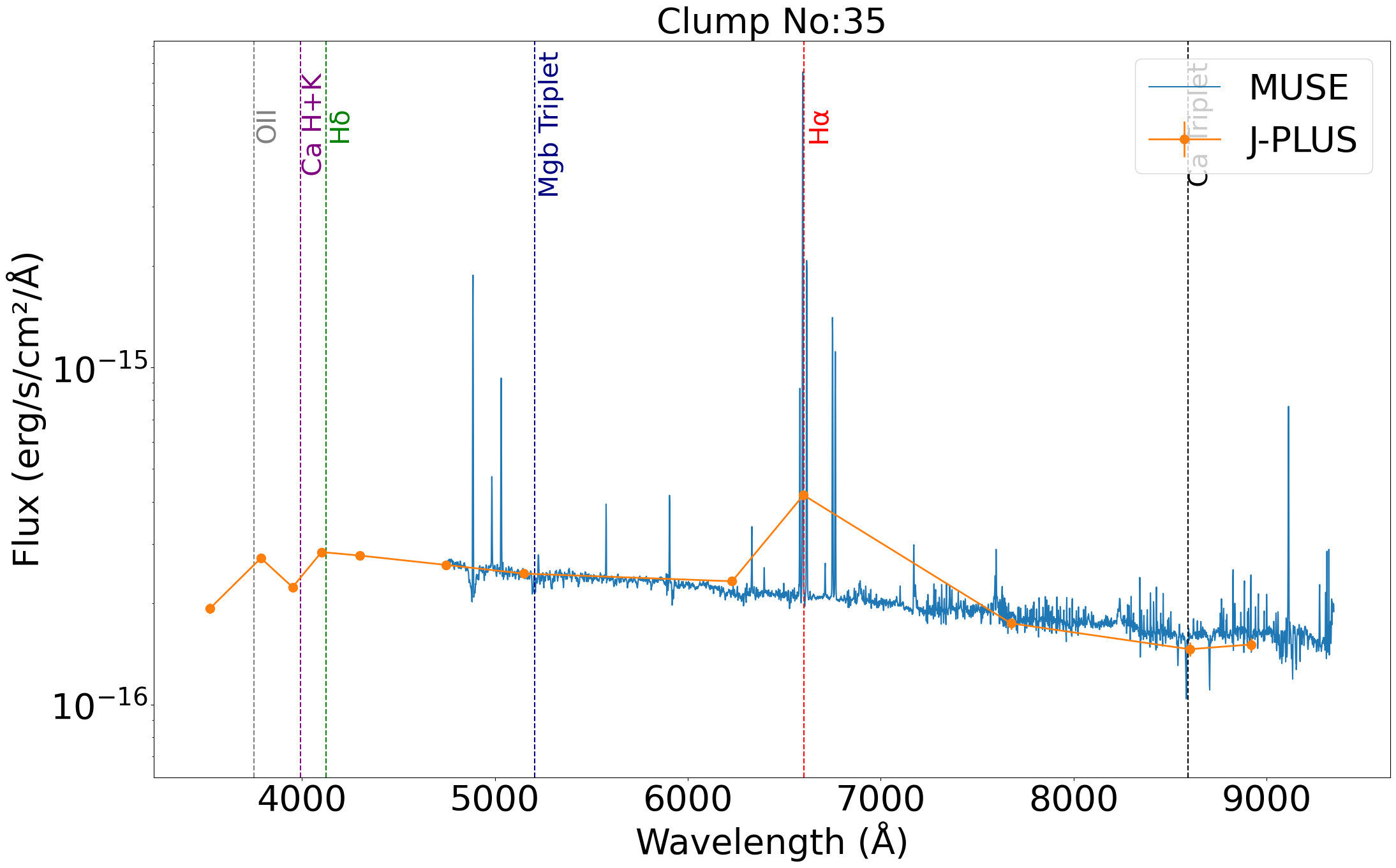}
    \includegraphics[width=9cm]
    {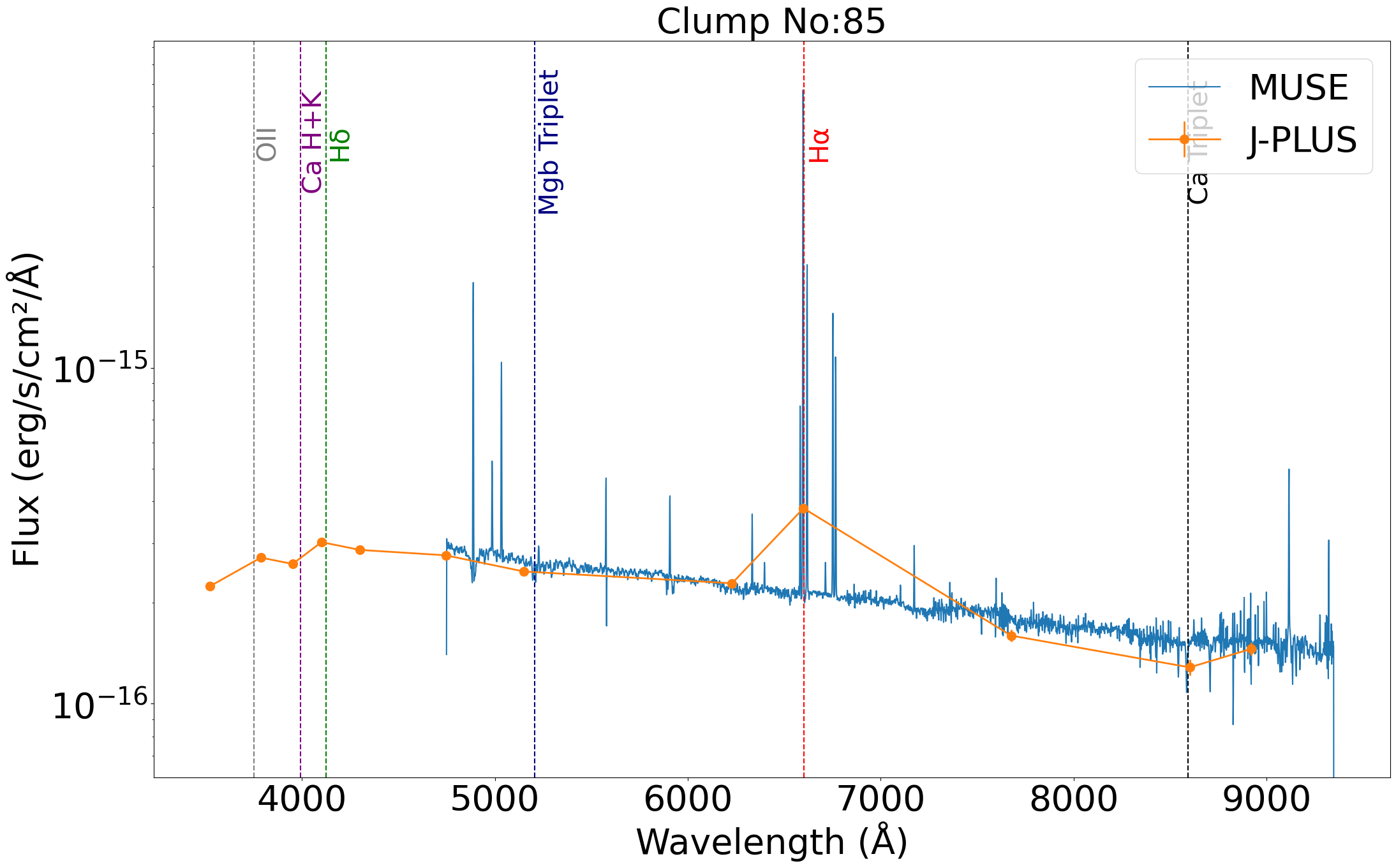}   
    \caption{Top: H$\alpha$+[NII] map and segmentation map of NGC 1087. Bottom: Examples of photospectra from J-PLUS in comparison with MUSE spectra of the clumps.}
    \label{fig:SED-muse}
\end{figure*}

\begin{figure*}
    \centering
    \includegraphics[width=5cm]{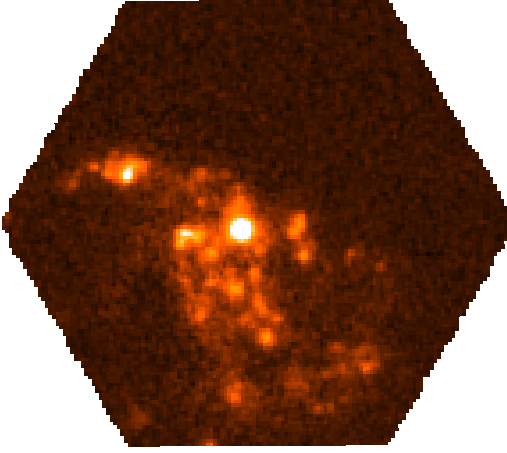}    \includegraphics[width=2cm]{Figs/gnuastro.pdf}
    \includegraphics[width=5cm]{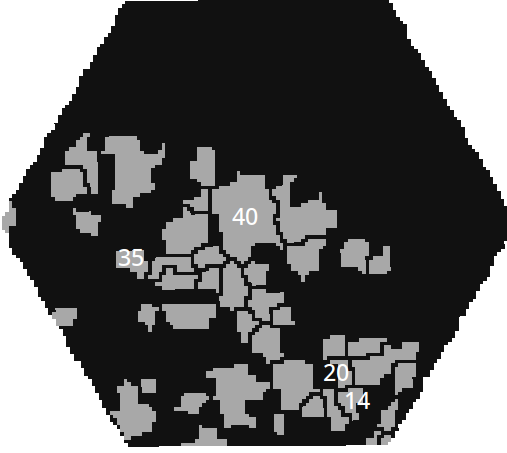}
    \includegraphics[width=9cm]{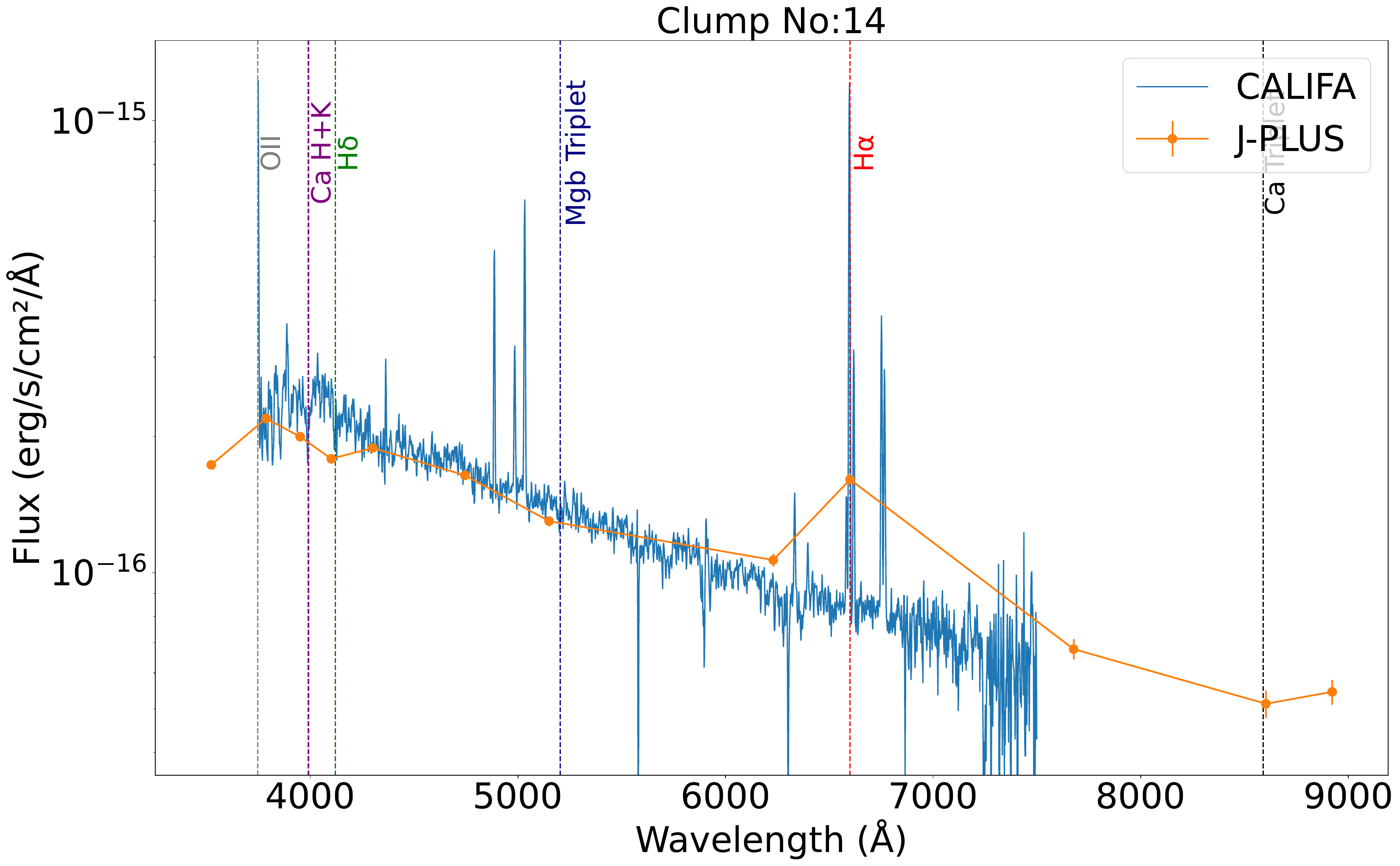}
    \includegraphics[width=9cm]{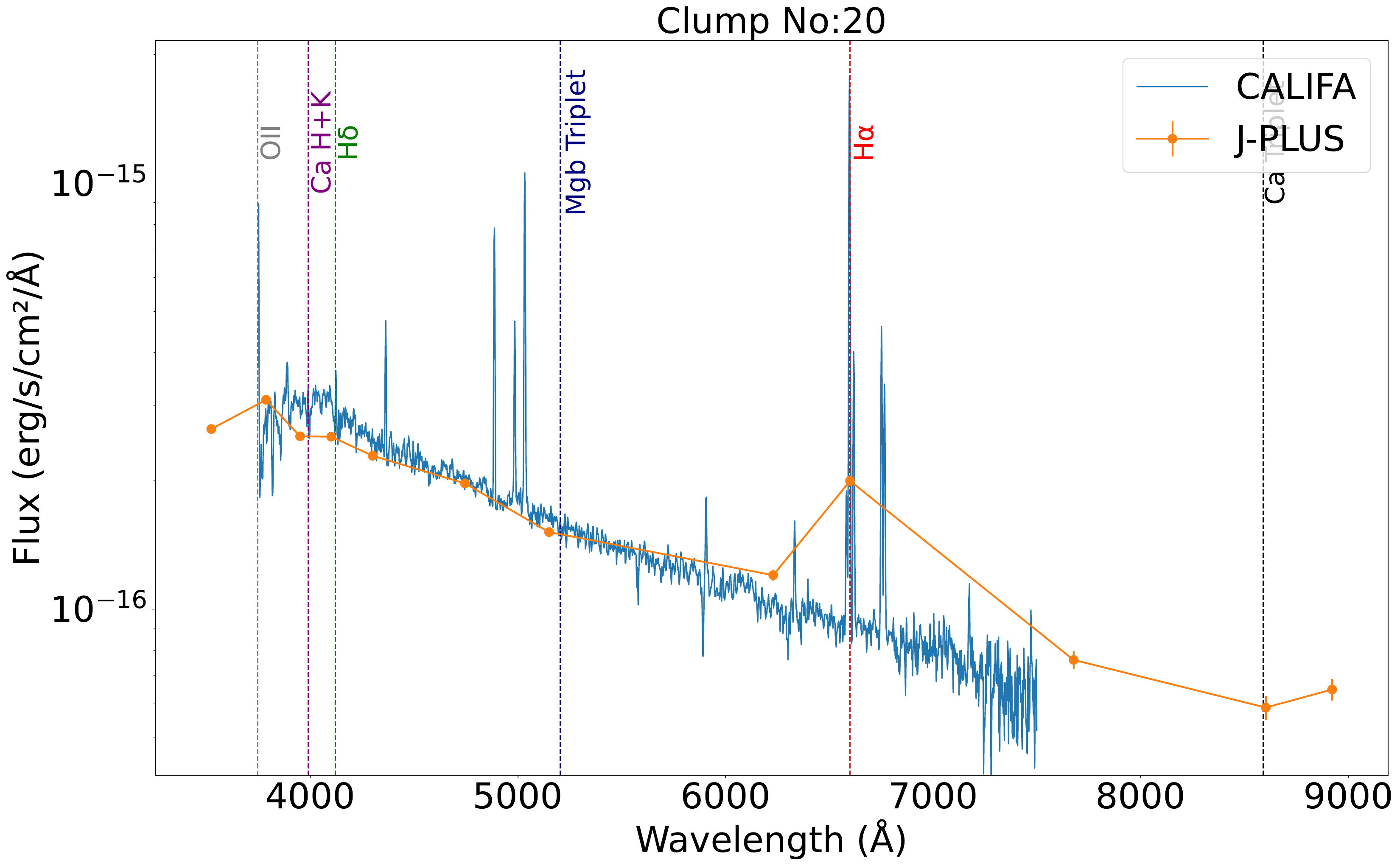}
    \includegraphics[width=9cm]{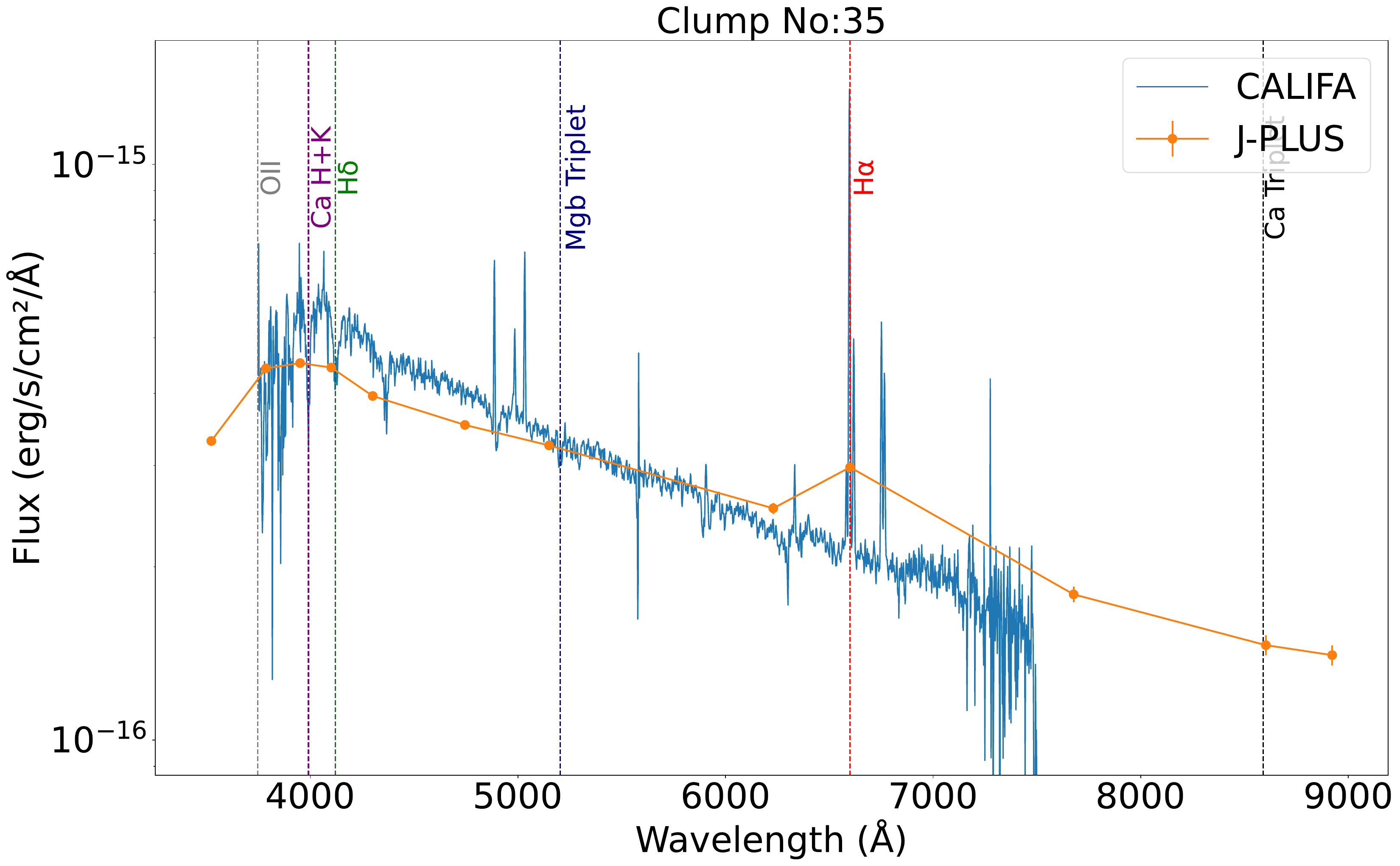} \includegraphics[width=9cm]{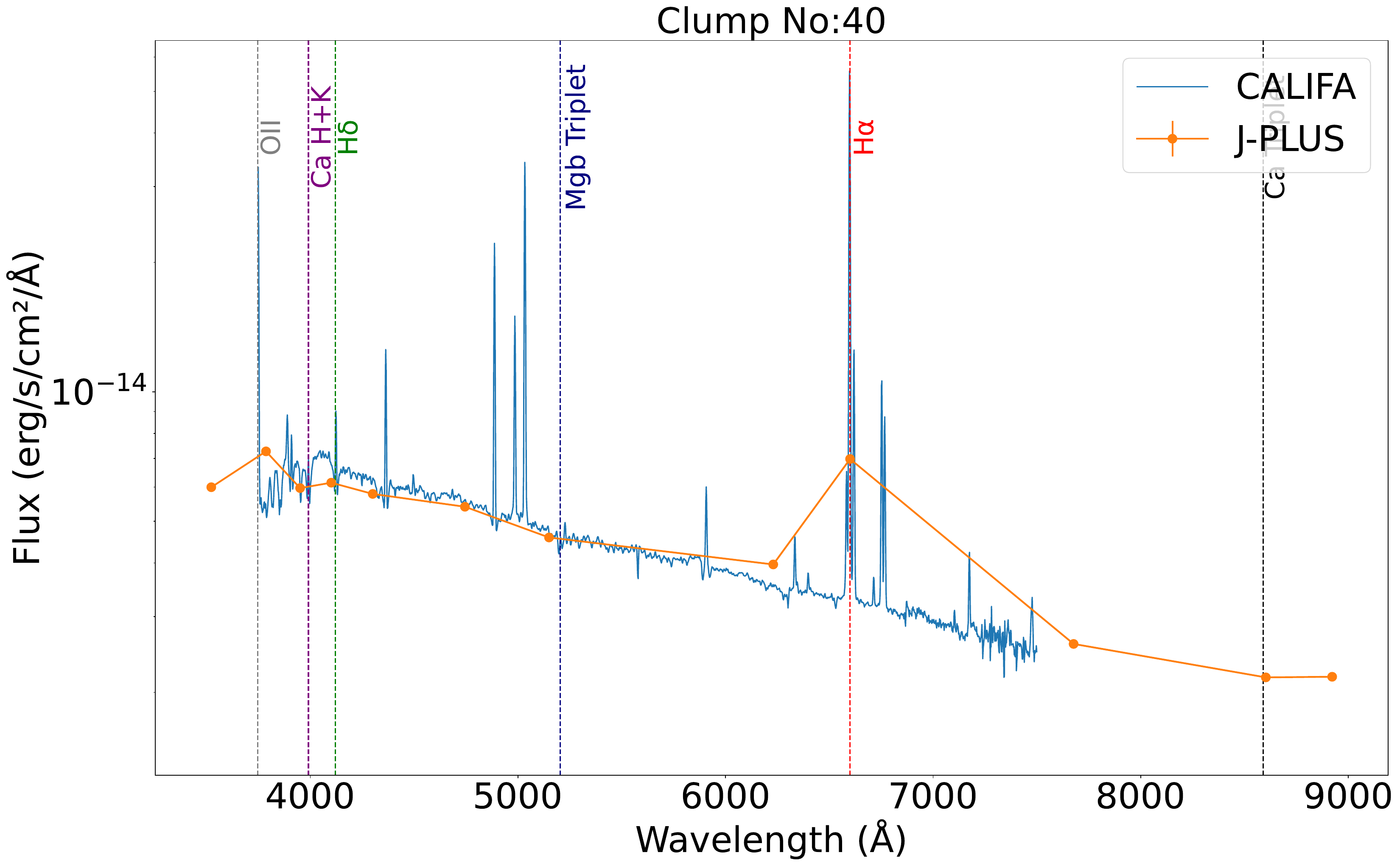}   
    \caption{Top: H$\alpha$+[NII] map and segmentation map of NGC 3395 (only the part of CALIFA size). Bottom: Examples of J-PLUS SEDs in comparison with CALIFA spectra of the clumps}
    \label{fig:SED-califa}
\end{figure*}

\begin{figure*}
    \centering
    \includegraphics[width=5cm]{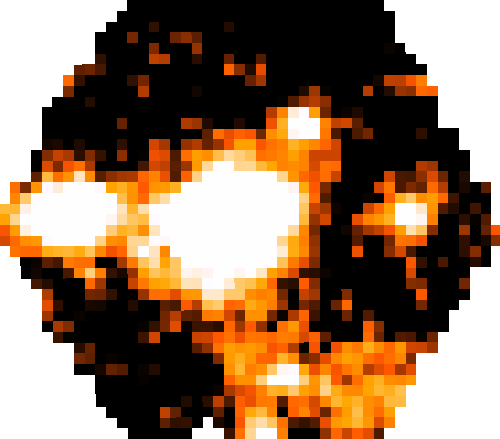}
    \includegraphics[width=2cm]{Figs/gnuastro.pdf}
    \includegraphics[width=5cm]{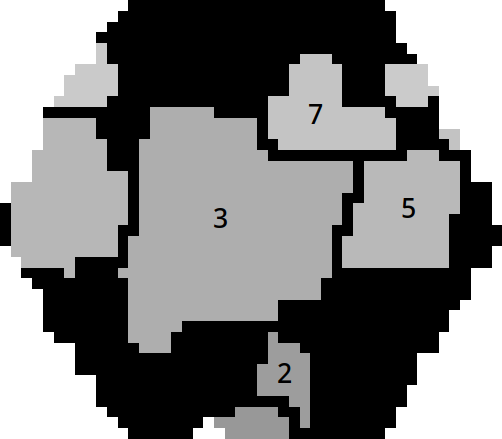}
  
    \includegraphics[width=9cm]{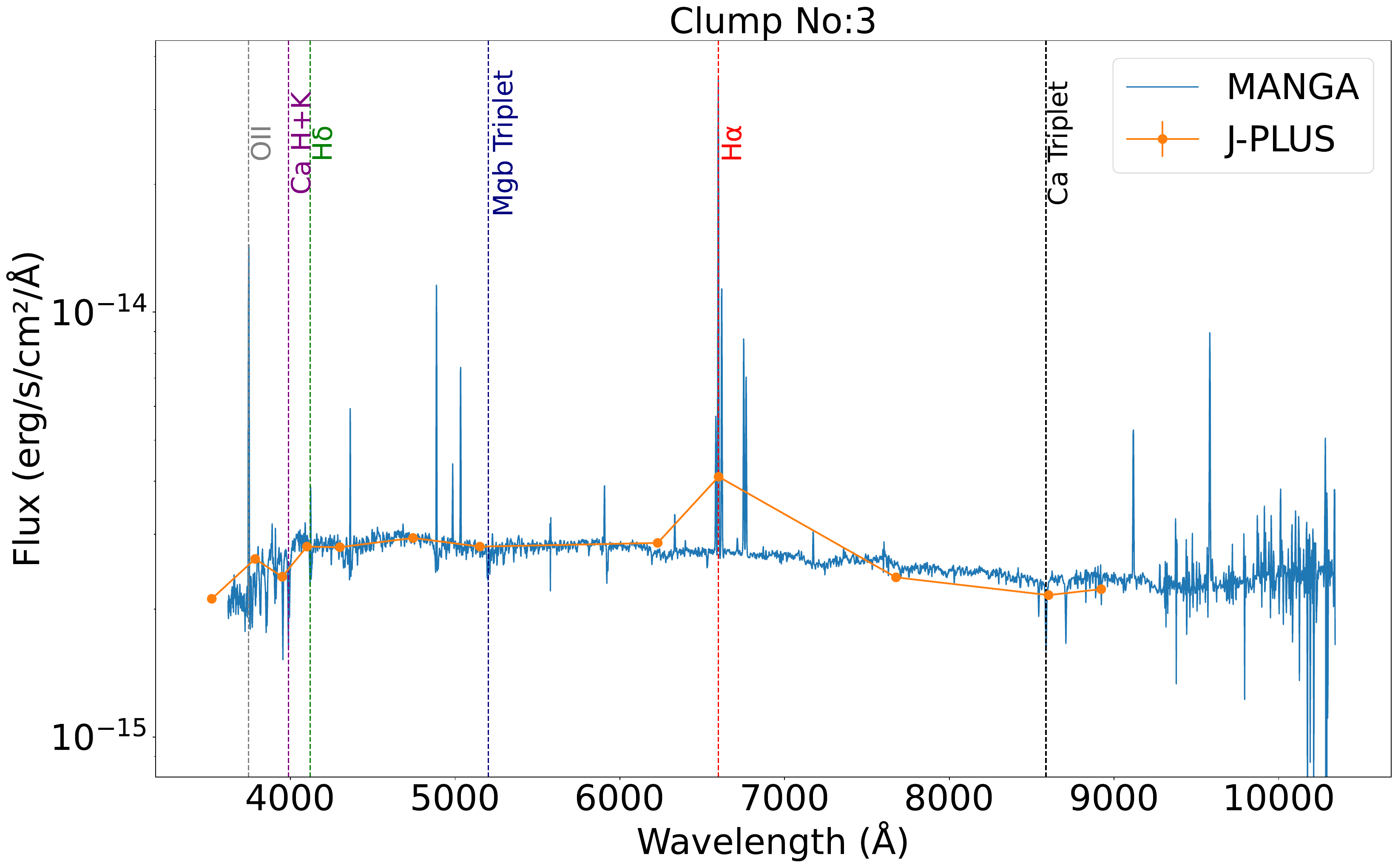}
    \includegraphics[width=9cm]{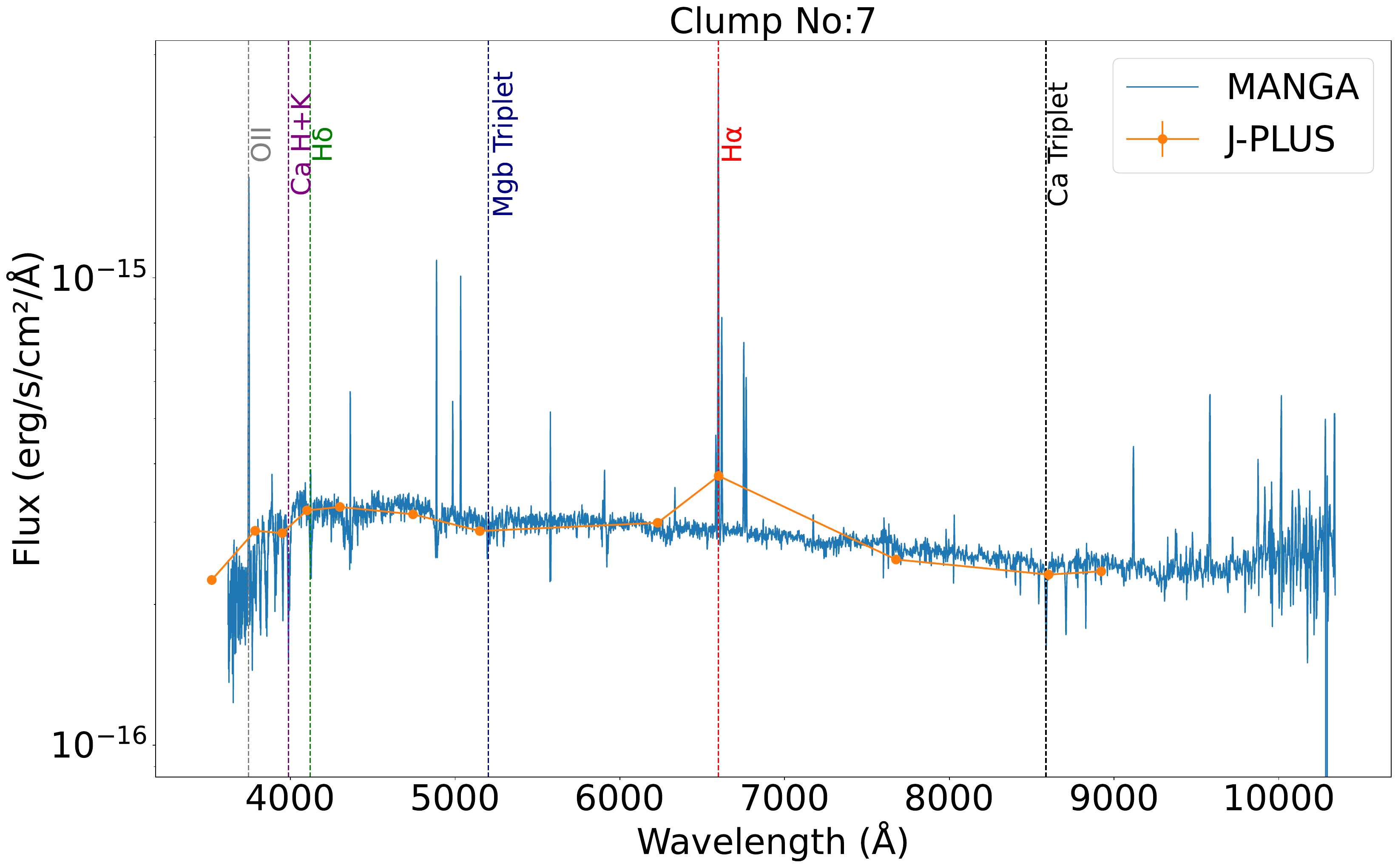}
    \includegraphics[width=9cm]{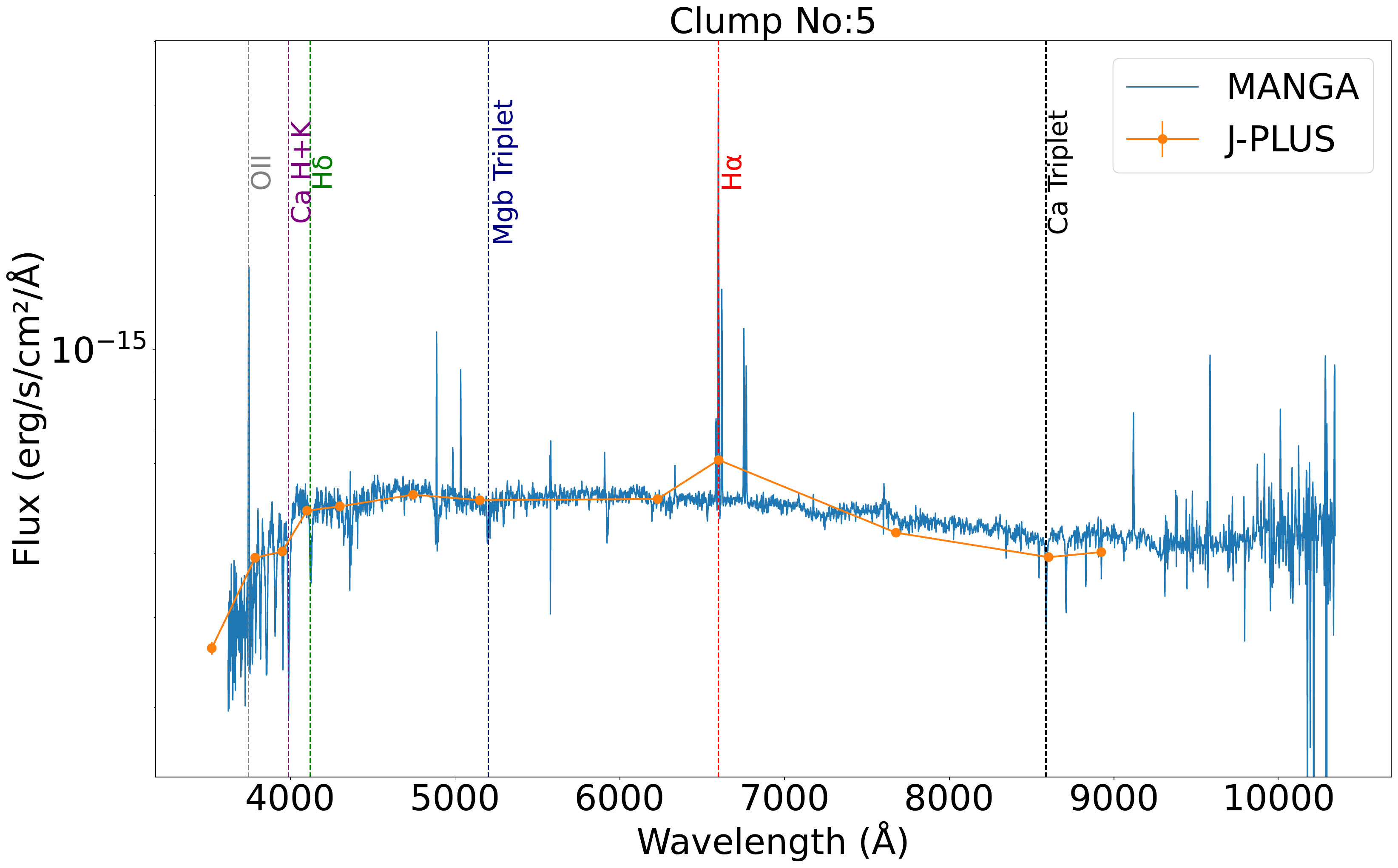}
    \includegraphics[width=9cm]{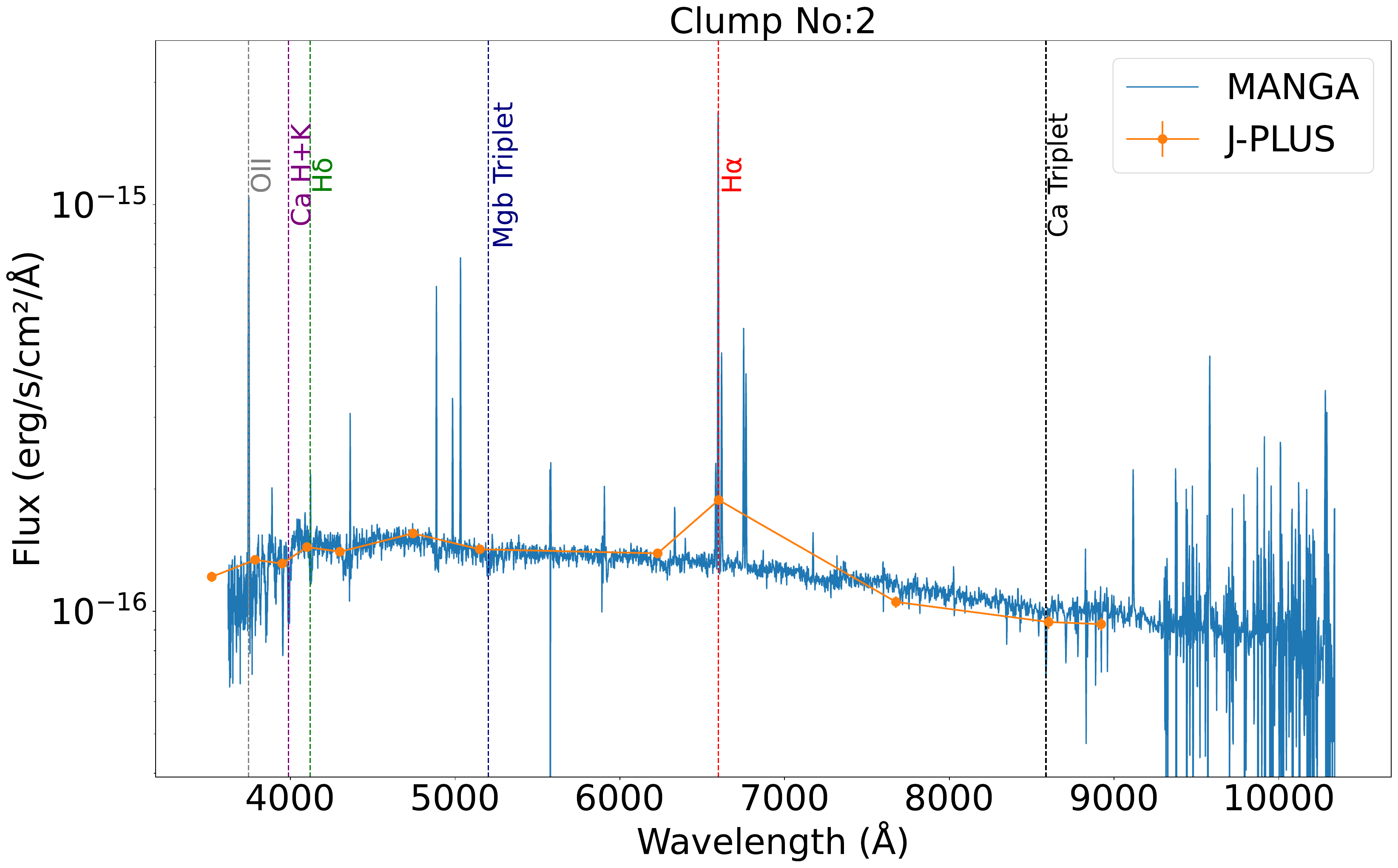}
    \caption{Top: H$\alpha$+[NII] map and segmentation map of MaNGA-8150-6103. Bottom: Examples of J-PLUS SEDs in comparison with MANGA spectra of the clumps.}
    \label{fig:SED-manga}
\end{figure*}
\subsubsection{Clump detection and SEDs} \label{sec:amodule-sed}
 Here we discuss the steps for the identification and SED construction of the emission line regions. \texttt{NoiseChisel} and \texttt{Segment} \citep{2015Akhlaghi,2019noisechisel} of Gnuastro are good tools for identifying the emission line regions in galaxies. \texttt{NoiseChisel} is used to separate the signals from the noise and estimate the background of the images. The sky-subtracted images are then used to identify the emission line regions in the galaxies using Gnuastro's \texttt{astsegment} program \citep{2019noisechisel}. It finds significant (compared to the ambient noise) local maxima over a region that is already detected to have a signal. \texttt{astsegment} works by growing pixels around the local maxima until the local minima are reached. Therefore, it is able to trace the shape of star-forming clumps accurately without any parametric assumptions. For more, see Section 3.2 of \cite{2015Akhlaghi} and Section 3 of \cite{2019noisechisel}.
 
 The segmentation map output provides the pixels associated to each clump. Using the position and area of the clumps detected from the J-PLUS H$\alpha$ emaps, we measured the total fluxes of clumps in all 12 filters for constructing their SEDs. \texttt{astmkcatalog} is used for making the catalog of the detected clumps from H$\alpha$+[NII] map in 12 filters. Then the SEDs are constructed using the fluxes from 12 filters.

Similarly, for getting the IFU spectra of the emission line regions detected in J-PLUS emaps, we used the same clump labeled images. For this, we first create a 3D cube of the segmentation map of J-PLUS with the same 3rd axis size in IFUs. Then using the Gnuastro's \texttt{MakeCatalog}, we measured the spectra catalog of all clumps. Then we compared the SEDs and the IFU spectra of every clump.

The SEDs and IFU spectra of some emission line regions of different galaxies are given from Fig.~\ref{fig:SED-muse} to Fig.~\ref{fig:SED-manga}. CALIFA cubes are already corrected for galactic extinction. Therefore, we have applied the galactic extinction correction for J-PLUS images using attenuation values in each filter \citep{2019Sanjuan} downloaded from J-PLUS the database. The extinction coefficients in J-PLUS were based on the recalibration of \cite{2011Schlafly} on \cite{1998Schlegel} maps. We removed this recalibration correction by multiplying the J-PLUS extinction coefficients by 1.16 to match the CALIFA extinction (see \citealt{2011Schlafly} for more details).

\section{Results and discussions} \label{sec:results}
\subsection{Comparison of H$\alpha$+[NII] emaps of J-PLUS with other IFUs}
The spatially resolved H$\alpha$ maps trace the star formation at $<$ 10 Myr, ionization mechanisms, mass assembly, and other galaxy properties. The precise continuum subtraction is essential for creating emaps that accurately estimate these physical properties of the observed regions. Here we discuss our findings from the comparison of the H$\alpha$ maps constructed from \texttt{J-SHE} pipeline with other IFUs.

\subsubsection{PHANGS-MUSE emaps} \label{sec:phangsemaps}
For the validation with PHANGS-MUSE, the two spiral galaxies that overlap with J-PLUS DR3 are used. Their H$\alpha$+[NII] map shows a smooth and continuous H$\alpha$ emission (Fig.~\ref{fig:rprof-ngc1087}) and the star-forming regions are mostly distributed along the spiral arms. The radial profiles of NGC 1087 from J-PLUS and PHANGS-MUSE demonstrate that the 3F method effectively reproduces emaps comparable to those derived from PHANGS-MUSE.
However, in the case of NGC 628, the flux in the cube is slightly lower compared to J-PLUS, possibly due to the over-subtraction of the sky in the PHANGS-MUSE observation of NGC 628 \citep{2022Emsellem}. This is because the NGC628 coadd is fully covered by the central part of the galaxy and does not have any pixels from outside of the galaxy's main disk. This problem did not occur for NGC1087 because the MUSE coadd extended sufficiently outside of the main disk to allow accurate estimation of the sky level. Also in the MUSE-PHANGS data, there is a significant drop in the flux near the center of NGC628 (near R $\sim$ 20 pixels), in contrast with the J-PLUS data and clearly indicates over subtraction due to an artifact arising from a bright star near the center of the galaxy.

\subsubsection{CALIFA emaps}
In the CALIFA sample, galaxies have different morphological types. 
The wavelength coverage of CALIFA only covered a part of the iSDSS filter, 64\% of the iSDSS J-PLUS filter curve is beyond the edge of the CALIFA wavelength range (see Fig.~\ref{fig:jfilt-ifuspec}). Therefore, we only used the CALIFA emap from the 3D method to validate the J-PLUS emaps. The H$\alpha$ + [NII] map shows that CALIFA galaxies exhibit different features of H$\alpha$ + [NII] emission, which can be observed along the rings, within spiral arms, and as clumpy structures. From Fig.~\ref{fig: emap-ngc628-ngc3395} and in Appendix Fig.~\ref{fig: califa} we show emaps of some star-forming CALIFA galaxies. 

\subsubsection{MaNGA emaps}
MaNGA provides high-spectral resolution with a larger spectral coverage for individual galaxies but covers a smaller FoV compared to other IFUs studied here. Due to its smaller field of view at the redshift range studied here, most of the MaNGA observations cover the center of galaxies or part of the galaxy regions. Fig.~\ref{fig: emap-manga-8309-3703} clearly shows the advantage of J-PLUS FoV compared to that of MaNGA. For this case, J-PLUS can provide the spatially resolved H$\alpha$ structure of the full extent of the galaxy in a single observation, while MaNGA only covers the central region of this galaxy. We have demonstrated emap comparison of some of the MaNGA galaxies in Fig.~\ref{fig: manga}. We note that the cases where the absorption present within the filter windows affects the estimation of the emap and the flux will be underestimated. 
\subsection{Radial profile flux ratios}
The ratio of J-PLUS to IFU emap radial profile can help in identifying regions of significant divergence and provides a straightforward way to compare the two emaps. This can reveal if one method consistently overestimates or underestimates a measurement. A ratio close to 1 indicates good agreement between the two methods and values significantly different from 1 suggest discrepancies. Here we have analyzed how the radial profile ratios vary with radius. We have plotted the ratio of the J-PLUS (2D) profile to the IFUs profile in the lower panel of Fig.~\ref{fig:rprof-ngc1087} to Fig.~\ref{fig:rprof-manga-8309-3703}.

The variation in the radial profile of a given galaxy obtained from different instruments can indeed be influenced by several factors. Instruments may have different background noise levels, different levels of sensitivity, depths, and different PSFs. In the comparison of PHANGS-MUSE with J-PLUS, the radial profile ratios of NGC1087 (left panel in Fig.~\ref{fig:rprof-ngc1087}) show that the values are consistently close to 1 in the two methods, suggesting that both methods yield similar results. But for NGC 628 (right panel in Fig.~\ref{fig:rprof-ngc1087}), the discrepancy is due to the difference in the sky subtraction (see sect.\ref{sec:phangsemaps}). Background noise can introduce large scatter in the radial profile ratios.

The scatter in the ratio profile also depends on the morphology. The spiral galaxies with continuous H$\alpha$ emission show strong agreement and a ratio of one in the radial profile in two different instruments. Galaxies with a clumpy or ring structure show some discrepancy in the region where there is very low flux because of the different background contribution, which leads to a large scatter in the ratio of radial profiles. For example, for CALIFA galaxies, (e.g., NGC2780, UGC03899, and NGC3600) in the appendix: Fig.~\ref{fig: califa} show a large scatter in the ratios, where there is less S/N. The ratio of the profile with high S/N shows values close to one with a maximum scatter of $\sim$0.2 dex.

We also noticed that the galaxies hosting AGN at the center (e.g., NGC 5443) show a variation in the flux at the center in J-PLUS (high) compared to CALIFA. This might be due to the optical variability of AGN and that the surveys were not observed at the same time.

\subsection{H$\alpha$ emission line regions}
The H$\alpha$ emission line is crucial to identify the star-forming regions and estimate their SFRs. The SEDs of these spatially resolved star-forming regions help us to investigate the stellar population within galaxies, and their properties such as age, metallicity, and luminosity provide insights into the evolution in galaxies. Because J-PLUS is a large-area photometric survey, it provides an efficient way to map spatially resolved star-forming regions in a large sample of galaxies, offering an alternative to spectroscopy, which is more time consuming.  In order to validate our method of building J-PLUS SEDs of H$\alpha$ emission line regions, we have compared the SEDs with the spectra from IFUs discussed in Section \ref{sec:data}. Later we are planning to fit the SEDs of these clumps to study the stellar population properties.

\subsubsection{Comparing J-PLUS SEDs with IFU spectra}
The upper panels of Figure \ref{fig:SED-muse} show the H$\alpha$ + [NII] map and the Gnuastro clump map of PHANGS-MUSE NGC 1087, while the lower panel shows the SEDs and IFU-spectra of some examples of emission line (mostly star-forming) regions. We have detected 108 clumps in this galaxy. The J-PLUS SEDs and the PHANGS-MUSE spectra show good agreement. The emission lines are clearly visible in the SEDs. 

The example for the CALIFA galaxy NGC 3395 is given in Fig.~\ref{fig:SED-califa}.
In CALIFA, some of the clumps show a slight variation in the blue part, probably due to the difference of Galactic extinction correction and also due to the offset in astrometry. The clumps in PHANGS-MUSE and MaNGA (Fig.~\ref{fig:SED-manga}) show a perfect match between the spectra and SEDs. Although PHANGS-MUSE and CALIFA have higher spectral resolution, the advantage of J-PLUS is the wide spectral coverage. Compared to PHANGS-MUSE, J-PLUS has extended wavelength coverage in the blue part of the spectra, which mainly has a Ca + H + K 3970 \angstrom~ absorption line, OII 3727 \angstrom~ emission-line doublet. On the other hand, compared to CALIFA, J-PLUS has a wider wavelength coverage in the redder part of the spectra, including the Ca Triplet 8544.44~\angstrom. J-PLUS has similar spectral coverage as MANGA and shows strong agreement between the SED and the spectra of the clumps.

\subsubsection{Ratio of J-PLUS and IFU H$\alpha$+[NII] fluxes}
\begin{figure}
    \centering
    \includegraphics[width=9cm]{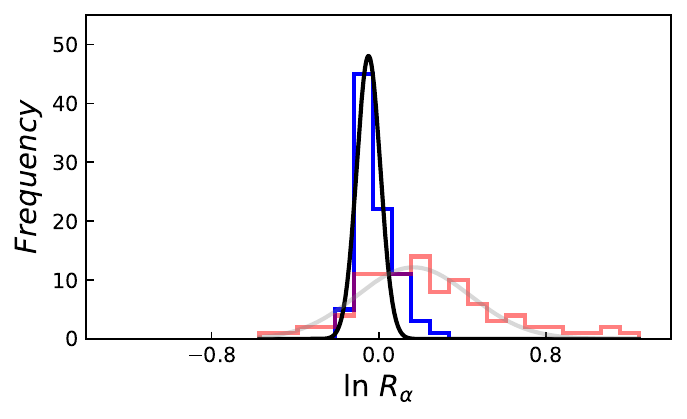}
   \caption{Distribution of the ratio of the H$\alpha$ flux (with the Gaussian fit) of emission line regions in J-PLUS to MUSE-PHANGS (blue) and MaNGA (red) IFUs.}
    \label{fig:clumpratio-muse}
\end{figure}

\begin{figure}
    \centering
    \hspace{-0.005\textwidth}
    \includegraphics[width=9.0cm]{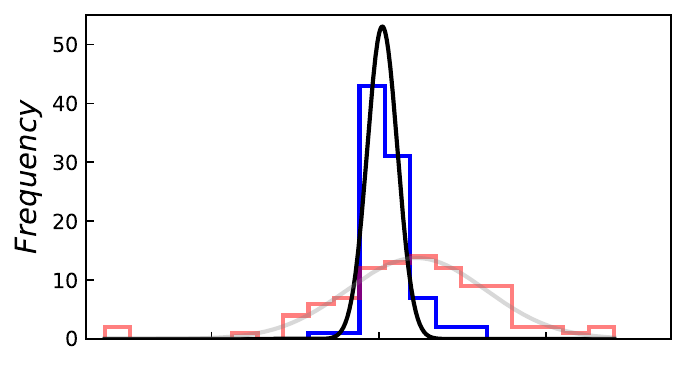}
    
    \vspace{-0.014\textwidth}
    \includegraphics[width=9cm]{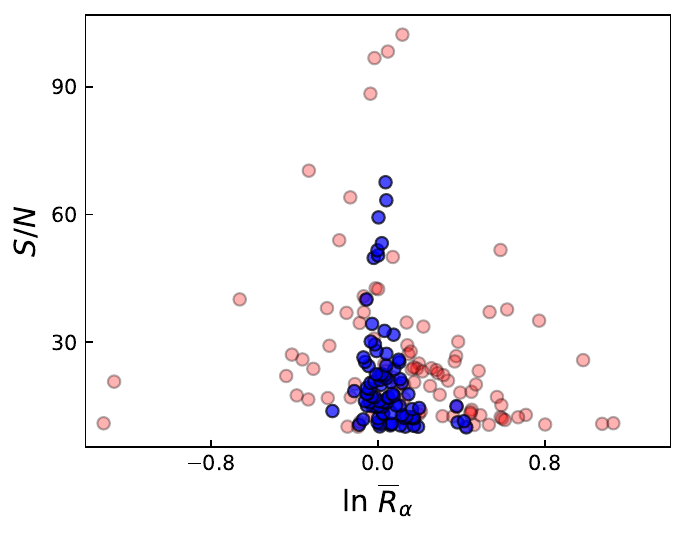}
    \caption{Upper: The distribution of $\ln \overline{R}_{\alpha}$ with the Gaussian fit for MUSE-PHANGS (blue) and MaNGA (red). Lower: $\ln \overline{R}_{\alpha}$ vs S/N plot for MUSE-PHANGS (blue) and MaNGA (red) emission line regions. }
    \label{fig:HaR-ratio}
\end{figure}

To quantify the goodness of our extraction method, we analyzed the ratio of the J-PLUS and IFU H$\alpha$+[NII] fluxes of detected emission line clumps, noted $R_{\alpha}$.
IFU emaps created from synthetic J-PLUS NBs (2D method) is used for this analysis.

For PHANGS-MUSE, we used 108 clumps detected in NGC 1087. We chose to analyze only NGC 1087 due to the oversubtraction of the sky in NGC 628, which will affect the reliability of the final results. We only consider clumps with S/N greater than $10$ to avoid low signal to noise areas. Fig.~\ref{fig:clumpratio-muse} shows the histogram of logarithm of $R_{\alpha}$ for $87$ high-quality clumps detected in NGC1087 (blue). The median value of the Gaussian fit is  $-0.04$.

We also computed the ratio from the $r-$band fluxes of the clumps, finding a similar difference between J-PLUS and PHANGS-MUSE as in the H$\alpha$ flux ratios (Fig.~\ref{fig:clumpratio-muse}). The $r-$band ratio, noted $R_{r}$, encodes a variety of observational differences between both datasets, such as flux calibration, sky subtraction, or effective FWHM. We corrected these effects in the H$\alpha$ ratio by just dividing each clump value with its $r-$band ratio, obtaining a normalized H$\alpha$ ratio ($\overline{R}_{\alpha}$). The distribution of $\ln \overline{R}_{\alpha}$, which is a proxy for fractional differences, is presented in Fig.~\ref{fig:HaR-ratio} (upper panel). We find a much better agreement between both datasets, with a median value of $0.02$. We fitted a Gaussian function to the distribution, finding $\ln \overline{R}_{\alpha} = 0.02 \pm 0.07$. That denotes that J-PLUS is recovering the H$\alpha$ flux with $2$\% bias and $7$\% dispersion. We also plotted $\ln \overline{R}_{\alpha}$ with S/N of the clumps. The scatter is more for lower S/N clumps than for higher S/N clumps.

In the case of MaNGA, a small part of each galaxy is covered and a few star-forming regions are detected. For the complete sample of MaNGA galaxies, we have detected a total of $169$ clumps. Again, we discarded the clumps with S/N less than 10 and also removed those detected in J-PLUS and not present in MaNGA. Explanation for those cases are beyond the scope of this paper. A more detailed study of these clumps will be discussed in a forthcoming article. The red histogram in Fig.~\ref{fig:clumpratio-muse} shows the distribution of ${R}_{\alpha}$, with a Gaussian fit yielding $\ln {R}_{\alpha} = 0.16 \pm 0.27$. After normalizing by the r-ratio, the resulting value of $\ln \overline{R}_{\alpha}$ value is $0.18 \pm 0.33$. We also calculated the ratios of the i and J0660 filters between J-PLUS and the synthetic J-PLUS filters from MaNGA, obtaining fitting values of $0.02 \pm 0.11$, $0.04 \pm 0.10$ and $0.00 \pm 0.14$ respectively. 
The difference in continuum subtracted H$\alpha$ fluxes is expected due to the different PSFs between J-PLUS (1.1") and MaNGA(2.5"). The much larger MaNGA PSF FWHM is due to the hexagonal grid of optical fiber configuration (2" core diameter fibers separated by 0.5") of the MaNGA survey which introduce systematic errors and underestimations of the flux measurements.

Since 2D emaps are not created for CALIFA due to incomplete coverage of the iSDSS filter's wavelength range, we exclude them from this analysis. Also, there is a slight difference in astrometry in CALIFA galaxies compared to the J-PLUS. We also noticed that there are some clumps that is detected in J-plus and not present in the CALIFA. Some of these were supernovae, which is already reported. The investigation of these objects are beyond the scope of this study. In future paper we will discuss more detail about the clumps in J-PLUS galaxies with their physics. 

\section{Conclusions} \label{sec:conclusions}
In this study, we developed and validated the J-SHE pipeline for constructing spatially resolved H$\alpha$+[NII] emission maps and extract the SEDs of every emission line region in nearby galaxies (z $<$ 0.0165) in J-PLUS DR3 in a very automated way. The main findings drawn from our study are summarized below.
\begin{itemize}
    \item The wide FoV, spatial resolution, wavelength coverage, and narrow band filters of J-PLUS are suitable for spatially resolved spectral studies of nearby galaxies.
    
    \item Emaps of H$\alpha$+[NII] were derived using two filters to trace the continuum ($r~and~i$) and one to trace the emission line ($J0660$). The methodology was validated with $158$ galaxies in common with the PHANGS-MUSE, CALIFA, and MaNGA IFU surveys.

    \item The J-PLUS emaps provide similar emission line regions and radial profiles when compared to the IFU surveys when analyzed using synthetic photometry to compute the J-PLUS extraction procedure. 

    \item The H$\alpha$+[NII] images from J-PLUS allowed us to extract photospectra for the star-forming clumps in the studied galaxy disks and their properties. This is less time-consuming as compared to spectroscopic surveys and can provide the SEDs of all the clumps within the wide field of view.

    \item The J-PLUS narrow bands cover key spectral lines such as [OII]3727\AA, Ca H+K, H$\alpha$, Ca triplet, and [OIII]4959, 5007\AA \ (z>0.006) providing valuable insights into various stages of stellar evolution and star formation processes in galaxies in resolved scales. The J-PLUS SEDs of the emission line regions clearly showcase these spectral features, mirroring the spectra extracted from IFU data in the same regions.

    \item The uniqueness of J-PLUS in comparison to other IFUs, is large FoV that allows us to map the complete extent of galaxy structure, galaxy interactions, galaxy groups, and galaxy clusters in a single FoV. J-Surveys can capture this wider contexts, providing insights into large-scale structures and galaxy interactions.
    This broader coverage can be advantageous for studying the global properties of galaxies and their environments. 

    \item The measured H$\alpha$+[NII] fluxes in the star-forming clumps from J-PLUS and the IFU data show a difference of 2\% and a dispersion of 7\% after correcting by differences in r-band. This validates the extraction and measurement procedures.
\end{itemize}

In subsequent papers, we will also provide the catalog of the SED information of all emission line regions in all nearby galaxies observed in J-PLUS DR3. This will help to study the spatially resolved stellar population and emission line properties of galaxies in statistical way. In the next project, we derive the resolved star formation main sequence using all the nearby galaxies having the spectroscopic redshift.

As a final note, the Javalambre Physics of the Accelerating Universe Astrophysical Survey \citep[J-PAS;][] {2014Benitez, 2016Dupke} will observe thousands of square degrees in the northern sky and will provide data for thousands of galaxies with significantly improved spatial resolution thanks to its $56$ NB filters in the optical. Its continuous set of filters functions like a low-resolution IFU over large areas, enabling a deeper understanding of galaxy formation and evolution on resolved scales.  

\section{Data availability} 
Table. A1 is only available in electronic form at the CDS via anonymous ftp to cdsarc.u-strasbg.fr (130.79.128.5) or via http://cdsweb.u-strasbg.fr/cgi-bin/qcat?J/A+A/.

\begin{acknowledgements}

RPT thanks the Spanish Ministry of Science and Innovation
(MCIN/AEI/10.13039/501100011033 y FEDER, Una manera de hacer Europa)
with grant No. PID2021-124918NA-C43, the Governments of Spain and Arag\'on through their general budgets and the Fondo de Inversiones de Teruel. RPT also thanks the European Union - NextGenerationEU through the Recovery and Resilience Facility program Planes Complementarios con las CCAA de Astrof\'{\i}sica y F\'{\i}sica de Altas Energ\'{\i}as - LA4. RPT acknowledge David Fernández Gil and Francisco Arizo Borillo for their valuable inputs during the analysis of this work.

ACS acknowledges funding from the brazilian agencies: Conselho Nacional de Desenvolvimento Cient\'ifico e Tecnol\'ogico (CNPq) and the Rio Grande do Sul Research Foundation (FAPERGS) through grants 
CNPq-314301/2021-6, FAPERGS- 24/2551-0001548-5.

AE acknowledges the financial support from the Spanish Ministry of Science and Innovation and the European Union - NextGenerationEU through the Recovery and Resilience Facility project ICTS-MRR-2021-03-CEFCA.

AAC acknowledges financial support from the Severo Ochoa grant CEX2021- 001131-S funded by MCIN/AEI/10.13039/501100011033 and the Spanish project PID2023-153123NB-I00, funded by MCIN/AEI.

Based on observations made with the JAST80 telescope and T80Cam camera for the J-PLUS project at the Observatorio Astrof\'{\i}sico de Javalambre (OAJ), in Teruel, owned, managed, and operated by the Centro de Estudios de F\'{\i}sica del  Cosmos de Arag\'on (CEFCA). We acknowledge the OAJ Data Processing and Archiving Unit (UPAD) for reducing the OAJ data used in this work.

Funding for the J-PLUS Project has been provided by the Governments of Spain and Arag\'on through the Fondo de Inversiones de Teruel; the Aragonese Government through the Research Groups E96, E103, E16\_17R, E16\_20R, and E16\_23R; the Spanish Ministry of Science and Innovation (MCIN/AEI/10.13039/501100011033 y FEDER, Una manera de hacer Europa) with grants PID2021-124918NB-C41, PID2021-124918NB-C42, PID2021-124918NA-C43, and PID2021-124918NB-C44; the Spanish Ministry of Science, Innovation and Universities (MCIU/AEI/FEDER, UE) with grants PGC2018-097585-B-C21 and PGC2018-097585-B-C22; the Spanish Ministry of Economy and Competitiveness (MINECO) under AYA2015-66211-C2-1-P, AYA2015-66211-C2-2, AYA2012-30789, and ICTS-2009-14; and European FEDER funding (FCDD10-4E-867, FCDD13-4E-2685). The Brazilian agencies FINEP, FAPESP, and the National Observatory of Brazil have also contributed to this project.

This work was partly done using GNU Astronomy Utilities (Gnuastro, ascl.net/1801.009) version 0.23.11-45775. Work on Gnuastro has been funded by the Japanese Ministry of Education, Culture, Sports, Science, and Technology (MEXT) scholarship and its Grant-in-Aid for Scientific Research (21244012, 24253003), the European Research Council (ERC) advanced grant 339659-MUSICOS, the Spanish Ministry of Economy and Competitiveness (MINECO, grant number AYA2016-76219-P) and the NextGenerationEU grant through the Recovery and Resilience Facility project ICTS-MRR-2021-03-CEFCA.

This work was carried out as part of the
PHANGS-MUSE collaboration. Based on observations collected at the European Southern Observatory under ESO programmes 1100.B-0651,
095.C-0473, and 094.C-0623 (PHANGS–MUSE; PI Schinnerer), as
well as 094.B-0321 (MAGNUM; PI Marconi), 099.B-0242, 0100.B-
0116, 098.B-0551 (MAD; PI Carollo) and 097.B-0640 (TIMER; PI
Gadotti). 

This study uses data provided by the Calar Alto Legacy Integral Field Area (CALIFA) survey (http://califa.caha.es/).
Based on observations collected at the Centro Astronómico Hispano Alemán (CAHA) at Calar Alto, operated jointly by the Max-Planck-Institut fűr Astronomie and the Instituto de Astrofísica de Andalucía (CSIC).

Funding for the Sloan Digital Sky 
Survey IV has been provided by the 
Alfred P. Sloan Foundation, the U.S. 
Department of Energy Office of 
Science, and the Participating 
Institutions. 

SDSS-IV acknowledges support and 
resources from the Center for High 
Performance Computing  at the 
University of Utah. The SDSS 
website is www.sdss4.org.

SDSS-IV is managed by the 
Astrophysical Research Consortium 
for the Participating Institutions 
of the SDSS Collaboration including 
the Brazilian Participation Group, 
the Carnegie Institution for Science, 
Carnegie Mellon University, Center for 
Astrophysics Harvard \& 
Smithsonian, the Chilean Participation 
Group, the French Participation Group, 
Instituto de Astrof\'isica de 
Canarias, The Johns Hopkins 
University, Kavli Institute for the 
Physics and Mathematics of the 
Universe (IPMU) / University of 
Tokyo, the Korean Participation Group, 
Lawrence Berkeley National Laboratory, 
Leibniz Institut f\"ur Astrophysik 
Potsdam (AIP),  Max-Planck-Institut 
f\"ur Astronomie (MPIA Heidelberg), 
Max-Planck-Institut f\"ur 
Astrophysik (MPA Garching), 
Max-Planck-Institut f\"ur 
Extraterrestrische Physik (MPE), 
National Astronomical Observatories of 
China, New Mexico State University, 
New York University, University of 
Notre Dame, Observat\'ario 
Nacional / MCTI, The Ohio State 
University, Pennsylvania State 
University, Shanghai 
Astronomical Observatory, United 
Kingdom Participation Group, 
Universidad Nacional Aut\'onoma 
de M\'exico, University of Arizona, 
University of Colorado Boulder, 
University of Oxford, University of 
Portsmouth, University of Utah, 
University of Virginia, University 
of Washington, University of 
Wisconsin, Vanderbilt University, 
and Yale University.
\end{acknowledgements}
\bibliography{ref}{}

\begin{thebibliography}{46}
\expandafter\ifx\csname natexlab\endcsname\relax\def\natexlab#1{#1}\fi

\bibitem[{{Akhlaghi}(2019{\natexlab{a}})}]{2019noisechisel}
{Akhlaghi}, M. 2019{\natexlab{a}}, arXiv e-prints, arXiv:1909.11230

\bibitem[{{Akhlaghi}(2019{\natexlab{b}})}]{2019Akhlaghi}
{Akhlaghi}, M. 2019{\natexlab{b}}, in Astronomical Society of the Pacific Conference Series, Vol. 521, Astronomical Data Analysis Software and Systems XXVI, ed. M.~{Molinaro}, K.~{Shortridge}, \& F.~{Pasian}, 299

\bibitem[{{Akhlaghi}(2024)}]{gnuastrobook}
{Akhlaghi}, M. 2024, GNU Astronomy Utilities manual version 0.23, DOI:10.5281/zenodo.12738457 (Free Software Foundation)

\bibitem[{{Akhlaghi} \& {Ichikawa}(2015)}]{2015Akhlaghi}
{Akhlaghi}, M. \& {Ichikawa}, T. 2015, \apjs, 220, 1

\bibitem[{Bacon {et~al.}(2010)Bacon, Accardo, Adjali, Anwand, Bauer, Biswas, Blaizot, Boudon, Brau-Nogue, Brinchmann, Caillier, Capoani, Carollo, Contini, Couderc, Daguisé, Deiries, Delabre, Dreizler, Dubois, Dupieux, Dupuy, Emsellem, Fechner, Fleischmann, François, Gallou, Gharsa, Glindemann, Gojak, Guiderdoni, Hansali, Hahn, Jarno, Kelz, Koehler, Kosmalski, Laurent, Le~Floch, Lilly, Lizon, Loupias, Manescau, Monstein, Nicklas, Olaya, Pares, Pasquini, Pécontal-Rousset, Pelló, Petit, Popow, Reiss, Remillieux, Renault, Roth, Rupprecht, Serre, Schaye, Soucail, Steinmetz, Streicher, Stuik, Valentin, Vernet, Weilbacher, Wisotzki, \& Yerle}]{2010Bacon}
Bacon, R., Accardo, M., Adjali, L., {et~al.} 2010, in Society of Photo-Optical Instrumentation Engineers (SPIE) Conference Series, Vol. 7735, Ground-based and Airborne Instrumentation for Astronomy III, ed. I.~S. {McLean}, S.~K. {Ramsay}, \& H.~{Takami}, 773508

\bibitem[{{Bacon} {et~al.}(1995){Bacon}, {Adam}, {Baranne}, {Courtes}, {Dubet}, {Dubois}, {Emsellem}, {Ferruit}, {Georgelin}, {Monnet}, {Pecontal}, {Rousset}, \& {Say}}]{1995Bacon}
{Bacon}, R., {Adam}, G., {Baranne}, A., {et~al.} 1995, \aaps, 113, 347

\bibitem[{{Bacon} {et~al.}(2014){Bacon}, {Vernet}, {Borisova}, {Bouch{\'e}}, {Brinchmann}, {Carollo}, {Carton}, {Caruana}, {Cerda}, {Contini}, {Franx}, {Girard}, {Guerou}, {Haddad}, {Hau}, {Herenz}, {Herrera}, {Husemann}, {Husser}, {Jarno}, {Kamann}, {Krajnovic}, {Lilly}, {Mainieri}, {Martinsson}, {Palsa}, {Patricio}, {P{\'e}contal}, {Pello}, {Piqueras}, {Richard}, {Sandin}, {Schroetter}, {Selman}, {Shirazi}, {Smette}, {Soto}, {Streicher}, {Urrutia}, {Weilbacher}, {Wisotzki}, \& {Zins}}]{2014Bacon}
{Bacon}, R., {Vernet}, J., {Borisova}, E., {et~al.} 2014, The Messenger, 157, 13

\bibitem[{{Barrera-Ballesteros} {et~al.}(2023){Barrera-Ballesteros}, {S{\'a}nchez}, {Espinosa-Ponce}, {L{\'o}pez-Cob{\'a}}, {Carigi}, {Lugo-Aranda}, {Lacerda}, {Bruzual}, {Hernandez-Toledo}, {Boardman}, {Drory}, {Lane}, \& {Brownstein}}]{2023Barrera-Ballesteros}
{Barrera-Ballesteros}, J.~K., {S{\'a}nchez}, S.~F., {Espinosa-Ponce}, C., {et~al.} 2023, \rmxaa, 59, 213

\bibitem[{{Benitez} {et~al.}(2014){Benitez}, {Dupke}, {Moles}, {Sodre}, {Cenarro}, {Marin-Franch}, {Taylor}, {Cristobal}, {Fernandez-Soto}, {Mendes de Oliveira}, {Cepa-Nogue}, {Abramo}, {Alcaniz}, {Overzier}, {Hernandez-Monteagudo}, {Alfaro}, {Kanaan}, {Carvano}, {Reis}, {Martinez Gonzalez}, {Ascaso}, {Ballesteros}, {Xavier}, {Varela}, {Ederoclite}, {Vazquez Ramio}, {Broadhurst}, {Cypriano}, {Angulo}, {Diego}, {Zandivarez}, {Diaz}, {Melchior}, {Umetsu}, {Spinelli}, {Zitrin}, {Coe}, {Yepes}, {Vielva}, {Sahni}, {Marcos-Caballero}, {Kitaura}, {Maroto}, {Masip}, {Tsujikawa}, {Carneiro}, {Gonzalez Nuevo}, {Carvalho}, {Reboucas}, {Carvalho}, {Abdalla}, {Bernui}, {Pigozzo}, {Ferreira}, {Chandrachani Devi}, {Bengaly}, {Campista}, {Amorim}, {Asari}, {Bongiovanni}, {Bonoli}, {Bruzual}, {Cardiel}, {Cava}, {Cid Fernandes}, {Coelho}, {Cortesi}, {Delgado}, {Diaz Garcia}, {Espinosa}, {Galliano}, {Gonzalez-Serrano}, {Falcon-Barroso}, {Fritz}, {Fernandes}, {Gorgas}, {Hoyos}, {Jimenez-Teja}, {Lopez-Aguerri}, {Lopez-San Juan},
  {Mateus}, {Molino}, {Novais}, {OMill}, {Oteo}, {Perez-Gonzalez}, {Poggianti}, {Proctor}, {Ricciardelli}, {Sanchez-Blazquez}, {Storchi-Bergmann}, {Telles}, {Schoennell}, {Trujillo}, {Vazdekis}, {Viironen}, {Daflon}, {Aparicio-Villegas}, {Rocha}, {Ribeiro}, {Borges}, {Martins}, {Marcolino}, {Martinez-Delgado}, {Perez-Torres}, {Siffert}, {Calvao}, {Sako}, {Kessler}, {Alvarez-Candal}, {De Pra}, {Roig}, {Lazzaro}, {Gorosabel}, {Lopes de Oliveira}, {Lima-Neto}, {Irwin}, {Liu}, {Alvarez}, {Balmes}, {Chueca}, {Costa-Duarte}, {da Costa}, {Dantas}, {Diaz}, {Fabregat}, {Ferrari}, {Gavela}, {Gracia}, {Gruel}, {Gutierrez}, {Guzman}, {Hernandez-Fernandez}, {Herranz}, {Hurtado-Gil}, {Jablonsky}, {Laporte}, {Le Tiran}, {Licandro}, {Lima}, {Martin}, {Martinez}, {Montero}, {Penteado}, {Pereira}, {Peris}, {Quilis}, {Sanchez-Portal}, {Soja}, {Solano}, {Torra}, \& {Valdivielso}}]{2014Benitez}
{Benitez}, N., {Dupke}, R., {Moles}, M., {et~al.} 2014, arXiv e-prints, arXiv:1403.5237

\bibitem[{{Ben{\'\i}tez} {et~al.}(2009){Ben{\'\i}tez}, {Gazta{\~n}aga}, {Miquel}, {Castander}, {Moles}, {Crocce}, {Fern{\'a}ndez-Soto}, {Fosalba}, {Ballesteros}, {Campa}, {Cardiel-Sas}, {Castilla}, {Crist{\'o}bal-Hornillos}, {Delfino}, {Fern{\'a}ndez}, {Fern{\'a}ndez-Sopuerta}, {Garc{\'\i}a-Bellido}, {Lobo}, {Mart{\'\i}nez}, {Ortiz}, {Pacheco}, {Paredes}, {Pons-Border{\'\i}a}, {S{\'a}nchez}, {S{\'a}nchez}, {Varela}, \& {de Vicente}}]{2009Ben}
{Ben{\'\i}tez}, N., {Gazta{\~n}aga}, E., {Miquel}, R., {et~al.} 2009, \apj, 691, 241

\bibitem[{{Bertin}(2011)}]{2011Bertin}
{Bertin}, E. 2011, in Astronomical Society of the Pacific Conference Series, Vol. 442, Astronomical Data Analysis Software and Systems XX, ed. I.~N. {Evans}, A.~{Accomazzi}, D.~J. {Mink}, \& A.~H. {Rots}, 435

\bibitem[{{Bertin} \& {Arnouts}(1996)}]{1996Bertin}
{Bertin}, E. \& {Arnouts}, S. 1996, \aaps, 117, 393

\bibitem[{{Blanton} {et~al.}(2017){Blanton}, {Bershady}, {Abolfathi}, {Albareti}, {Allende Prieto}, {Almeida}, {Alonso-Garc{\'\i}a}, {Anders}, {Anderson}, {Andrews}, {Aquino-Ort{\'\i}z}, {Arag{\'o}n-Salamanca}, {Argudo-Fern{\'a}ndez}, {Armengaud}, {Aubourg}, {Avila-Reese}, {Badenes}, {Bailey}, {Barger}, {Barrera-Ballesteros}, {Bartosz}, {Bates}, {Baumgarten}, {Bautista}, {Beaton}, {Beers}, {Belfiore}, {Bender}, {Berlind}, {Bernardi}, {Beutler}, {Bird}, {Bizyaev}, {Blanc}, {Blomqvist}, {Bolton}, {Boquien}, {Borissova}, {van den Bosch}, {Bovy}, {Brandt}, {Brinkmann}, {Brownstein}, {Bundy}, {Burgasser}, {Burtin}, {Busca}, {Cappellari}, {Delgado Carigi}, {Carlberg}, {Carnero Rosell}, {Carrera}, {Chanover}, {Cherinka}, {Cheung}, {G{\'o}mez Maqueo Chew}, {Chiappini}, {Choi}, {Chojnowski}, {Chuang}, {Chung}, {Cirolini}, {Clerc}, {Cohen}, {Comparat}, {da Costa}, {Cousinou}, {Covey}, {Crane}, {Croft}, {Cruz-Gonzalez}, {Garrido Cuadra}, {Cunha}, {Damke}, {Darling}, {Davies}, {Dawson}, {de la Macorra}, {Dell'Agli}, {De
  Lee}, {Delubac}, {Di Mille}, {Diamond-Stanic}, {Cano-D{\'\i}az}, {Donor}, {Downes}, {Drory}, {du Mas des Bourboux}, {Duckworth}, {Dwelly}, {Dyer}, {Ebelke}, {Eigenbrot}, {Eisenstein}, {Emsellem}, {Eracleous}, {Escoffier}, {Evans}, {Fan}, {Fern{\'a}ndez-Alvar}, {Fernandez-Trincado}, {Feuillet}, {Finoguenov}, {Fleming}, {Font-Ribera}, {Fredrickson}, {Freischlad}, {Frinchaboy}, {Fuentes}, {Galbany}, {Garcia-Dias}, {Garc{\'\i}a-Hern{\'a}ndez}, {Gaulme}, {Geisler}, {Gelfand}, {Gil-Mar{\'\i}n}, {Gillespie}, {Goddard}, {Gonzalez-Perez}, {Grabowski}, {Green}, {Grier}, {Gunn}, {Guo}, {Guy}, {Hagen}, {Hahn}, {Hall}, {Harding}, {Hasselquist}, {Hawley}, {Hearty}, {Gonzalez Hern{\'a}ndez}, {Ho}, {Hogg}, {Holley-Bockelmann}, {Holtzman}, {Holzer}, {Huehnerhoff}, {Hutchinson}, {Hwang}, {Ibarra-Medel}, {da Silva Ilha}, {Ivans}, {Ivory}, {Jackson}, {Jensen}, {Johnson}, {Jones}, {J{\"o}nsson}, {Jullo}, {Kamble}, {Kinemuchi}, {Kirkby}, {Kitaura}, {Klaene}, {Knapp}, {Kneib}, {Kollmeier}, {Lacerna}, {Lane}, {Lang}, {Law},
  {Lazarz}, {Lee}, {Le Goff}, {Liang}, {Li}, {Li}, {Lian}, {Lima}, {Lin}, {Lin}, {Bertran de Lis}, {Liu}, {de Icaza Lizaola}, {Long}, {Lucatello}, {Lundgren}, {MacDonald}, {Deconto Machado}, {MacLeod}, {Mahadevan}, {Geimba Maia}, {Maiolino}, {Majewski}, {Malanushenko}, {Malanushenko}, {Manchado}, {Mao}, {Maraston}, {Marques-Chaves}, {Masseron}, {Masters}, {McBride}, {McDermid}, {McGrath}, {McGreer}, {Medina Pe{\~n}a}, {Melendez}, {Merloni}, {Merrifield}, {Meszaros}, {Meza}, {Minchev}, {Minniti}, {Miyaji}, {More}, {Mulchaey}, {M{\"u}ller-S{\'a}nchez}, {Muna}, {Munoz}, {Myers}, {Nair}, {Nandra}, {Correa do Nascimento}, {Negrete}, {Ness}, {Newman}, {Nichol}, {Nidever}, {Nitschelm}, {Ntelis}, {O'Connell}, {Oelkers}, {Oravetz}, {Oravetz}, {Pace}, {Padilla}, {Palanque-Delabrouille}, {Alonso Palicio}, {Pan}, {Parejko}, {Parikh}, {P{\^a}ris}, {Park}, {Patten}, {Peirani}, {Pellejero-Ibanez}, {Penny}, {Percival}, {Perez-Fournon}, {Petitjean}, {Pieri}, {Pinsonneault}, {Pisani}, {Poleski}, {Prada}, {Prakash}, {Queiroz},
  {Raddick}, {Raichoor}, {Barboza Rembold}, {Richstein}, {Riffel}, {Riffel}, {Rix}, {Robin}, {Rockosi}, {Rodr{\'\i}guez-Torres}, {Roman-Lopes}, {Rom{\'a}n-Z{\'u}{\~n}iga}, {Rosado}, {Ross}, {Rossi}, {Ruan}, {Ruggeri}, {Rykoff}, {Salazar-Albornoz}, {Salvato}, {S{\'a}nchez}, {Aguado}, {S{\'a}nchez-Gallego}, {Santana}, {Santiago}, {Sayres}, {Schiavon}, {da Silva Schimoia}, {Schlafly}, {Schlegel}, {Schneider}, {Schultheis}, {Schuster}, {Schwope}, {Seo}, {Shao}, {Shen}, {Shetrone}, {Shull}, {Simon}, {Skinner}, {Skrutskie}, {Slosar}, {Smith}, {Sobeck}, {Sobreira}, {Somers}, {Souto}, {Stark}, {Stassun}, {Stauffer}, {Steinmetz}, {Storchi-Bergmann}, {Streblyanska}, {Stringfellow}, {Su{\'a}rez}, {Sun}, {Suzuki}, {Szigeti}, {Taghizadeh-Popp}, {Tang}, {Tao}, {Tayar}, {Tembe}, {Teske}, {Thakar}, {Thomas}, {Thompson}, {Tinker}, {Tissera}, {Tojeiro}, {Hernandez Toledo}, {de la Torre}, {Tremonti}, {Troup}, {Valenzuela}, {Martinez Valpuesta}, {Vargas-Gonz{\'a}lez}, {Vargas-Maga{\~n}a}, {Vazquez}, {Villanova}, {Vivek}, {Vogt},
  {Wake}, {Walterbos}, {Wang}, {Weaver}, {Weijmans}, {Weinberg}, {Westfall}, {Whelan}, {Wild}, {Wilson}, {Wood-Vasey}, {Wylezalek}, {Xiao}, {Yan}, {Yang}, {Ybarra}, {Y{\`e}che}, {Zakamska}, {Zamora}, {Zarrouk}, {Zasowski}, {Zhang}, {Zhao}, {Zheng}, {Zheng}, {Zhou}, {Zhou}, {Zhu}, {Zoccali}, \& {Zou}}]{2017Blanton}
{Blanton}, M.~R., {Bershady}, M.~A., {Abolfathi}, B., {et~al.} 2017, \aj, 154, 28

\bibitem[{{Bundy} {et~al.}(2015){Bundy}, {Bershady}, {Law}, {Yan}, {Drory}, {MacDonald}, {Wake}, {Cherinka}, {S{\'a}nchez-Gallego}, {Weijmans}, {Thomas}, {Tremonti}, {Masters}, {Coccato}, {Diamond-Stanic}, {Arag{\'o}n-Salamanca}, {Avila-Reese}, {Badenes}, {Falc{\'o}n-Barroso}, {Belfiore}, {Bizyaev}, {Blanc}, {Bland-Hawthorn}, {Blanton}, {Brownstein}, {Byler}, {Cappellari}, {Conroy}, {Dutton}, {Emsellem}, {Etherington}, {Frinchaboy}, {Fu}, {Gunn}, {Harding}, {Johnston}, {Kauffmann}, {Kinemuchi}, {Klaene}, {Knapen}, {Leauthaud}, {Li}, {Lin}, {Maiolino}, {Malanushenko}, {Malanushenko}, {Mao}, {Maraston}, {McDermid}, {Merrifield}, {Nichol}, {Oravetz}, {Pan}, {Parejko}, {Sanchez}, {Schlegel}, {Simmons}, {Steele}, {Steinmetz}, {Thanjavur}, {Thompson}, {Tinker}, {van den Bosch}, {Westfall}, {Wilkinson}, {Wright}, {Xiao}, \& {Zhang}}]{2015Bundy}
{Bundy}, K., {Bershady}, M.~A., {Law}, D.~R., {et~al.} 2015, \apj, 798, 7

\bibitem[{{Cardelli} {et~al.}(1989){Cardelli}, {Clayton}, \& {Mathis}}]{1989Cardelli}
{Cardelli}, J.~A., {Clayton}, G.~C., \& {Mathis}, J.~S. 1989, \apj, 345, 245

\bibitem[{{Cenarro} {et~al.}(2019){Cenarro}, {Moles}, {Crist{\'o}bal-Hornillos}, {Mar{\'\i}n-Franch}, {Ederoclite}, {Varela}, {L{\'o}pez-Sanjuan}, {Hern{\'a}ndez-Monteagudo}, {Angulo}, {V{\'a}zquez Rami{\'o}}, {Viironen}, {Bonoli}, {Orsi}, {Hurier}, {San Roman}, {Greisel}, {Vilella-Rojo}, {D{\'\i}az-Garc{\'\i}a}, {Logro{\~n}o-Garc{\'\i}a}, {Gurung-L{\'o}pez}, {Spinoso}, {Izquierdo-Villalba}, {Aguerri}, {Allende Prieto}, {Bonatto}, {Carvano}, {Chies-Santos}, {Daflon}, {Dupke}, {Falc{\'o}n-Barroso}, {Gon{\c{c}}alves}, {Jim{\'e}nez-Teja}, {Molino}, {Placco}, {Solano}, {Whitten}, {Abril}, {Ant{\'o}n}, {Bello}, {Bielsa de Toledo}, {Castillo-Ram{\'\i}rez}, {Chueca}, {Civera}, {D{\'\i}az-Mart{\'\i}n}, {Dom{\'\i}nguez-Mart{\'\i}nez}, {Garzar{\'a}n-Calderaro}, {Hern{\'a}ndez-Fuertes}, {Iglesias-Marzoa}, {I{\~n}iguez}, {Jim{\'e}nez Ruiz}, {Kruuse}, {Lamadrid}, {Lasso-Cabrera}, {L{\'o}pez-Alegre}, {L{\'o}pez-Sainz}, {Ma{\'\i}cas}, {Moreno-Signes}, {Muniesa}, {Rodr{\'\i}guez-Llano}, {Rueda-Teruel}, {Rueda-Teruel},
  {Soriano-Lagu{\'\i}a}, {Tilve}, {Valdivielso}, {Yanes-D{\'\i}az}, {Alcaniz}, {Mendes de Oliveira}, {Sodr{\'e}}, {Coelho}, {Lopes de Oliveira}, {Tamm}, {Xavier}, {Abramo}, {Akras}, {Alfaro}, {Alvarez-Candal}, {Ascaso}, {Beasley}, {Beers}, {Borges Fernandes}, {Bruzual}, {Buzzo}, {Carrasco}, {Cepa}, {Cortesi}, {Costa-Duarte}, {De Pr{\'a}}, {Favole}, {Galarza}, {Galbany}, {Garcia}, {Gonz{\'a}lez Delgado}, {Gonz{\'a}lez-Serrano}, {Guti{\'e}rrez-Soto}, {Hernandez-Jimenez}, {Kanaan}, {Kuncarayakti}, {Landim}, {Laur}, {Licandro}, {Lima Neto}, {Lyman}, {Ma{\'\i}z Apell{\'a}niz}, {Miralda-Escud{\'e}}, {Morate}, {Nogueira-Cavalcante}, {Novais}, {Oncins}, {Oteo}, {Overzier}, {Pereira}, {Rebassa-Mansergas}, {Reis}, {Roig}, {Sako}, {Salvador-Rusi{\~n}ol}, {Sampedro}, {S{\'a}nchez-Bl{\'a}zquez}, {Santos}, {Schmidtobreick}, {Siffert}, {Telles}, \& {Vilchez}}]{2019Cenarro}
{Cenarro}, A.~J., {Moles}, M., {Crist{\'o}bal-Hornillos}, D., {et~al.} 2019, \aap, 622, A176

\bibitem[{{Cenarro} {et~al.}(2014){Cenarro}, {Moles}, {Mar{\'\i}n-Franch}, {Crist{\'o}bal-Hornillos}, {Yanes D{\'\i}az}, {Ederoclite}, {Varela}, {V{\'a}zquez Rami{\'o}}, {Valdivielso}, {Ben{\'\i}tez}, {Cepa}, {Dupke}, {Fern{\'a}ndez-Soto}, {Mendes de Oliveira}, {Sodr{\'e}}, {Taylor}, {Rueda-Teruel}, {Rueda-Teruel}, {Luis-Simoes}, {Chueca}, {Ant{\'o}n}, {Bello}, {D{\'\i}az-Mart{\'\i}n}, {Guill{\'e}n-Civera}, {Hern{\'a}ndez-Fuertes}, {Iglesias-Marzoa}, {Jim{\'e}nez-Mej{\'\i}as}, {Lasso-Cabrera}, {L{\'o}pez-Alegre}, {L{\'o}pez-Sainz}, {Rodr{\'\i}guez-Hern{\'a}ndez}, {Su{\'a}rez}, {Lamadrid}, {Ma{\'\i}cas}, {Abril-Iba{\~n}ez}, {Tilve}, \& {Rodr{\'\i}guez-Llano}}]{2014Cenarro}
{Cenarro}, A.~J., {Moles}, M., {Mar{\'\i}n-Franch}, A., {et~al.} 2014, in Society of Photo-Optical Instrumentation Engineers (SPIE) Conference Series, Vol. 9149, Observatory Operations: Strategies, Processes, and Systems V, ed. A.~B. {Peck}, C.~R. {Benn}, \& R.~L. {Seaman}, 91491I

\bibitem[{{Croom} {et~al.}(2012){Croom}, {Lawrence}, {Bland-Hawthorn}, {Bryant}, {Fogarty}, {Richards}, {Goodwin}, {Farrell}, {Miziarski}, {Heald}, {Jones}, {Lee}, {Colless}, {Brough}, {Hopkins}, {Bauer}, {Birchall}, {Ellis}, {Horton}, {Leon-Saval}, {Lewis}, {L{\'o}pez-S{\'a}nchez}, {Min}, {Trinh}, \& {Trowland}}]{2012Croom}
{Croom}, S.~M., {Lawrence}, J.~S., {Bland-Hawthorn}, J., {et~al.} 2012, \mnras, 421, 872

\bibitem[{{Dupke} {et~al.}(2016){Dupke}, {Benitez}, {Moles}, {Sodre}, {Irwin}, \& {J-PAS Collaboration}}]{2016Dupke}
{Dupke}, R.~A., {Benitez}, N., {Moles}, M., {et~al.} 2016, in American Astronomical Society Meeting Abstracts, Vol. 227, American Astronomical Society Meeting Abstracts \#227, 349.17

\bibitem[{{Emsellem} {et~al.}(2022){Emsellem}, {Schinnerer}, {Santoro}, {Belfiore}, {Pessa}, {McElroy}, {Blanc}, {Congiu}, {Groves}, {Ho}, {Kreckel}, {Razza}, {Sanchez-Blazquez}, {Egorov}, {Faesi}, {Klessen}, {Leroy}, {Meidt}, {Querejeta}, {Rosolowsky}, {Scheuermann}, {Anand}, {Barnes}, {Be{\v{s}}li{\'c}}, {Bigiel}, {Boquien}, {Cao}, {Chevance}, {Dale}, {Eibensteiner}, {Glover}, {Grasha}, {Henshaw}, {Hughes}, {Koch}, {Kruijssen}, {Lee}, {Liu}, {Pan}, {Pety}, {Saito}, {Sandstrom}, {Schruba}, {Sun}, {Thilker}, {Usero}, {Watkins}, \& {Williams}}]{2022Emsellem}
{Emsellem}, E., {Schinnerer}, E., {Santoro}, F., {et~al.} 2022, \aap, 659, A191

\bibitem[{{Infante-Sainz} {et~al.}(2024){Infante-Sainz}, {Akhlaghi}, \& {Eskandarlou}}]{2024Infante-Sainz}
{Infante-Sainz}, R., {Akhlaghi}, M., \& {Eskandarlou}, S. 2024, Research Notes of the American Astronomical Society, 8, 22

\bibitem[{{Jakobsen} {et~al.}(2022){Jakobsen}, {Ferruit}, {Alves de Oliveira}, {Arribas}, {Bagnasco}, {Barho}, {Beck}, {Birkmann}, {B{\"o}ker}, {Bunker}, {Charlot}, {de Jong}, {de Marchi}, {Ehrenwinkler}, {Falcolini}, {Fels}, {Franx}, {Franz}, {Funke}, {Giardino}, {Gnata}, {Holota}, {Honnen}, {Jensen}, {Jentsch}, {Johnson}, {Jollet}, {Karl}, {Kling}, {K{\"o}hler}, {Kolm}, {Kumari}, {Lander}, {Lemke}, {L{\'o}pez-Caniego}, {L{\"u}tzgendorf}, {Maiolino}, {Manjavacas}, {Marston}, {Maschmann}, {Maurer}, {Messerschmidt}, {Moseley}, {Mosner}, {Mott}, {Muzerolle}, {Pirzkal}, {Pittet}, {Plitzke}, {Posselt}, {Rapp}, {Rauscher}, {Rawle}, {Rix}, {R{\"o}del}, {Rumler}, {Sabbi}, {Salvignol}, {Schmid}, {Sirianni}, {Smith}, {Strada}, {te Plate}, {Valenti}, {Wettemann}, {Wiehe}, {Wiesmayer}, {Willott}, {Wright}, {Zeidler}, \& {Zincke}}]{2022Jakobsen}
{Jakobsen}, P., {Ferruit}, P., {Alves de Oliveira}, C., {et~al.} 2022, \aap, 661, A80

\bibitem[{{James} {et~al.}(2004){James}, {Shane}, {Beckman}, {Cardwell}, {Collins}, {Etherton}, {de Jong}, {Fathi}, {Knapen}, {Peletier}, {Percival}, {Pollacco}, {Seigar}, {Stedman}, \& {Steele}}]{2004James}
{James}, P.~A., {Shane}, N.~S., {Beckman}, J.~E., {et~al.} 2004, \aap, 414, 23

\bibitem[{{Kennicutt}(1998)}]{Kennicutt1998}
{Kennicutt}, Jr., R.~C. 1998, \araa, 36, 189

\bibitem[{{Kreckel} {et~al.}(2021){Kreckel}, {Emsellem}, {Schinnerer}, {Santoro}, {Belfiore}, {Pessa}, \& {Phangs Collaboration}}]{2021Kreckel}
{Kreckel}, K., {Emsellem}, E., {Schinnerer}, E., {et~al.} 2021, in American Astronomical Society Meeting Abstracts, Vol. 238, American Astronomical Society Meeting Abstracts, 310.05

\bibitem[{{Le F{\`e}vre} {et~al.}(2003){Le F{\`e}vre}, {Saisse}, {Mancini}, {Brau-Nogue}, {Caputi}, {Castinel}, {D'Odorico}, {Garilli}, {Kissler-Patig}, {Lucuix}, {Mancini}, {Pauget}, {Sciarretta}, {Scodeggio}, {Tresse}, \& {Vettolani}}]{2003Oliver}
{Le F{\`e}vre}, O., {Saisse}, M., {Mancini}, D., {et~al.} 2003, in Society of Photo-Optical Instrumentation Engineers (SPIE) Conference Series, Vol. 4841, Instrument Design and Performance for Optical/Infrared Ground-based Telescopes, ed. M.~{Iye} \& A.~F.~M. {Moorwood}, 1670--1681

\bibitem[{{Lee} {et~al.}(2023){Lee}, {Sandstrom}, {Leroy}, {Thilker}, {Schinnerer}, {Rosolowsky}, {Larson}, {Egorov}, {Williams}, {Schmidt}, {Emsellem}, {Anand}, {Barnes}, {Belfiore}, {Be{\v{s}}li{\'c}}, {Bigiel}, {Blanc}, {Bolatto}, {Boquien}, {den Brok}, {Cao}, {Chandar}, {Chastenet}, {Chevance}, {Chiang}, {Congiu}, {Dale}, {Deger}, {Eibensteiner}, {Faesi}, {Glover}, {Grasha}, {Groves}, {Hassani}, {Henny}, {Henshaw}, {Hoyer}, {Hughes}, {Jeffreson}, {Jim{\'e}nez-Donaire}, {Kim}, {Kim}, {Klessen}, {Koch}, {Kreckel}, {Kruijssen}, {Li}, {Liu}, {Lopez}, {Maschmann}, {Chen}, {Meidt}, {Murphy}, {Neumann}, {Neumayer}, {Pan}, {Pessa}, {Pety}, {Querejeta}, {Pinna}, {Rodr{\'\i}guez}, {Saito}, {S{\'a}nchez-Bl{\'a}zquez}, {Santoro}, {Sardone}, {Smith}, {Sormani}, {Scheuermann}, {Stuber}, {Sutter}, {Sun}, {Teng}, {Tre{\ss}}, {Usero}, {Watkins}, {Whitmore}, \& {Razza}}]{2023Lee}
{Lee}, J.~C., {Sandstrom}, K.~M., {Leroy}, A.~K., {et~al.} 2023, \apjl, 944, L17

\bibitem[{{Lee} {et~al.}(2022){Lee}, {Whitmore}, {Thilker}, {Deger}, {Larson}, {Ubeda}, {Anand}, {Boquien}, {Chandar}, {Dale}, {Emsellem}, {Leroy}, {Rosolowsky}, {Schinnerer}, {Schmidt}, {Lilly}, {Turner}, {Van Dyk}, {White}, {Barnes}, {Belfiore}, {Bigiel}, {Blanc}, {Cao}, {Chevance}, {Congiu}, {Egorov}, {Glover}, {Grasha}, {Groves}, {Henshaw}, {Hughes}, {Klessen}, {Koch}, {Kreckel}, {Kruijssen}, {Liu}, {Lopez}, {Mayker}, {Meidt}, {Murphy}, {Pan}, {Pety}, {Querejeta}, {Razza}, {Saito}, {S{\'a}nchez-Bl{\'a}zquez}, {Santoro}, {Sardone}, {Scheuermann}, {Schruba}, {Sun}, {Usero}, {Watkins}, \& {Williams}}]{2022Lee}
{Lee}, J.~C., {Whitmore}, B.~C., {Thilker}, D.~A., {et~al.} 2022, \apjs, 258, 10

\bibitem[{{Leroy} {et~al.}(2021){Leroy}, {Schinnerer}, {Hughes}, {Rosolowsky}, {Pety}, {Schruba}, {Usero}, {Blanc}, {Chevance}, {Emsellem}, {Faesi}, {Herrera}, {Liu}, {Meidt}, {Querejeta}, {Saito}, {Sandstrom}, {Sun}, {Williams}, {Anand}, {Barnes}, {Behrens}, {Belfiore}, {Benincasa}, {Be{\v{s}}li{\'c}}, {Bigiel}, {Bolatto}, {den Brok}, {Cao}, {Chandar}, {Chastenet}, {Chiang}, {Congiu}, {Dale}, {Deger}, {Eibensteiner}, {Egorov}, {Garc{\'\i}a-Rodr{\'\i}guez}, {Glover}, {Grasha}, {Henshaw}, {Ho}, {Kepley}, {Kim}, {Klessen}, {Kreckel}, {Koch}, {Kruijssen}, {Larson}, {Lee}, {Lopez}, {Machado}, {Mayker}, {McElroy}, {Murphy}, {Ostriker}, {Pan}, {Pessa}, {Puschnig}, {Razza}, {S{\'a}nchez-Bl{\'a}zquez}, {Santoro}, {Sardone}, {Scheuermann}, {Sliwa}, {Sormani}, {Stuber}, {Thilker}, {Turner}, {Utomo}, {Watkins}, \& {Whitmore}}]{2021Leroy}
{Leroy}, A.~K., {Schinnerer}, E., {Hughes}, A., {et~al.} 2021, \apjs, 257, 43

\bibitem[{{Logro{\~n}o-Garc{\'\i}a} {et~al.}(2019){Logro{\~n}o-Garc{\'\i}a}, {Vilella-Rojo}, {L{\'o}pez-Sanjuan}, {Varela}, {Viironen}, {Muniesa}, {Cenarro}, {Crist{\'o}bal-Hornillos}, {Ederoclite}, {Mar{\'\i}n-Franch}, {Moles}, {V{\'a}zquez Rami{\'o}}, {Bonoli}, {D{\'\i}az-Garc{\'\i}a}, {Orsi}, {San Roman}, {Akras}, {Chies-Santos}, {Coelho}, {Daflon}, {Costa-Duarte}, {Dupke}, {Galbany}, {Gonz{\'a}lez Delgado}, {Hernandez-Jimenez}, {Lopes de Oliveira}, {Mendes de Oliveira}, {Oteo}, {Gon{\c{c}}alves}, {S{\'a}nchez-Portal}, {Schmidtobreick}, \& {Sodr{\'e}}}]{2019Garcia}
{Logro{\~n}o-Garc{\'\i}a}, R., {Vilella-Rojo}, G., {L{\'o}pez-Sanjuan}, C., {et~al.} 2019, \aap, 622, A180

\bibitem[{{L{\'o}pez-Sanjuan} {et~al.}(2019){L{\'o}pez-Sanjuan}, {Varela}, {Crist{\'o}bal-Hornillos}, {V{\'a}zquez Rami{\'o}}, {Carrasco}, {Tremblay}, {Whitten}, {Placco}, {Mar{\'\i}n-Franch}, {Cenarro}, {Ederoclite}, {Alfaro}, {Coelho}, {Civera}, {Hern{\'a}ndez-Fuertes}, {Jim{\'e}nez-Esteban}, {Jim{\'e}nez-Teja}, {Ma{\'\i}z Apell{\'a}niz}, {Sobral}, {V{\'\i}lchez}, {Alcaniz}, {Angulo}, {Dupke}, {Hern{\'a}ndez-Monteagudo}, {Mendes de Oliveira}, {Moles}, \& {Sodr{\'e}}}]{2019Sanjuan}
{L{\'o}pez-Sanjuan}, C., {Varela}, J., {Crist{\'o}bal-Hornillos}, D., {et~al.} 2019, \aap, 631, A119

\bibitem[{{L{\'o}pez-Sanjuan} {et~al.}(2024){L{\'o}pez-Sanjuan}, {V{\'a}zquez Rami{\'o}}, {Xiao}, {Yuan}, {Carrasco}, {Varela}, {Crist{\'o}bal-Hornillos}, {Tremblay}, {Ederoclite}, {Mar{\'\i}n-Franch}, {Cenarro}, {Coelho}, {Daflon}, {del Pino}, {Dom{\'\i}nguez S{\'a}nchez}, {Fern{\'a}ndez-Ontiveros}, {Hern{\'a}n-Caballero}, {Jim{\'e}nez-Esteban}, {Alcaniz}, {Angulo}, {Dupke}, {Hern{\'a}ndez-Monteagudo}, {Moles}, \& {Sodr{\'e}}}]{2024lopez}
{L{\'o}pez-Sanjuan}, C., {V{\'a}zquez Rami{\'o}}, H., {Xiao}, K., {et~al.} 2024, \aap, 683, A29

\bibitem[{{Lumbreras-Calle} {et~al.}(2022){Lumbreras-Calle}, {L{\'o}pez-Sanjuan}, {Sobral}, {Fern{\'a}ndez-Ontiveros}, {V{\'\i}lchez}, {Hern{\'a}n-Caballero}, {Akhlaghi}, {D{\'\i}az-Garc{\'\i}a}, {Alcaniz}, {Angulo}, {Cenarro}, {Crist{\'o}bal-Hornillos}, {Dupke}, {Ederoclite}, {Hern{\'a}ndez-Monteagudo}, {Mar{\'\i}n-Franch}, {Moles}, {Sodr{\'e}}, {V{\'a}zquez Rami{\'o}}, \& {Varela}}]{2022Lumbreras-Calle}
{Lumbreras-Calle}, A., {L{\'o}pez-Sanjuan}, C., {Sobral}, D., {et~al.} 2022, \aap, 668, A60

\bibitem[{{Miller} {et~al.}(2000){Miller}, {Bureau}, {Verolme}, {de Zeeuw}, {Bacon}, {Copin}, {Emsellem}, {Davies}, {Peletier}, {Allington-Smith}, {Carollo}, \& {Monnet}}]{2000Miller}
{Miller}, B.~W., {Bureau}, M., {Verolme}, E., {et~al.} 2000, in Astronomical Society of the Pacific Conference Series, Vol. 195, Imaging the Universe in Three Dimensions, ed. W.~{van Breugel} \& J.~{Bland-Hawthorn}, 158

\bibitem[{{Moles} {et~al.}(2008){Moles}, {Ben{\'\i}tez}, {Aguerri}, {Alfaro}, {Broadhurst}, {Cabrera-Ca{\~n}o}, {Castander}, {Cepa}, {Cervi{\~n}o}, {Crist{\'o}bal-Hornillos}, {Fern{\'a}ndez-Soto}, {Gonz{\'a}lez Delgado}, {Infante}, {M{\'a}rquez}, {Mart{\'\i}nez}, {Masegosa}, {del Olmo}, {Perea}, {Prada}, {Quintana}, \& {S{\'a}nchez}}]{2008Moles}
{Moles}, M., {Ben{\'\i}tez}, N., {Aguerri}, J.~A.~L., {et~al.} 2008, \aj, 136, 1325

\bibitem[{{Novais} \& {Sodr{\'e}}(2019)}]{2019Novais}
{Novais}, P.~M. \& {Sodr{\'e}}, L. 2019, \mnras, 482, 2717

\bibitem[{{Piqueras} {et~al.}(2019){Piqueras}, {Conseil}, {Shepherd}, {Bacon}, {Leclercq}, \& {Richard}}]{2019Piqueras}
{Piqueras}, L., {Conseil}, S., {Shepherd}, M., {et~al.} 2019, in Astronomical Society of the Pacific Conference Series, Vol. 521, Astronomical Data Analysis Software and Systems XXVI, ed. M.~{Molinaro}, K.~{Shortridge}, \& F.~{Pasian}, 545

\bibitem[{{S{\'a}nchez} {et~al.}(2023){S{\'a}nchez}, {Galbany}, {Walcher}, {Garc{\'\i}a-Benito}, \& {Barrera-Ballesteros}}]{2023Sanchez}
{S{\'a}nchez}, S.~F., {Galbany}, L., {Walcher}, C.~J., {Garc{\'\i}a-Benito}, R., \& {Barrera-Ballesteros}, J.~K. 2023, \mnras, 526, 5555

\bibitem[{{S{\'a}nchez} {et~al.}(2012){S{\'a}nchez}, {Kennicutt}, {Gil de Paz}, {van de Ven}, {V{\'\i}lchez}, {Wisotzki}, {Walcher}, {Mast}, {Aguerri}, {Albiol-P{\'e}rez}, {Alonso-Herrero}, {Alves}, {Bakos}, {Bart{\'a}kov{\'a}}, {Bland-Hawthorn}, {Boselli}, {Bomans}, {Castillo-Morales}, {Cortijo-Ferrero}, {de Lorenzo-C{\'a}ceres}, {Del Olmo}, {Dettmar}, {D{\'\i}az}, {Ellis}, {Falc{\'o}n-Barroso}, {Flores}, {Gallazzi}, {Garc{\'\i}a-Lorenzo}, {Gonz{\'a}lez Delgado}, {Gruel}, {Haines}, {Hao}, {Husemann}, {Igl{\'e}sias-P{\'a}ramo}, {Jahnke}, {Johnson}, {Jungwiert}, {Kalinova}, {Kehrig}, {Kupko}, {L{\'o}pez-S{\'a}nchez}, {Lyubenova}, {Marino}, {M{\'a}rmol-Queralt{\'o}}, {M{\'a}rquez}, {Masegosa}, {Meidt}, {Mendez-Abreu}, {Monreal-Ibero}, {Montijo}, {Mour{\~a}o}, {Palacios-Navarro}, {Papaderos}, {Pasquali}, {Peletier}, {P{\'e}rez}, {P{\'e}rez}, {Quirrenbach}, {Rela{\~n}o}, {Rosales-Ortega}, {Roth}, {Ruiz-Lara}, {S{\'a}nchez-Bl{\'a}zquez}, {Sengupta}, {Singh}, {Stanishev}, {Trager}, {Vazdekis}, {Viironen}, {Wild},
  {Zibetti}, \& {Ziegler}}]{2012Sanches}
{S{\'a}nchez}, S.~F., {Kennicutt}, R.~C., {Gil de Paz}, A., {et~al.} 2012, \aap, 538, A8

\bibitem[{{Schlafly} \& {Finkbeiner}(2011)}]{2011Schlafly}
{Schlafly}, E.~F. \& {Finkbeiner}, D.~P. 2011, \apj, 737, 103

\bibitem[{{Schlegel} {et~al.}(1998){Schlegel}, {Finkbeiner}, \& {Davis}}]{1998Schlegel}
{Schlegel}, D.~J., {Finkbeiner}, D.~P., \& {Davis}, M. 1998, \apj, 500, 525

\bibitem[{{Sobral} {et~al.}(2015){Sobral}, {Matthee}, {Best}, {Smail}, {Khostovan}, {Milvang-Jensen}, {Kim}, {Stott}, {Calhau}, {Nayyeri}, \& {Mobasher}}]{2015Sobral}
{Sobral}, D., {Matthee}, J., {Best}, P.~N., {et~al.} 2015, \mnras, 451, 2303

\bibitem[{{Sobral} {et~al.}(2013){Sobral}, {Smail}, {Best}, {Geach}, {Matsuda}, {Stott}, {Cirasuolo}, \& {Kurk}}]{2013Sobral}
{Sobral}, D., {Smail}, I., {Best}, P.~N., {et~al.} 2013, \mnras, 428, 1128

\bibitem[{{Stroe} \& {Sobral}(2015)}]{2015Stroe}
{Stroe}, A. \& {Sobral}, D. 2015, \mnras, 453, 242

\bibitem[{{Vilella-Rojo} {et~al.}(2021){Vilella-Rojo}, {Logro{\~n}o-Garc{\'\i}a}, {L{\'o}pez-Sanjuan}, {Viironen}, {Varela}, {Moles}, {Cenarro}, {Crist{\'o}bal-Hornillos}, {Ederoclite}, {Hern{\'a}ndez-Monteagudo}, {Mar{\'\i}n-Franch}, {V{\'a}zquez Rami{\'o}}, {Galbany}, {Gonz{\'a}lez Delgado}, {Hern{\'a}n-Caballero}, {Lumbreras-Calle}, {S{\'a}nchez-Bl{\'a}zquez}, {Sobral}, {V{\'\i}lchez}, {Alcaniz}, {Angulo}, {Dupke}, \& {Sodr{\'e}}}]{2021Vilella-Rojo}
{Vilella-Rojo}, G., {Logro{\~n}o-Garc{\'\i}a}, R., {L{\'o}pez-Sanjuan}, C., {et~al.} 2021, \aap, 650, A68

\bibitem[{{Vilella-Rojo} {et~al.}(2015){Vilella-Rojo}, {Viironen}, {L{\'o}pez-Sanjuan}, {Cenarro}, {Varela}, {D{\'\i}az-Garc{\'\i}a}, {Crist{\'o}bal-Hornillos}, {Ederoclite}, {Mar{\'\i}n-Franch}, \& {Moles}}]{2015Vilella-Rojo}
{Vilella-Rojo}, G., {Viironen}, K., {L{\'o}pez-Sanjuan}, C., {et~al.} 2015, \aap, 580, A47

\end{thebibliography}
\bibliographystyle{aa}

\onecolumn
\begin{appendix} 

\section{Additional figures}
\begin{figure*}[h!]
    \includegraphics[width=9cm]{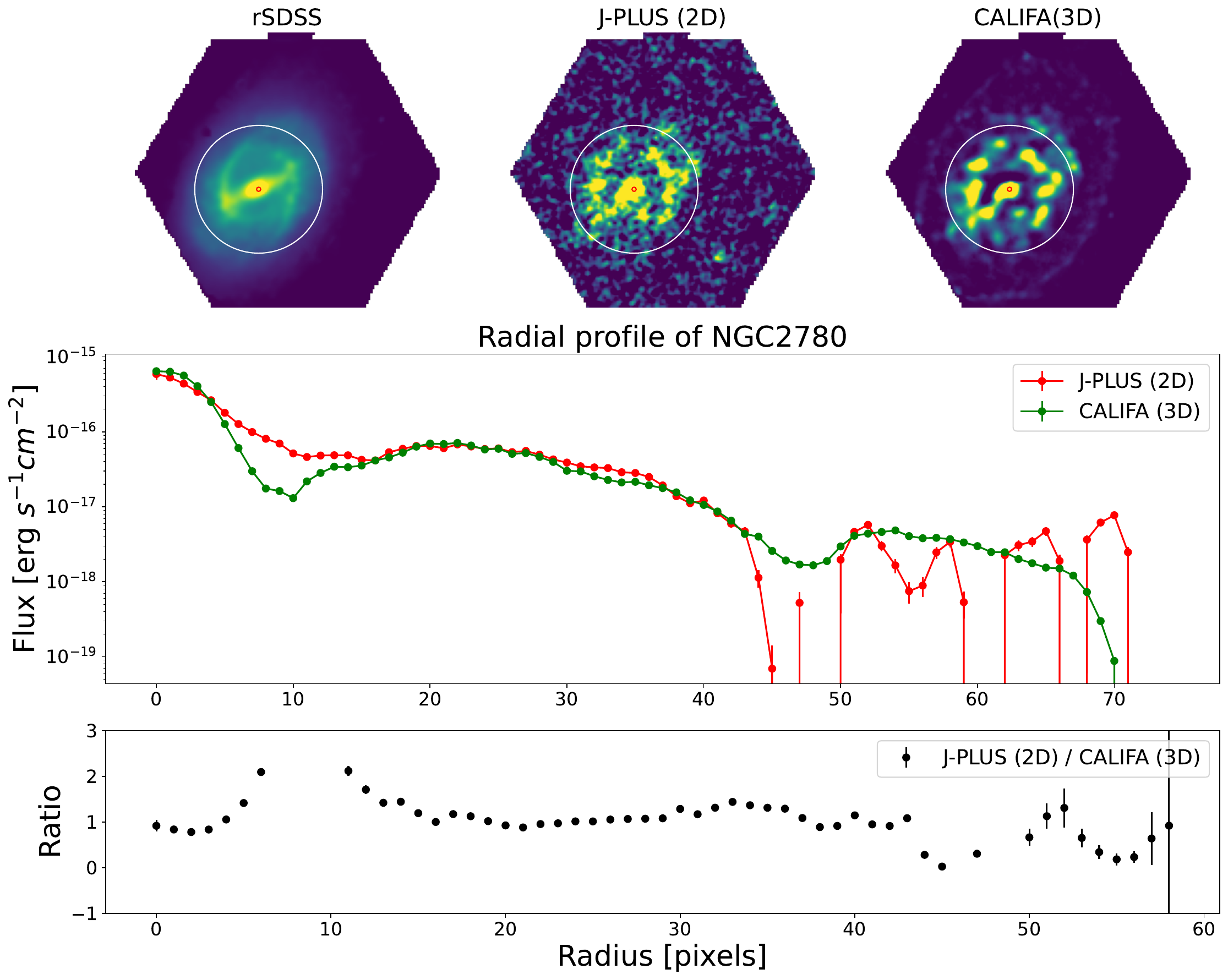}
    \includegraphics[width=9cm]{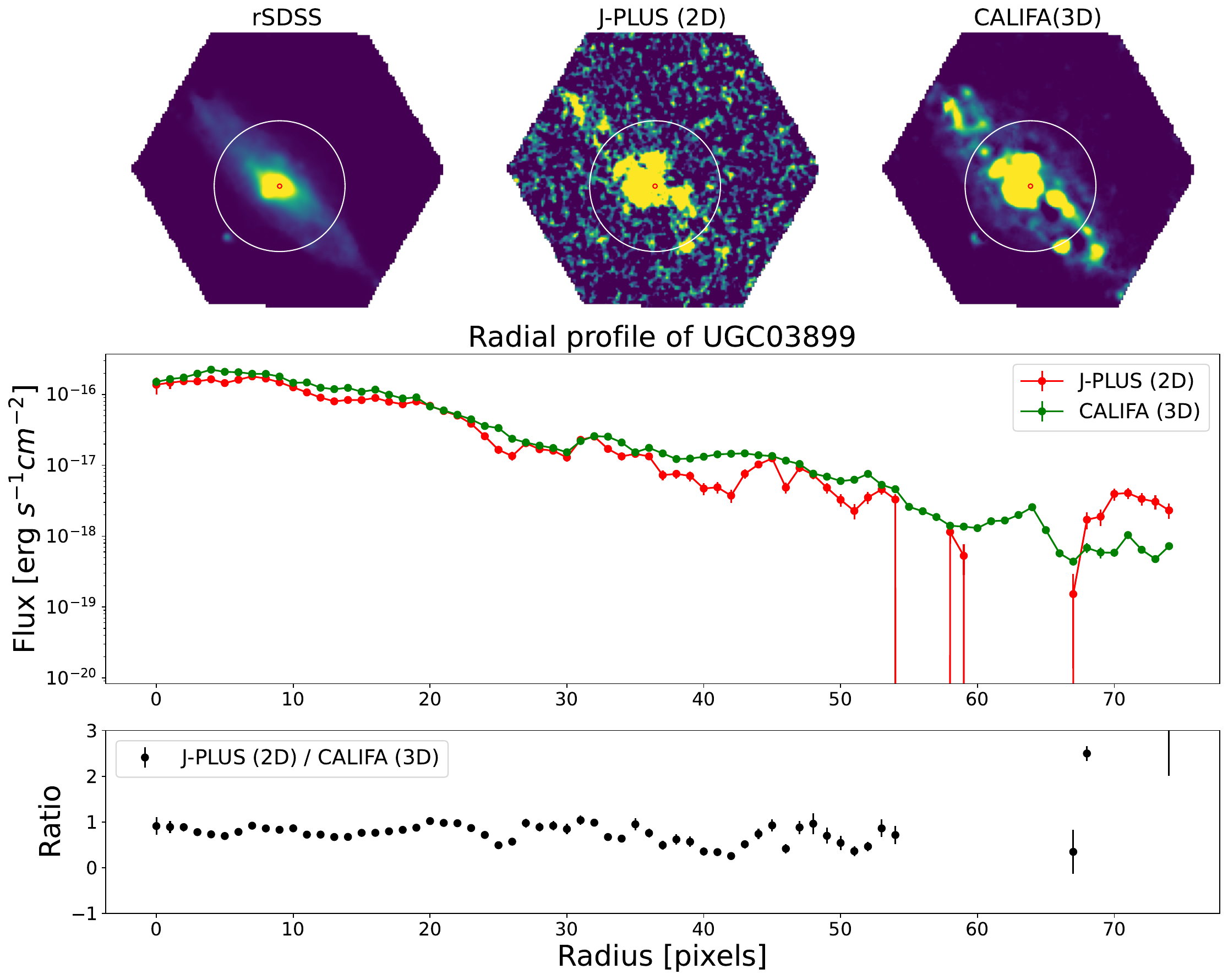}      
    \includegraphics[width=9cm]{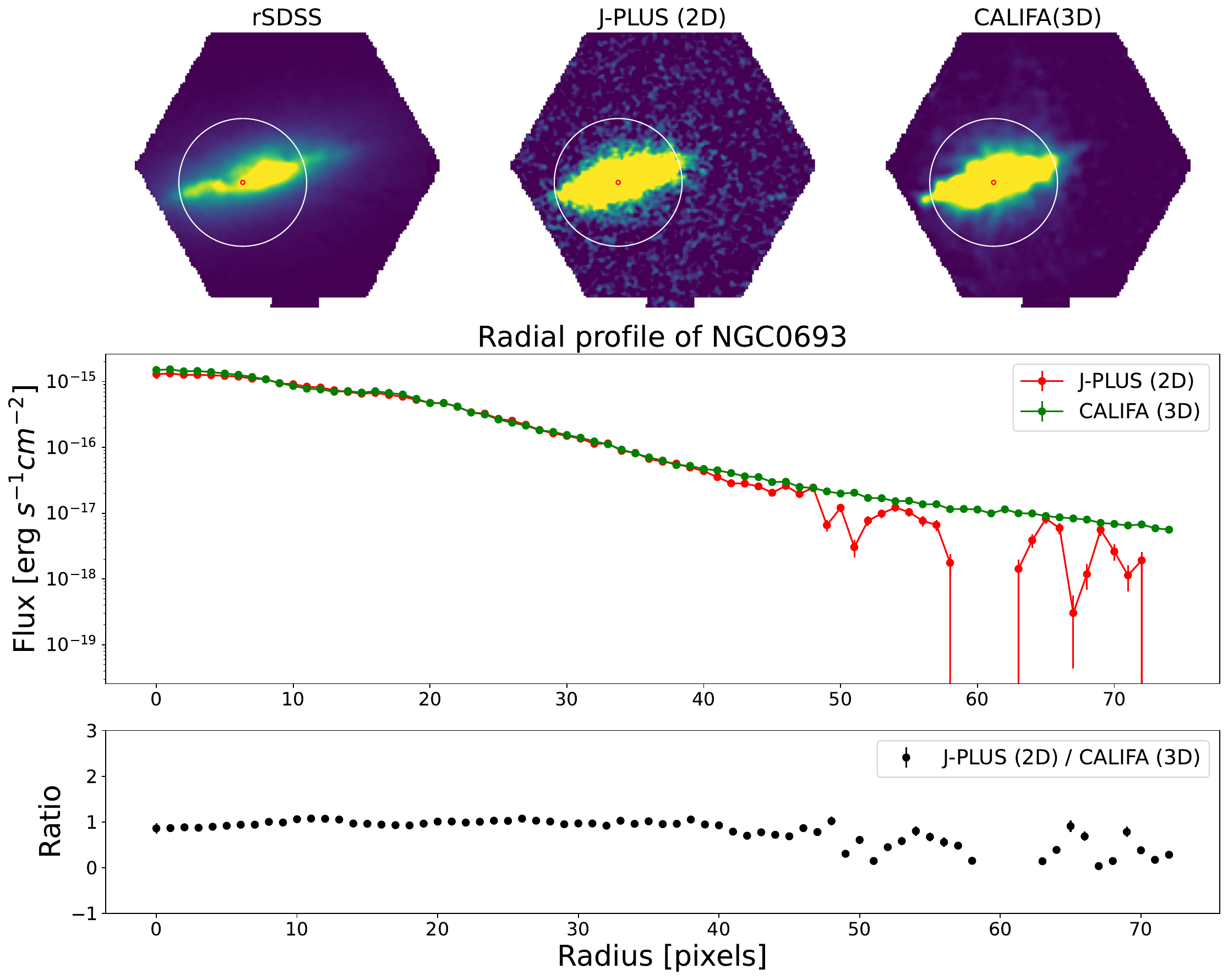}
    \includegraphics[width=9cm]{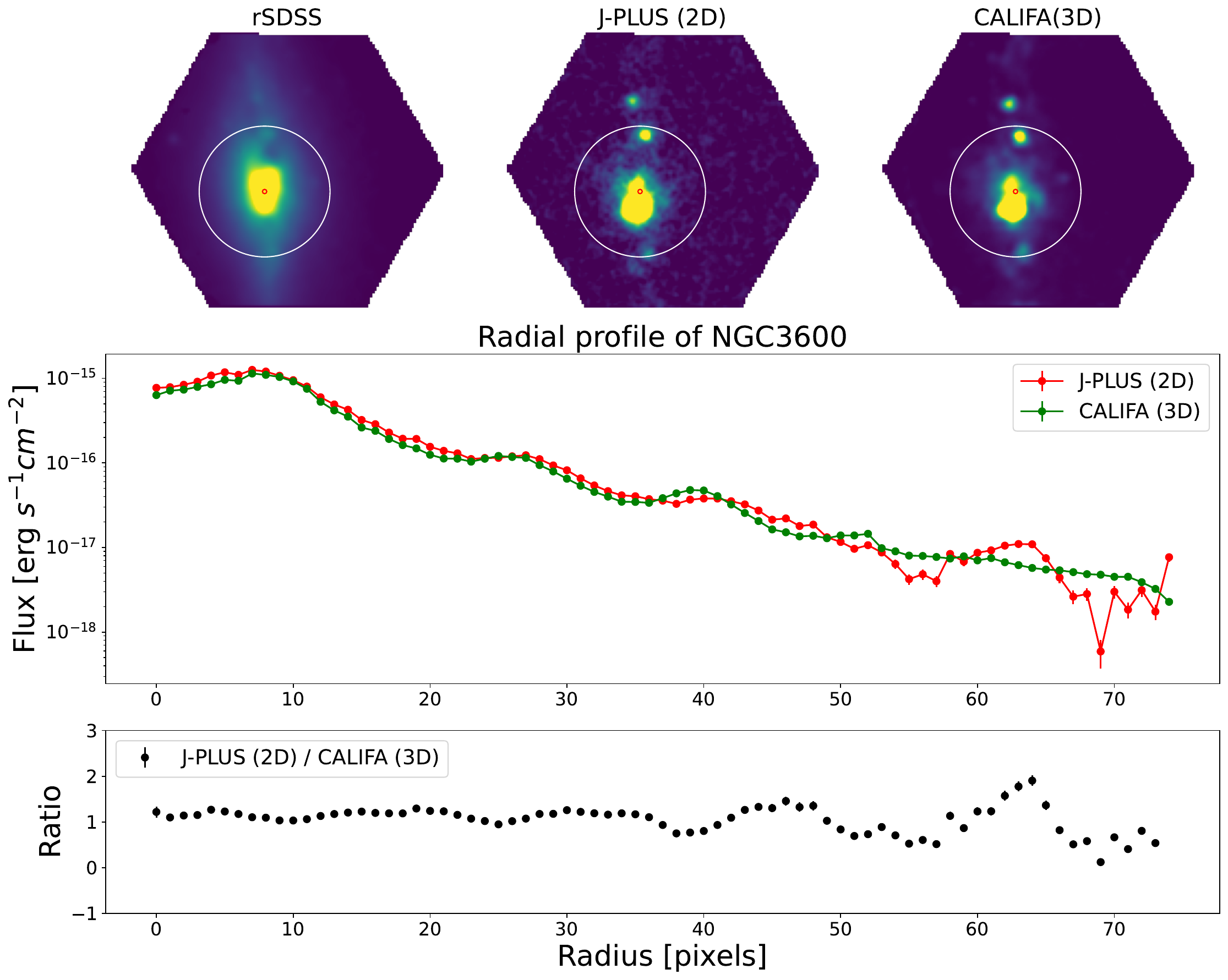}      
    \caption{Upper: rSDSS, J-PLUS (2D), and CALIFA (3D) H$\alpha$+[NII] map of galaxies. Middle: Radial profiles of H$\alpha$+[NII] maps from J-PLUS (2D), and CALIFA (3D). Lower: The ratio of the radial profiles between J-PLUS and CALIFA H$\alpha$+[NII] maps.}
   \label{fig: califa}
\end{figure*}
\vspace{-0.5cm}
\begin{figure*}
    \centering    
    \includegraphics[width=9cm]{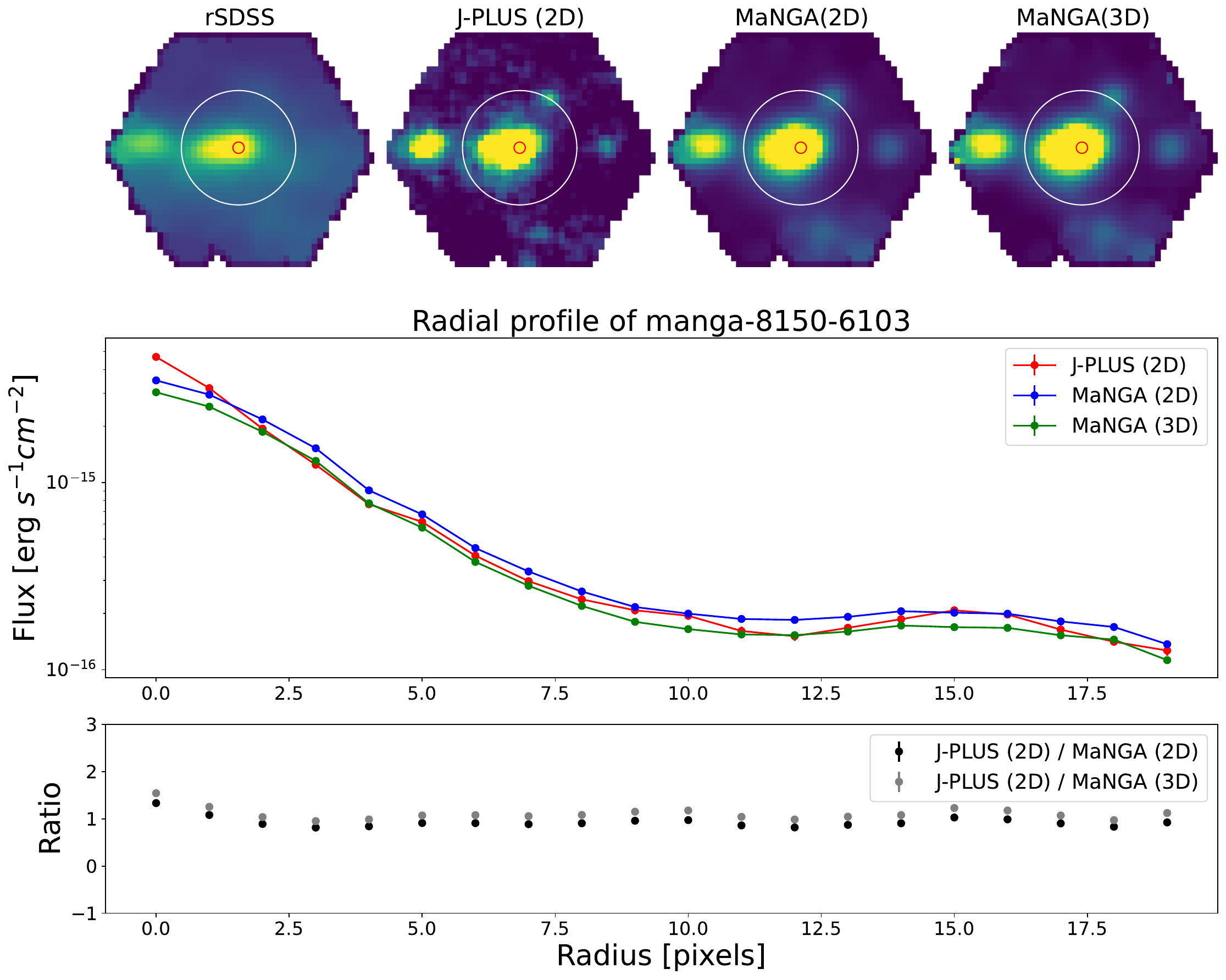} 
    \includegraphics[width=9cm]{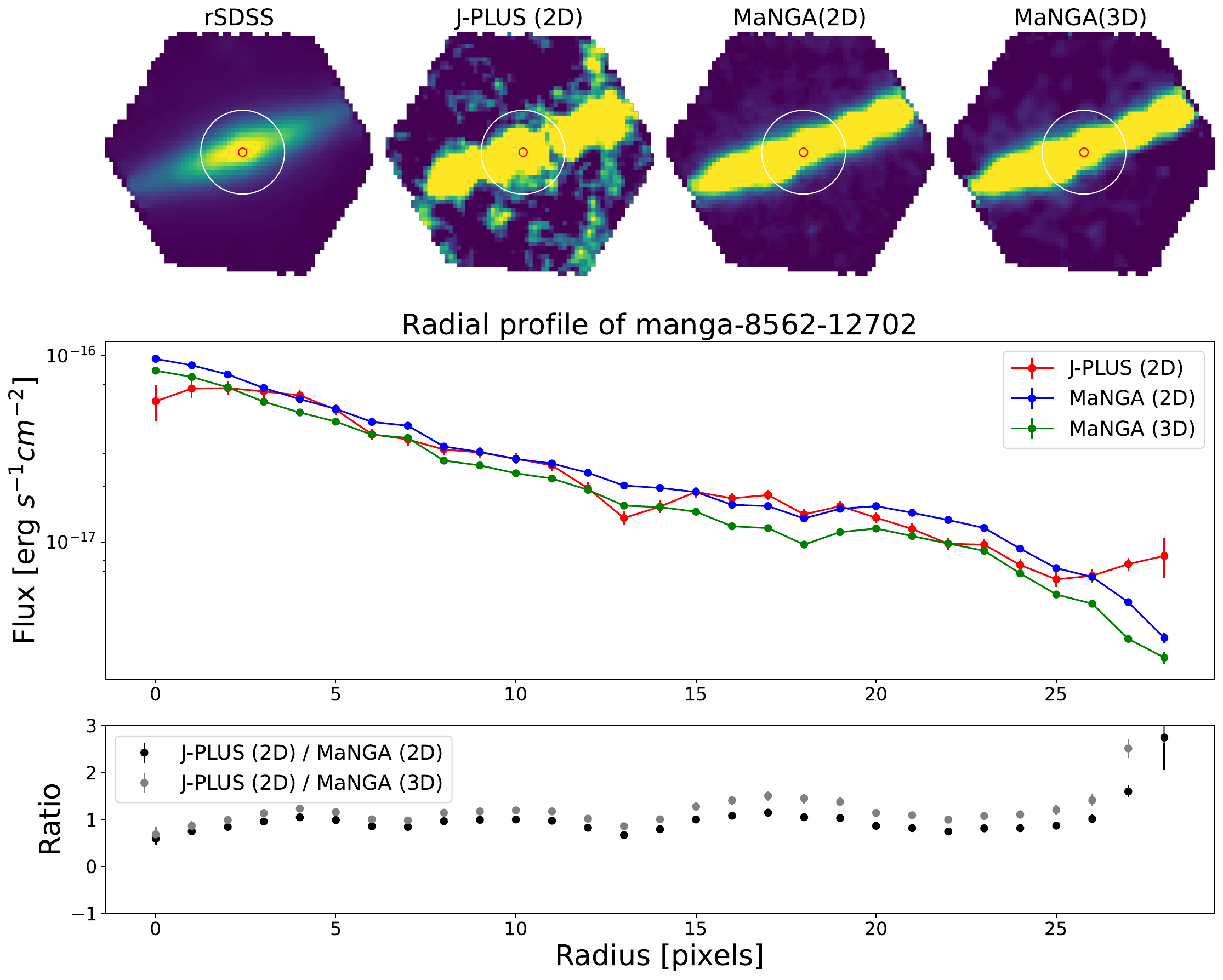} 
    \includegraphics[width=9cm]{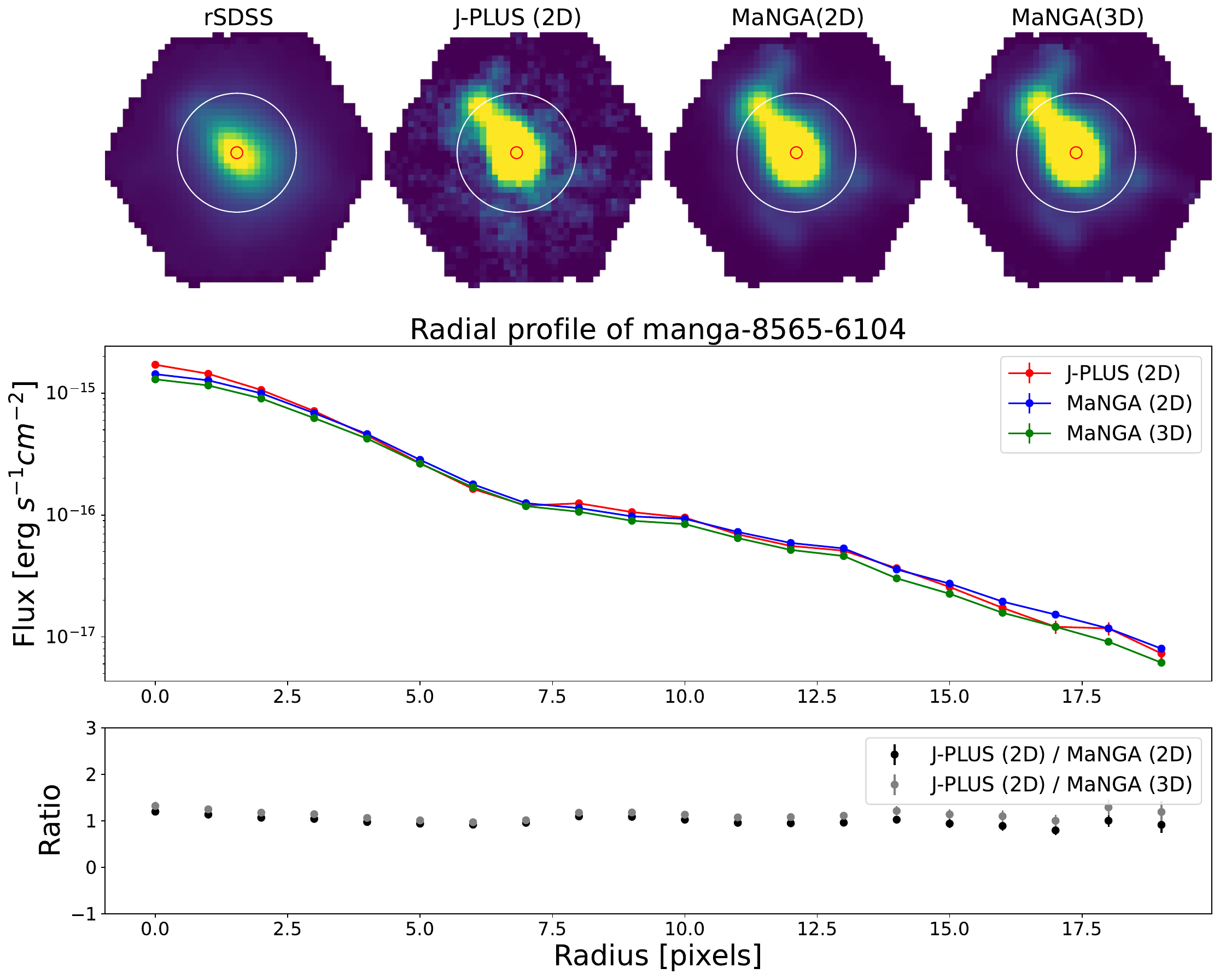}  
    \includegraphics[width=9cm]{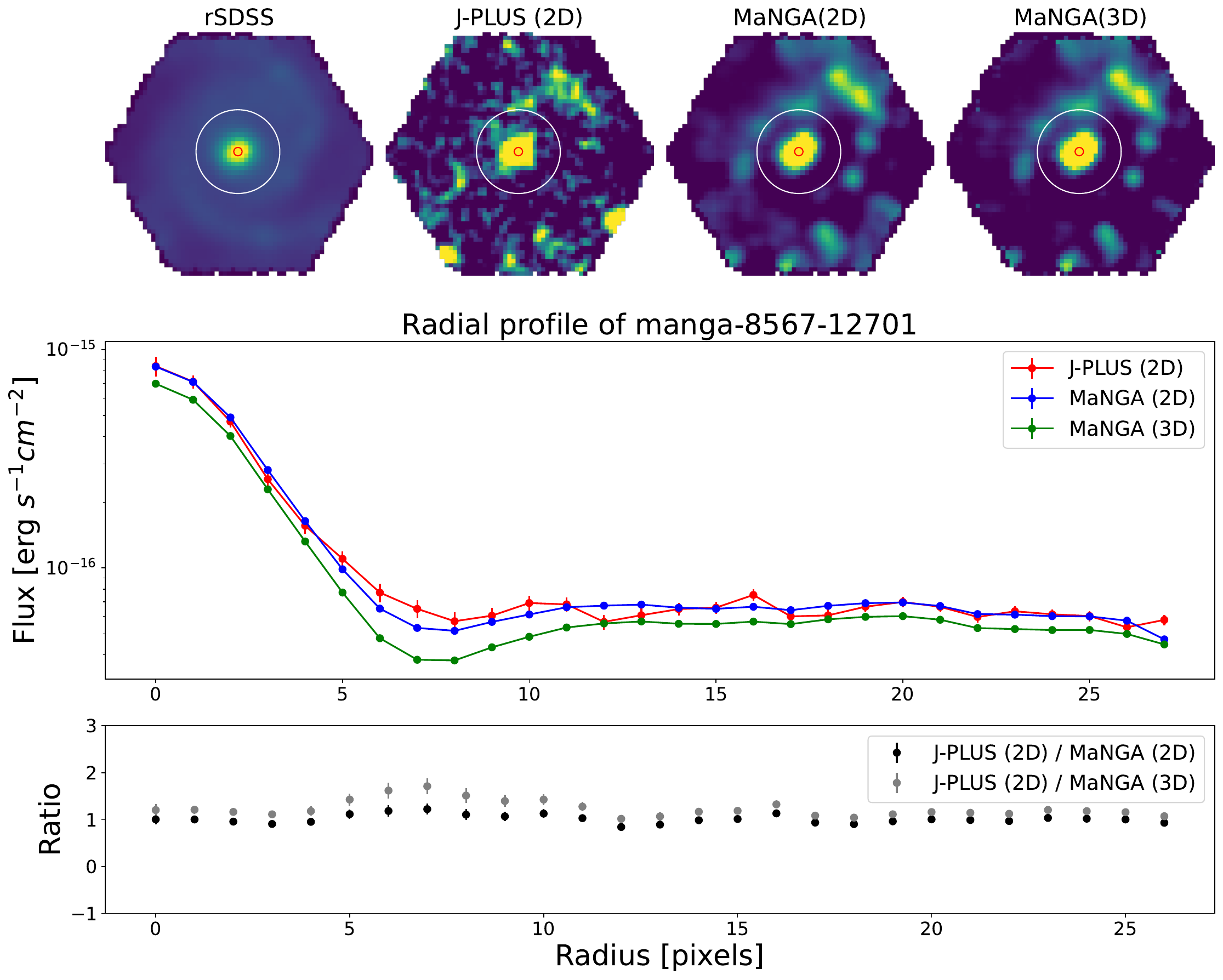} 
    \caption{Upper: rSDSS, J-PLUS (2D), MaNGA (2D) and MaNGA (3D) H$\alpha$+[NII] map of galaxies. Middle: Radial profiles of H$\alpha$+[NII] maps from J-PLUS (2D), MaNGA (2D) and MaNGA (3D). Lower: The ratio of the radial profiles between J-PLUS and MaNGA H$\alpha$+[NII] maps.}
   \label{fig: manga}
\end{figure*}
\end{appendix}
\end{document}